\documentclass[aps,prd,reprint,amsmath,amssymb,showpacs,superscriptaddress,nofootinbib]{revtex4-2}
\usepackage{epsfig,slashed,color,bm,natbib,comment,enumitem,ulem,cancel}
\def\vect#1{\mbox{\boldmath $#1$}}

\def\matri#1{{\mbox{\sf  #1}}}

\def\pd#1#2{\frac{\partial #1}{\partial #2}}
\def\bpd#1#2{\frac{\bar{\partial} #1}{\partial #2}}

\def\tfrac#1#2{{\textstyle\frac{#1}{#2}}}
\def\mom#1#2{\pd{L}{(\partial_{#1}\vpsi_{#2})}}
\def\momc#1#2{\pd{{\cal L}}{(\partial_{#1}\chi_{#2})}}

\newcommand{\vpsi}{\varphi}

\def\zero#1{{^0}\!#1}
\def\czero#1{{^0}#1}

\newcommand{\comma}{, }
\newcommand{\be}{\begin{equation}}
\newcommand{\ee}{\end{equation}}
\newcommand{\bea}{\begin{eqnarray}}
\newcommand{\eea}{\end{eqnarray}}
\begin{document}
\title{Fresh perspective on gauging the conformal group}
\author{M.P.~\surname{Hobson}}
\email{mph@mrao.cam.ac.uk}
\affiliation{Astrophysics Group, Cavendish Laboratory, JJ Thomson Avenue,
Cambridge CB3 0HE\comma UK}
\author{A.N.~\surname{Lasenby}}
\email{a.n.lasenby@mrao.cam.ac.uk}
\affiliation{Astrophysics Group, Cavendish Laboratory, JJ Thomson Avenue,
Cambridge CB3 0HE\comma UK}
\affiliation{Kavli Institute for Cosmology, Madingley Road, Cambridge
  CB3 0HA, UK\\ \phantom{AAA}}
\date{received 27 January 2021; accepted 4 April 2021}

\begin{abstract}
We consider the construction of gauge theories of gravity that are
invariant under local conformal transformations. We first clarify the
geometric nature of global conformal transformations, in both their
infinitesimal and finite forms, and the consequences of global
conformal invariance for field theories, before reconsidering existing
approaches for gauging the conformal group, namely auxiliary conformal
gauge theory and biconformal gauge theory, neither of which is
generally accepted as a complete solution. We then demonstrate that,
provided any matter fields belong to an irreducible
representation of the Lorentz group, the recently proposed extended
Weyl gauge theory (eWGT) may be considered as an alternative method
for gauging the conformal group, since eWGT is invariant under the
full set of local conformal transformations, including inversions, as
well as possessing conservation laws that provide a natural local
generalisation of those satisfied by field theories with global
conformal invariance, and also having an `ungauged' limit that
corresponds to global conformal transformations. By contrast, although
standard Weyl gauge theory also enjoys the first of these properties,
it does not share the other two, and so cannot be considered a valid
gauge theory of the conformal group.
\end{abstract}


\maketitle

\section{Introduction}
\label{sec:intro}

In the classical description of a physical system, any property has
meaning only relative to the same property of some other reference
system, and not in any absolute sense\footnote{This may not hold for
  the quantum description in the presence of quantum anomalies.}. Thus,
any measurement corresponds to calculating the ratio of two quantities
with the same units. By using `natural units', all physical quantities
can be expressed in terms of length, and so the description of
physical systems should be invariant under the group of
transformations that leave the ratios of lengths unchanged, namely the
global conformal transformations. These include Poincar\'e
transformations, which preserve length, together with global scale
changes and special conformal transformations (SCTs), all of which are
connected to the identity, and so may be considered in their
infinitesimal forms.
In addition, conformal transformations also include inversions, which
are both finite and discrete, and hence excluded from the
infinitesimal transformations.

The freedom to make an arbitrary choice of units at any point in space
and time further suggests that the description of physical systems
should, in fact, be invariant under local conformal transformations,
which therefore motivates the study of gauging the conformal group.
This is usually performed by considering only the infinitesimal
transformations, hence excluding inversions, and allowing the constant
group parameters to become arbitrary functions of position. Field
theories constructed to be invariant under these local transformations
are known as conformal gauge theories, and have been widely studied
since the 1970s as potential modified gravity theories.

One finds, however, that this standard approach to gauging the
conformal group and the resulting class of auxiliary conformal gauge
theories (ACGTs)
 \cite{Crispim77,Crispim78,Kaku77,Kaku78,Lord85,Wheeler91} suffer from
serious theoretical difficulties. Most notably, SCTs are not
represented in the final structure of AGCTs, since the corresponding
gauge field can be algebraically eliminated from the theory. More
precisely, one may show that for any self-consistent ACGT action,
the resulting field equation for the SCT gauge field can be solved and
substituted back into the action to obtain an effective action that is
independent of this gauge field, which is thus termed an {\it
  auxiliary} field (hence the name for this class of theories). Thus,
in this approach, it appears that the symmetry reduces back to the
local Weyl group.


These difficulties have motivated an alternative approach, known as
biconformal
gauging \cite{Ivanov82a,Ivanov82b,Wheeler98,Wehner99,Hazboun12,Wheeler13,Wheeler14,
  Wheeler19}, which is built on the observation that the reduction
to the local Weyl group that occurs in ACGTs is associated with
the breaking of the symmetry that exists between the generators of
translations and SCTs in the conformal algebra. Biconformal gauging
preserves this symmetry by construction, although again considers only
transformations that are connected to the identity. The
resulting biconformal gauge theories (BCGT) are successful in
circumventing many of the problems encountered in the standard
approach and have some very interesting and promising
features. Nonetheless, the resulting requirement of an eight-dimensional
base manifold complicates their physical interpretation.

Neither of these approaches is thus currently generally accepted, and
so the role of the conformal group in the construction of gauge
theories of gravity remains uncertain. In this paper, we therefore
consider an alternative approach to gauging the conformal group, which
is motivated in part by consideration of finite conformal
transformations, which are therefore not necessarily connected
to the identity and so include inversions.

Our reasons for including inversions explicitly are twofold. First,
from a physical perspective, it is well known that both the Faraday
action for the electromagnetic field and the Dirac action for a
massless spinor field are invariant not only under the elements of the
conformal group that are connected to the identity, but {\it also} under
inversions \cite{Cunningham10,Bateman10a,Bateman10b,Schouten36}. Second,
from a mathematical viewpoint, if one considers finite conformal
transformations, rather than infinitesimal ones, then the inversion
operation effectively {\it replaces} the SCT as the fourth distinct element
of the conformal group, since the SCT is merely the composition of an
inversion, a finite translation and a second inversion. Indeed, this
correspondence extends to the action of the elements of the finite
conformal group on fields, provided the latter belong to an
irreducible representation of the Lorentz group.  Moreover, the
inversion is {\it itself} the composition of a scaling and reflection, both
of which are position dependent in prescribed ways. Thus, when one
gauges the finite conformal group, the only transformations to
consider beyond those of the local Weyl group are {\it gauged reflections},
which have not been addressed previously, to our knowledge. Since
reflections are merely improper Lorentz transformations, however, they
may be localised straightforwardly by gauging the {\it full} Lorentz group,
rather than only the restricted Lorentz group that is usually
considered.  Once again, this approach extends to the action of finite
conformal transformations on fields that belong to an
irreducible representation of the Lorentz group.

These considerations suggest an alternative means of circumventing the
difficulties associated with the gauging of SCTs discussed above. In
particular, it follows that both Weyl gauge theory (WGT) and the
recently proposed extended Weyl gauge theory (eWGT) \cite{eWGTpaper}
{\it already} accommodate all the gauged symmetries of the full finite
conformal group, without the need to introduce any more gauge fields,
provided that each occurrence of a restricted Lorentz
transformation in the finite transformation laws for the covariant
derivative and gauge fields, respectively, instead denotes an element
of the full Lorentz group. In this way, both WGT and eWGT actions
constructed in the usual way are invariant under (finite) local
conformal transformations. As we will show, however, {\it only} eWGT
also possesses conservation laws that provide a natural local
generalisation of those satisfied by field theories with global
conformal invariance, and has an `ungauged' limit that corresponds to
global conformal transformations. This suggests that eWGT may be
considered as a valid alternative gauge theory of the conformal group,
whereas WGT cannot be considered as such.

The remainder of this paper is arranged as follows. In
Section~\ref{sec:gci}, we briefly outline global conformal invariance,
and clarify the geometric nature of conformal transformations, in both
their infinitesimal and finite forms; in the latter we focus
particularly on the role played by inversions. We also clarify the
requirements that global conformal invariance places on field
theories, which is sometimes unclear in the literature. In
Section~\ref{sec:lci}, we briefly discuss the principles underlying
the gauging of a spacetime symmetry group, and in particular compare
Kibble's original method with the more recently employed quotient
manifold method. We then reconsider previous approaches to gauging the
conformal group, focussing first on applying Kibble's approach
directly to construct ACGT, which is rarely presented in the
literature, and summarising its theoretical shortcomings mentioned
above. We also discuss the concept of `ungauging'
\cite{Hehl78,Lord86b}, by which one seeks to identify the symmetry
group underlying any given gauge theory, and propose some
modifications to the existing approach before applying it to ACGT. We
then give a very brief outline of BCGT. In
Section~\ref{sec:newapproach}, we summarise eWGT, focussing in
particular on the forms of the covariant derivative, field strengths,
action, field equations and conservation laws, all of which illustrate
its very different structure from other gravitational gauge
theories. We then demonstrate our central point that, with the
above-mentioned modest extension to the transformation laws of the
gauge fields, both eWGT and WGT are invariant under the full set of
finite local conformal transformations, including inversions, but that
only eWGT also possesses conservation laws that provide a natural
local generalisation of those satisfied by field theories with global
conformal invariance, and has an `ungauged' limit that corresponds to
global conformal transformations. We conclude in
Section~\ref{sec:conc}. In addition, in Appendix~\ref{app:a}, we
include a simple derivation of the finite forms for the action on the
coordinates of every element of the conformal group (not just those
connected to the identity) {\it directly} from their defining
requirement, without integrating infinitesimal forms. Finally, in
Appendix~\ref{app:b}, we present a brief outline of the consequences
of general global and local symmetries for field theories, focussing
in particular on Noether's first and second theorems, the latter being
discussed surprisingly rarely in the literature.


\section{Global conformal invariance}
\label{sec:gci}

In Minkowski spacetime, conformal coordinate transformations $x^\mu
\to x^{\prime\mu}$ from some Cartesian inertial coordinate system
$x^\mu$ are those that leave the light cone (and hence causal
structure) invariant, such that
\be
ds^2 = \eta_{\mu\nu}\,dx^\mu\,dx^\nu = 
\Omega^2(x(x'))\eta_{\mu\nu}\,dx^{\prime\mu}\,dx^{\prime\nu},
\label{eq:ctdef}
\ee
for some (real) function $\Omega(x)$. Indeed, more generally,
conformal transformations preserve the `angle' or inner product
between any two vectors, which is equivalent to the invariance of the
ratio of their lengths.

\subsection{Infinitesimal global conformal transformations}
\label{sec:infgct}

We first briefly discuss infinitesimal global conformal
transformations, since these are the more commonly considered form in
the literature. This allows us both to establish our notation and to
clarify the geometric meaning of the transformations in a manner that
enables a transparent extension to their finite forms presented in
Section~\ref{sec:finitegct} below.

For an infinitesimal coordinate transformation $x^{\prime\mu} = x^\mu
+ \xi^\mu(x)$ to satisfy (\ref{eq:ctdef}), it is readily established that
one requires
\be
\partial_{(\alpha}\xi_{\beta)} = \tfrac{1}{n}(\partial_\mu\xi^\mu)\eta_{\alpha\beta},
\label{killingeq}
\ee
which is the conformal Killing equation in $n$-dimensional Minkowski
spacetime, although hereinafter we will concentrate exclusively on the
case $n=4$. One may show in the usual (albeit rather intricate) manner
that the most general solution for $\xi^\mu(x)$ has the form
\be
\xi^\mu(x) = a^\mu + {\omega^\mu}_\nu x^\nu + \rho x^\mu +
c^\mu x^2 -2 c\cdot x\, x^\mu,
\label{eq:gct}
\ee
where the 15 infinitesimal parameters $a^\mu$, $\omega^{\mu\nu} =
-\omega^{\nu\mu}$, $\rho$ and $c^\mu$ are {\it constants}, i.e.\ not
functions of spacetime position, and we use the shorthand notation
$x^2 \equiv \eta_{\mu\nu} x^\mu x^\nu$ and $c\cdot x \equiv
\eta_{\mu\nu}c^\mu x^\nu$, in which
$\eta_{\mu\nu}=\mbox{diag}(1,-1,-1,-1)$ is the Minkowski metric in
Cartesian inertial coordinates. If the four parameters $c^\mu$ defining
the so-called special conformal transformation (SCT) vanish, then
(\ref{eq:gct}) reduces to an infinitesimal global Weyl
transformation. Moreover, if the parameter $\rho$ defining the
dilation (or scale transformation) also vanishes, then (\ref{eq:gct})
further reduces to an infinitesimal global Poincar\'e transformation,
consisting of a restricted Lorentz rotation defined
by the six parameters $\omega^{\mu\nu}$ and a spacetime translation
defined by the four parameters $a^\mu$. The group of global Poincar\'e
transformations is the isometry group of Minkowski spacetime, such
that $\Omega^2(x)=1$ in (\ref{eq:ctdef}).

A more intuitive geometric interpretation of an infinitesimal global
conformal transformation may be arrived at directly using
(\ref{killingeq}), from which one may show that
\be
\partial_\beta\xi^\alpha =  {\varpi^\alpha}_{\beta}
+ \varrho\delta^\alpha_\beta,
\label{eq:gctinterp1}
\ee
where $\varpi^{\alpha\beta} = -\varpi^{\beta\alpha}$ and
$\varrho$ represent (in general) a position-dependent infinitesimal
rotation and dilation, respectively, which must satisfy the conditions
(assuming dimensionality $n \ge 3$)
\begin{subequations}
\label{eq:gctinterpall}
\bea
\partial_\mu\varpi^{\alpha\beta} -
2\delta^{[\alpha}_\mu\partial^{\beta]}\varrho &=& 0, 
\label{eq:gctinterp1a}\\
\partial_\alpha\partial_\beta\varrho & = & 0.
\label{eq:gctinterp2}
\eea
\end{subequations}
Successive integration of equations
(\ref{eq:gctinterp1}--\ref{eq:gctinterpall}), from the last to the
first, then yields $\varrho(x)=\rho-2c\cdot x$,
$\varpi^{\alpha\beta}(x)=\omega^{\alpha\beta} + 4c^{[\alpha}
  x^{\beta]}$ and the expression (\ref{eq:gct}) for $\xi^\alpha(x)$,
as before,
where $\rho$, $\omega^{\alpha\beta}$ and $c^\alpha$ are again
constants.

The action of an infinitesimal conformal transformation on some field
$\vpsi(x)$ defined on the spacetime may be determined by first
considering the 11-parameter (little) subgroup, say $H(1,3)$, of the
conformal group $C(1,3)$, obtained by setting $a^\mu=0$, which leaves
the origin $x^\mu=0$ invariant. Its generator matrices
$\Sigma_{\mu\nu}$, $\Delta$ and $\kappa_\mu$, corresponding to Lorentz
rotations, dilations and SCTs, respectively, satisfy the commutation
relations
\bea
\left[\Sigma_{\mu\nu},\Sigma_{\rho\sigma}\right]
&=& \eta_{\mu\sigma}\Sigma_{\nu\rho}-\eta_{\nu\sigma}\Sigma_{\mu\rho}+\eta_{\nu\rho}
\Sigma_{\mu\sigma}-\eta_{\mu\rho}\Sigma_{\nu\sigma},\nonumber \\
\left[\Sigma_{\mu\nu},\kappa_\rho\right] & = & \eta_{\nu\rho}\kappa_\mu -
\eta_{\mu\rho}\kappa_\nu,\qquad \left[\Sigma_{\mu\nu},\Delta\right]=0,\nonumber\\
\left[\Delta,\Delta\right] &=& 0,
\quad\left[\kappa_\mu,\kappa_\nu\right] = 0,\quad \left[\kappa_\mu,\Delta\right] =
\kappa_\mu.
\label{eq:hgenerators}
\eea
Using the method of induced representations, one may then show that
the action of a full infinitesimal conformal transformation on some
field $\vpsi(x)$ leads to a `form' variation
$\delta_0\vpsi(x)\equiv\vpsi'(x)-\vpsi(x)$ given by\footnote{For an
  infinitesimal transformation $x^{\prime\mu} = x^\mu + \xi^\mu(x)$,
  the `form' variation $\delta_0\vpsi(x)\equiv\vpsi'(x)-\vpsi(x)$ is
  related to the `total' variation
  $\delta\vpsi(x)\equiv\vpsi'(x')-\vpsi(x)$ by $\delta_0\vpsi(x) =
  \delta\vpsi(x) - \xi^\mu\partial_\mu\vpsi(x) = \delta\vpsi(x) +
  \xi^\mu P_\mu\vpsi(x)$.}
\be
\delta_0\vpsi(x) = (a^\mu P_\mu+\tfrac{1}{2}\omega^{\mu\nu}M_{\mu\nu}+\rho D +
c^\mu K_\mu)\vpsi(x),
\label{eq:gctformvar}
\ee
where the 15 generators of the conformal group $C(1,3)$ have the forms
\begin{subequations}
\label{eq:gcggenerators}
\bea
P_\mu &=& -\partial_\mu, 
\\
M_{\mu\nu} &=& x_\mu\partial_\nu - x_\nu\partial_\mu + \Sigma_{\mu\nu},
\\
D &=& -x\cdot\partial + \Delta,
\\
K_\mu &=& (2x_\mu\,x\cdot\partial -
x^2\partial_\mu)+2(x^\nu\Sigma_{\mu\nu}-x_\mu\Delta) + \kappa_\mu,\phantom{AA}
\label{eq:gcggeneratorsd}
\eea
\end{subequations}
which correspond to translations, Lorentz rotations, dilations and
SCTs, respectively. Note that the generator of the SCT can be
expressed in terms of (parts of) the other generators as $K_\mu =
x^2P_\mu + 2x^\nu\Sigma_{\mu\nu}-2x_\mu D + \kappa_\mu$. The
generators (\ref{eq:gcggenerators}) satisfy the commutation relations
\bea
\left[M_{\mu\nu},M_{\rho\sigma}\right]
&=& \eta_{\mu\sigma}M_{\nu\rho}-\eta_{\nu\sigma}M_{\mu\rho}+\eta_{\nu\rho}
M_{\mu\sigma}-\eta_{\mu\rho}M_{\nu\sigma},\nonumber \\
\left[M_{\mu\nu},P_\rho\right] 
& = & \eta_{\nu\rho}P_\mu - \eta_{\mu\rho}P_\nu,\nonumber \\
\left[M_{\mu\nu},K_\rho\right] & = & \eta_{\nu\rho}K_\mu -
\eta_{\mu\rho}K_\nu,\qquad \left[M_{\mu\nu},D\right]=0,\nonumber\\
\left[P_\mu,K_\nu\right] &=& 2(M_{\mu\nu}+\eta_{\mu\nu}D),\qquad
\left[D,D\right] = 0,\nonumber \\
\left[P_\mu,P_\nu\right] &=& 0,\qquad \left[P_\mu,D\right] = -P_\mu,\nonumber \\
\left[K_\mu,K_\nu\right] &=& 0,\qquad \left[K_\mu,D\right] =
K_\mu,
\label{eq:cgcomms}
\eea
which define the Lie algebra of the conformal group. Note that, as
expected, one recovers the Lie algebra of the Weyl group $W(1,3)$ by
ignoring commutators containing $K_\mu$, and if one also ignores
commutators containing $D$ one recovers the Lie algebra of the
Poincar\'e group $P(1,3)$.

From (\ref{eq:gct}) and (\ref{eq:gctformvar}), it is straightforward
to show that the form variation can also be written as
\be
\delta_0\vpsi(x) =
[\xi^\mu(x) P_\mu+\tfrac{1}{2}\varpi^{\mu\nu}(x)\Sigma_{\mu\nu}
+\varrho(x)\Delta +
c^\mu \kappa_\mu]\vpsi(x),
\label{eq:gctformvar2}
\ee
where $\xi^\mu(x)$ is given by (\ref{eq:gct}),
$\varpi^{\mu\nu}(x)=\omega^{\mu\nu} + 4c^{[\mu} x^{\nu]}$ and
$\varrho(x)=\rho-2c\cdot x$, which are clearly all functions of
spacetime position. The form variation (\ref{eq:gctformvar2}) is of
particular interest when the field $\vpsi(x)$ belongs to an
irreducible representation of the Lorentz group, since the action of
the conformal group is considerably simplified because the matrix
generators $\Delta$ and $\kappa_\mu$ have particularly simple
forms. First, according to Schur's lemma, any matrix that commutes
with the generators $\Sigma_{\mu\nu}$ must be a multiple of the
identity. Indeed, one has $\Delta = wI$, where $I$ is the identity and
$w$ is the Weyl weight (or scaling dimension) of the field
$\vpsi(x)$. Then, from $\left[\kappa_\mu,\Delta\right] = \kappa_\mu$,
one finds that $\kappa_\mu=0$. In this case, with $\Delta = wI$ and
$\kappa_\mu=0$, it is worth noting that the form variation
(\ref{eq:gctformvar2}) may be considered as a particular example of an
infinitesimal {\it local} Weyl transformation, consisting of the
combination of particular forms of position-dependent translation,
rotation and dilation; this is consistent with the geometric
interpretation of an infinitesimal global conformal transformation
expressed in (\ref{eq:gctinterp1}).

Fields that transform according to (\ref{eq:gctformvar2}) under a
conformal transformation are called primary fields. There also exist
non-primary fields, the most important example of which is the
derivative $\partial_\mu\vpsi(x)$ of a primary field. It is
straightforward to show that
\be \delta_0(\partial_\mu\vpsi) = \Theta\partial_\mu\vpsi -
({\varpi^\nu}_{\mu}+\varrho\delta^\nu_\mu)\partial_\nu\vpsi 
+2(c^\nu\Sigma_{\nu\mu}-c_\mu\Delta)\vpsi,
\label{eq:cgtformvard}
\ee
where $\Theta \equiv \xi^\alpha P_\alpha+ \tfrac{1}{2}\varpi^{\alpha\beta}\Sigma_{\alpha\beta}
+\varrho\Delta + c^\alpha \kappa_\alpha $ is the quantity
in square brackets on the RHS of (\ref{eq:gctformvar2}), with
generators appropriate to the nature of $\vpsi(x)$, and the final term
on the RHS of (\ref{eq:cgtformvard}) shows that $\partial_\mu\vpsi(x)$
is non-primary. Indeed, although the transformation law is linear,
this final term also means that it is inhomogeneous.  It is clear that
this behaviour results solely from the SCT; if $c^\mu=0$, and hence
$\varpi^{\mu\nu}=\omega^{\mu\nu}$ and $\varrho=\rho$, one
recovers a global Weyl transformation and (\ref{eq:cgtformvard})
becomes homogeneous and of an analogous form to the transformation law
(\ref{eq:gctformvar2}) of the original field $\vpsi(x)$ with
$c^\mu=0$, once the generators have been augmented to accommodate the
additional vector index on the partial derivative $\partial_\mu$.

\subsection{Finite global conformal transformations}
\label{sec:finitegct}

We now discuss the finite form of global conformal transformations,
which are central to our later considerations, but are not often
described in the gauge theory literature. We again seek to clarify the
geometric nature of the transformations, deriving novel (to our
knowledge) finite conditions that reduce to equations
(\ref{eq:gctinterp1}) and (\ref{eq:gctinterpall}) in the infinitesimal
limit, and focus in particular on the role played by
inversions.

The finite forms for the action on the coordinates of the elements of
the conformal group $C(1,3)$ corresponding to translations,
restricted Lorentz rotations (jointly Poincar\'e
transformations) and dilations (jointly Weyl transformations) are
easily found by obtaining the integral curves of the corresponding
infinitesimal expressions.  Again denoting the 15 now finite constant
parameters of the group by $a$, $\omega$, $\rho$ and $c$, these finite
forms are, respectively,
\be 
x^{\prime\mu} = x^\mu + a^\mu,\quad
x^{\prime\mu} = {\Lambda^\mu}_\nu(\omega) x^\nu,\quad
x^{\prime\mu} = e^\rho x^\mu,
\ee
where ${\Lambda^\mu}_\nu(\omega)$ is a restricted
Lorentz transformation matrix satisfying
$\eta_{\mu\nu}{\Lambda^\mu}_\rho{\Lambda^\nu}_\sigma =
\eta_{\rho\sigma}$ and $\mbox{det}\,{\Lambda^\mu}_\nu= 1$.

The same procedure can be used to find the finite form for the action
of a SCT, but more geometrical insight is obtained by first
introducing the inversion transformation, which may be taken to have
the form
\be
x^{\prime\mu} = \frac{x^\mu}{x^2},
\label{eq:inversion}
\ee
where $x^2 = \eta_{\mu\nu}x^\mu x^\nu \neq 0$. This discrete
transformation is clearly also an element of the full conformal group
(although not one connected to the identity), since the new (Minkowski
spacetime) metric is given by $\gamma'_{\mu\nu}(x) =
\eta_{\mu\nu}/(x^2)^2$. If one then considers the composite
transformation consisting of an inversion, followed by a finite
translation through $c^\mu$, followed by a second inversion, one
finds
\be
x^{\prime\mu} = \frac{x^\mu + c^\mu x^2}{1+2c\cdot x + c^2 x^2},
\label{eq:finitesct}
\ee
which reduces to the infinitesimal SCT in (\ref{eq:gct}) for small
$c^\mu$. Since every smooth conformal transformation of a
pseudo-Euclidean (Euclidean) space of dimension $n \ge 3$ can be
represented as a composition of isometry, dilation and
inversion \cite{Dubrovin79}, the expression (\ref{eq:finitesct}) must
represent a finite SCT. It is worth mentioning that, although both the
numerator and denominator of (\ref{eq:finitesct}) vanish for $x^\mu =
-c^\mu/c^2$, this point is mapped to infinity and hence the finite SCT
is not defined globally. Indeed, in order to define finite SCTs 
globally, one must consider a
conformally compactified Minkowski spacetime, which includes an
additional special point at infinity and its null cone, but we
will not consider this subtlety any further here.

Although not usually presented in the literature, one can in fact
derive the finite forms for the action on the coordinates of every
element of the conformal group (not just those connected to the
identity) {\it directly} from the defining requirement (\ref{eq:ctdef}),
without having to integrate infinitesimal forms. As just mentioned,
one may consider every smooth conformal transformation as a
composition of isometry, dilation and inversion. The coordinate
transformation matrix for the inversion (\ref{eq:inversion}) is given
by
\be {X^\mu}_\nu \equiv \pd{x^{\prime\mu}}{x^\nu} 
= \frac{1}{x^2}
\left(\delta^\mu_\nu-\frac{2x^\mu x_\nu}{x^2}\right) \equiv
\frac{1}{x^2}{I^\mu}_\nu(\hat{x}),
\label{eq:invtmat}
\ee
which one may identify \cite{deWitt03} as the composition of a
(position-dependent) dilation $1/x^2$ and 
reflection ${I^\mu}_\nu(\hat{x})$ in the hyperplane perpendicular to
the unit vector $\hat{x}\equiv x^\mu/\sqrt{|x^2|}$.  It is
straightforward to show that ${I^\mu}_\nu(\hat{x})$ is an improper
Lorentz transformation matrix, satisfying
$\eta_{\mu\nu}{I^\mu}_\rho(\hat{x}){I^\nu}_\sigma(\hat{x}) =
\eta_{\rho\sigma}$ and $\mbox{det}\,{I^\mu}_\nu(\hat{x})= -1$. Thus,
with no loss of generality, one may write the transformation matrix of
any smooth finite conformal transformation in the form
\be
{X^\mu}_\nu = \Omega(x){\Lambda^\mu}_\nu(x),
\label{eq:fgctmat}
\ee
where, in general, ${\Lambda^\mu}_\nu(x)$ represents a
position-dependent finite Lorentz transformation (either proper or
improper) and $\Omega(x)$ represents a position-dependent finite
dilation; indeed, one sees immediately that (\ref{eq:fgctmat})
satisfies the defining requirement (\ref{eq:ctdef}). As shown in
Appendix~\ref{app:a}, one may further use (\ref{eq:ctdef}) to derive
conditions on ${\Lambda^\mu}_\nu(x)$ and $\Omega(x)$, which may be
written as (for dimensionality $n \ge 3$)
\begin{subequations}
\label{eq:fgctrel}
\bea
{\Lambda_\gamma}^\alpha\partial_\mu\Lambda^{\gamma\beta} - 
2\delta^{[\alpha}_\mu\partial^{\beta]}\ln\Omega & = & 0, \label{eq:fgctrel1}\\
2\Omega\partial_\alpha\partial_\beta\Omega
+\eta_{\alpha\beta}(\partial_\gamma\Omega)(\partial^\gamma\Omega)-
4(\partial_\alpha\Omega)(\partial_\beta\Omega) & = & 0.\phantom{AA}\label{eq:fgctrel2}
\eea
\end{subequations}
It is straightforward to check that, on writing $x^{\prime\mu} \approx
x^\mu + \xi^\mu(x)$, ${\Lambda^\mu}_\nu(x) \approx \delta^\mu_\nu +
{\varpi^\mu}_{\nu}(x)$ and $\Omega(x)=e^{\varrho(x)}
\approx 1+\varrho(x)$, the expressions
(\ref{eq:fgctmat}--\ref{eq:fgctrel}) reduce correctly in the
infinitesimal limit to those given in
(\ref{eq:gctinterp1}--\ref{eq:gctinterpall}). Moreover, as discussed
further in Appendix~\ref{app:a}, the conditions
(\ref{eq:fgctrel}) may be used to determine
directly the action on coordinates of the four distinct
finite elements of the conformal group, namely position-independent
translations, rotations and scalings, together with
inversions (and hence also SCTs).

In the space of fields, the action of the finite elements of the
conformal group that are connected to the identity (thus excluding
inversions) is formally given by the exponential of the operator
appearing on the RHS of (\ref{eq:gctformvar}), such that for some
field $\vpsi(x)$ one has
\be 
\vpsi'(x) = \exp(a^\mu
P_\mu+\tfrac{1}{2}\omega^{\mu\nu}M_{\mu\nu}+\rho D 
+ c^\mu K_\mu)\vpsi(x), 
\ee
where the group parameters $a$, $\omega$, $\rho$ and $c$ are
now finite constants.

It is again of particular interest to consider the case of some physical
field $\vpsi(x)$ belonging to an irreducible representation of the
Lorentz group. Setting $c_\mu = 0$ for the moment, thereby neglecting
the finite SCTs and considering just the Weyl group, one may describe
the action of a finite transformation as
\be
\vpsi'(x') = e^{w\rho}\matri{S}(\omega)\vpsi(x),
\label{eq:gwt}
\ee
where $\matri{S}(\omega)$ is the matrix corresponding to the element
$\omega$ of the restricted Lorentz group (or SL(2,C)
group) in the representation to which $\vpsi(x)$ belongs (we have
suppressed Lorentz indices on these objects for notational
simplicity), and $w$ is the Weyl (or conformal) weight of the field
$\vpsi(x)$. Indeed, we have adopted the form (\ref{eq:gwt}) for the
action of finite global Weyl transformations on physical fields in our
previous work \cite{eWGTpaper}.

To determine the explicit form for the action of a finite SCT on some
such physical field $\vpsi(x)$, it is again convenient to consider
first the action of an inversion, for which the transformation matrix
is given by (\ref{eq:invtmat}). Thus, the action of an inversion on,
for example, a vector field $V^\mu(x)$ of Weyl (scaling) weight $w$,
is given by
\be
V^{\prime\mu}(x') = (x^2)^{-w}{I^\mu}_\nu(\hat{x})V^\nu(x).
\ee
Analogous transformation laws hold for higher-rank tensor fields.  One
may also show that the action of an inversion on a spinor field
$\psi(x)$ of Weyl weight $w$ is given by
\be
\psi'(x') = (x^2)^{-w}(\gamma\cdot\hat{x})\psi(x),
\ee
where $\gamma = \{\gamma^\mu\}$ denotes the set of Dirac matrices.  It
is worth noting that the quantities ${I^\mu}_\nu(\hat{x})$ and
$\gamma\cdot\hat{x}$ have previously been identified \cite{Ho11} as
matrices that decouple the Lorentz indices on tensor and spinor
fields, respectively, from SCTs, but surprisingly without giving their
geometric interpretation as a reflection in the hyperplane
perpendicular to the unit vector $\hat{x}$. It is straightforward to
show that these two decoupling matrices are related by the useful
formula
\be
(\gamma\cdot\hat{x})\gamma^\mu(\gamma\cdot\hat{x}) 
= -{I^\mu}_\nu(\hat{x})\gamma^\nu.
\ee

Recalling that a SCT is the composition of an inversion, followed by a
translation $c^\mu$, followed by a second inversion, it is now
straightforward to show that the action of a finite SCT on, for
example, a vector field or spinor field of Weyl weight $w$ is given
by, respectively,
\begin{subequations}
\label{eq:finitesctfield}
\bea
V^{\prime\mu}(x') & = & [\sigma(x,c)]^{-w}{I^\mu}_\alpha(\hat{x}')
{I^\alpha}_\nu(\hat{x})V^\nu(x), \\
\psi'(x') & = &
     [\sigma(x,c)]^{-w}(\gamma\cdot\hat{x}')(\gamma\cdot\hat{x})\psi(x),
\eea
\end{subequations}
where we have defined the (inverse) scaling $\sigma(x,c)\equiv
1+2c\cdot x + c^2x^2$ that appears in the denominator of the
coordinate transformation (\ref{eq:finitesct}) resulting from a finite
SCT. The transformations (\ref{eq:finitesctfield}) are thus the
composition of a reflection in the hyperplane perpendicular to
$\hat{x}$, a reflection in the hyperplane perpendicular to $\hat{x}'$
and a scaling.  Combining the two reflections, the action of a finite
SCT on the fields thus consists of a rotation in the hyperplane
defined by $\hat{x}$ and $\hat{x}'$ (through twice the angle between
the two directions) and a scaling.  It is worth noting that, since the
resulting rotation is composed of two reflections, it is of a special
(or constrained) 
form; in four dimensions not all (proper) Lorentz
rotations can be constructed in this way.

The above geometrical interpretation of a finite SCT is consistent
with the action of an infinitesimal SCT on a (primary) field that
belongs to an irreducible representation of the Lorentz group, which
is given by (\ref{eq:gctformvar2}) with $\Delta = wI$, $\kappa_\mu=0$,
$\omega^{\mu\nu}=0$ and $\rho=0$, and describes the combination of an
infinitesimal scaling and rotation (both position dependent). Indeed,
it is straightforward to show that in the limit of small $c^\mu$ the
transformations (\ref{eq:finitesctfield}) yield precisely the forms
$\varpi^{\mu\nu}(x)=4c^{[\mu} x^{\nu]}$ and
$\varrho(x)=-2c\cdot x$ for the infinitesimal parameters appearing
in (\ref{eq:gctformvar2}). Moreover, having introduced the reflection
operator ${I^\mu}_\nu(\hat{x})$, it is worth noting that the generator
$K_\mu$ for SCTs in (\ref{eq:gcggeneratorsd}) can be written as $K_\mu
=-x^2{I_\mu}^\nu(\hat{x})\partial_\nu+
2(x^\nu\Sigma_{\mu\nu}-x_\mu\Delta) + \kappa_\mu$.

Finally, we note that since the action both of inversions and SCTs on
fields that belong to an irreducible representation of the Lorentz
group consists of a scaling and a Lorentz rotation, albeit an improper
rotation for inversions, then the action of any element of the full
finite conformal group on such a field may be written in the form
(\ref{eq:gwt}), provided one extends the definition of
$\matri{S}(\omega)$ to include matrices corresponding to elements of
the full Lorentz group, and allows $\rho$ and $\omega$ to become particular functions of spacetime position.

\subsection{Global conformal invariant field theory}
\label{sec:globalcft}

As our final topic in the discussion of global conformal
  invariance, we describe its consequences for field theories, since
  this is sometimes unclear in the literature and it is key for our
  later considerations to identify the appropriate conservation laws
  to compare with those obtained once the conformal group is
  gauged. The conservation laws discussed below are inevitably
  determined by considering only the infinitesimal forms of the
  transformations, and hence include only those simply connected to
  the identity. Nonetheless, we also briefly consider the consequences
  of invariance under inversions.

Consider a Minkowski spacetime ${\cal M}$, labelled using Cartesian
inertial coordinates, in which the dynamics of some set of fields
$\vpsi_i(x)$ ($i=1,2,\ldots$) is described by the action
\begin{equation}
S = \int L(\vpsi_i,\partial_\mu\vpsi_i)\,d^4x,
\label{srmatteraction}
\end{equation}
such as that considered in Appendix~\ref{app:b1}. The index $i$ here
again merely labels different matter fields, rather than denoting the
tensor or spinor components of individual fields (which are suppressed
throughout). It is also worth noting that these fields may include a
scalar compensator field (often denoted also by $\phi$) with Weyl
weight $w=-1$, which may be used to replace mass parameters in the
standard forms of Lagrangians for massive matter fields to achieve
global conformal invariance (for example by making the substitution
$m\bar{\psi}\psi \to \mu\phi\bar{\psi}\psi$ in the action for a
massive Dirac field $\psi$, where $\mu$ is a dimensionless parameter
but $\mu\phi$ has dimensions of mass in natural
units) \cite{Dirac73,eWGTpaper}.

The consequences of invariance of the action (\ref{srmatteraction})
under an infinitesimal global conformal transformation (connected to
the identity) may be determined by substituting the forms
(\ref{eq:gct}) and (\ref{eq:gctformvar}) into the general expressions
(\ref{eq:genactioninv}) and (\ref{eq:lagformvar}). Recalling that
the operators $\delta_0$ and
$\partial_\mu$ commute, and equating to zero the coefficients
multiplying the constant parameters $a^\mu$, $\omega^{\mu\nu}$, $\rho$
and $c^\mu$, respectively, leads to the conditions
%
\begin{subequations}
\label{eq:gctconds}
\bea
\partial_\mu L -\pd{L}{\vpsi_i}\partial_\mu\vpsi_i - 
\mom{\alpha}{i}\partial_\mu\partial_\alpha\vpsi_i&=& 0,
\nonumber \\[-1mm]
& & \nonumber\\[-5mm]
\label{eq:gctcondsa}\\
\pd{L}{\vpsi_i}\Sigma_{\mu\nu}\vpsi_i
\!+\! \mom{\alpha}{i}[\Sigma_{\mu\nu}\partial_\alpha\vpsi_i
+(\eta_{\alpha\mu}\partial_\nu\!-\!\eta_{\alpha\nu}\partial_\mu)\vpsi_i] & =
& 0, 
\nonumber \\[-1mm]
& & \nonumber\\[-5mm]
\label{eq:gctcondsb}\\
\pd{L}{\vpsi_i}\Delta\vpsi_i+
\mom{\alpha}{i}(\Delta-I)\partial_\alpha\vpsi_i+4L & = & 0, 
\nonumber \\[-1mm]
& & \nonumber\\[-5mm]
\label{eq:gctcondsc}\\
\pd{L}{\vpsi_i}\kappa_\mu\vpsi_i
+ \mom{\alpha}{i}
[\kappa_\mu\partial_\alpha\vpsi_i
+2(\Sigma_{\mu\alpha}-\eta_{\mu\alpha}\Delta)\vpsi_i] & = &0, 
\nonumber \\[-1mm]
& & \nonumber\\[-5mm]
\label{eq:gctcondsd}
\eea
\end{subequations}
%
which hold up to a total divergence of any quantity that vanishes on
the boundary of the integration region in (\ref{srmatteraction}). The first condition is equivalent to requiring that $L$
has no explicit dependence on spacetime position $x$, and this
condition has been used to derive the second and third
conditions. Moreover, the first three conditions have all been used to
derive the final condition. In particular, this means that for the
action to be invariant under SCTs (which is necessary for conformal
invariance), it must be Poincar\'e and scale invariant, in addition to
satisfying the condition (\ref{eq:gctcondsd}). Conversely, an action
that is invariant under Poincar\'e transformations and SCTs is
necessarily scale invariant.

Again adopting the forms (\ref{eq:gct}) and (\ref{eq:gctformvar}),
one finds that the general expression (\ref{eq:noetherjdef}) for the
Noether current becomes
\be
J^\mu = -
a^\alpha {t^\mu}_\alpha + \tfrac{1}{2}\omega^{\alpha\beta}{M^\mu}_{\alpha\beta}  + \rho {D}^\mu + c^\alpha
        {{K}^\mu}_\alpha,
\label{eq:noetherj}
\ee
where the coefficients of the parameters of the conformal
transformation are defined by
\begin{subequations}
\label{eq:confcurrents}
\bea
{{t}^\mu}_\alpha &\equiv& \mom{\mu}{i}\partial_\alpha\vpsi_i-\delta^\mu_\alpha
L, \\
{{M}^\mu}_{\alpha\beta} 
& \equiv & x_\alpha {t^\mu}_\beta - x_\beta
{t^\mu}_\alpha + {s^\mu}_{\alpha\beta}, \\
{D}^\mu & \equiv & -x^\alpha {t^\mu}_\alpha + j^\mu, \\
{{K}^\mu}_\alpha & \equiv & 
(2x_\alpha x^\beta\!-\!\delta^\beta_\alpha x^2){t^\mu}_\beta
\!+\! 2x^\beta({s^\mu}_{\alpha\beta}\!-\!\eta_{\alpha\beta}j^\mu) \!+\!
{k^\mu}_\alpha, \nonumber \\[-1mm]
& & \nonumber\\[-5mm]
\label{eq:confcurrentsd}
\eea
\end{subequations}
which are the (total) canonical energy-momentum, angular momentum,
dilation current and special conformal current, respectively, of the
fields $\vpsi_i$, and we have also defined the quantities
\begin{subequations}
\label{eq:srcurrentsdef}
\bea
{s^\mu}_{\alpha\beta} &\equiv& 
\mom{\mu}{i}
\Sigma_{\alpha\beta}\vpsi_i,  \\
j^\mu  &\equiv&  \mom{\mu}{i}
\Delta\vpsi_i,\\
{k^\mu}_\alpha  &\equiv&  \mom{\mu}{i}\kappa_\alpha\vpsi_i,
\eea
\end{subequations}
which are the (total) canonical spin angular momentum, intrinsic
dilation current and intrinsic special conformal current of
the fields. In (\ref{eq:confcurrentsd}), it is worth
noting that the first term on the RHS can be written as
$-x^2{I_\alpha}^\beta(\hat{x}){t^\mu}_\beta$.

If the field equations $\delta L/\delta\vpsi_i=0$ are satisfied, then
invariance of the action (\ref{srmatteraction}) reduces to the
conservation law $\partial_\mu J^\mu=0$. Since the parameters of a
global conformal transformation in (\ref{eq:noetherj}) are constants,
one thus obtains separate conservation laws of the form
(\ref{eq:globalconslaws}), given by
\begin{subequations}
\label{eq:gctcons}
\bea
\partial_\mu {t^\mu}_\alpha & = & 0, \label{eq:srtransinv}\\
\partial_\mu{s^\mu}_{\alpha\beta} + 2t_{[\alpha\beta]} & = & 
0,\label{eq:srrotinv}\\
\partial_\mu j^\mu - {t^\mu}_\mu & = & 0,\label{eq:srscaleinv}\\
\partial_\mu {k^\mu}_\alpha + 2({s^\mu}_{\alpha\mu}-j_\alpha)&=&0,
\label{eq:gctconsd}
\eea
\end{subequations}
which again hold up to a total divergence, and may be considered as
the ``on-shell'' specialisation of the conditions
(\ref{eq:gctconds}). As previously, the first condition has been used
to derive the second and third conditions, and the first three
conditions have all been used to derive the final condition. It is
worth noting that the conditions (\ref{eq:gctconds}) and
(\ref{eq:gctcons}) in fact hold for {\it any subset} of terms in the
Lagrangian $L$ for which the resulting action is conformally invariant.

As mentioned earlier, if the fields $\vpsi_i(x)$ belong to irreducible
representations of the Lorentz group, which is usually the case for
physical fields, then $\Delta = wI$ and $\kappa_\mu=0$, and hence
${k^\mu}_\alpha$ also vanishes. In this case, it is usual to
define the ``field virial'' \cite{Coleman71}
\be
\mathfrak{V}_\alpha 
\equiv \mom{\beta}{i}(\Sigma_{\alpha\beta}-\eta_{\alpha\beta}w)\vpsi_i
= {s^\beta}_{\alpha\beta} - j_\alpha.
\label{eq:virialdef}
\ee
From (\ref{eq:gctconsd}), one sees that for (any
subset of) an action that is Poincar\'e and scale invariant also to be
conformally invariant, the field virial $\mathfrak{V}_\alpha$ must
vanish up to a total divergence. Remarkably, this last condition is
found to hold for all renormalizable field theories involving
particles with spin 0, 1/2 and 1, even though scale invariance (and
hence conformal invariance) is, in general, broken in such
theories \cite{Blagojevic02}.

As is well known, the conservation laws (\ref{eq:srtransinv}) and
(\ref{eq:srrotinv}), resulting from translational and Lorentz
invariance, respectively, can both be expressed in terms of the single
Belinfante energy-momentum tensor \cite{Belinfante40} 
\be
t^{\mu\nu}_{\rm B} = t^{\mu\nu}
+ \tfrac{1}{2}\partial_\lambda(s^{\mu\nu\lambda} + s^{\nu\mu\lambda} -
s^{\lambda\nu\mu}),\label{eq:belinfanteem}
\ee
as the two properties $\partial_\mu t_{\rm
  B}^{\mu\nu} = 0$ and $t_{\rm B}^{\mu\nu} = t_{\rm
  B}^{\nu\mu}$. Moreover, for theories describing fields $\vpsi_i(x)$
that belong to irreducible representations of the Lorentz group and
where the field virial (\ref{eq:virialdef}) vanishes up to a total
divergence, scale invariance is thus equivalent to conformal
invariance, and one may further define the improved energy-momentum
tensor \cite{Callan70} 
\be
\theta^{\mu\nu} = t_{\rm B}^{\mu\nu} -
\tfrac{1}{6}(\partial^\mu\partial^\nu - \eta^{\mu\nu}\square^2)\sum_i
\vpsi_i^2,\label{eq:improvedem}
\ee
such that the remaining conditions in (\ref{eq:gctcons}) can be
expressed in terms of this single quantity as 
$\partial_\mu \theta^{\mu\nu} = 0$, $\theta^{\mu\nu} =
\theta^{\nu\mu}$ and ${\theta^\mu}_\mu =0$.

Since the discussion thus far relies on infinitesimal transformations,
it applies only to elements of the conformal group that are
continuously connected to the identity, and hence neglects invariance
of the action (\ref{srmatteraction}) under inversions, which are
intrinsically both finite and discrete. As mentioned in the
Introduction, this issue is of particular interest since it is
straightforward to show that both the Faraday action for the
electromagnetic field and the Dirac action for a massless spinor field
are invariant not only under the continuous elements of the conformal
group considered above, but also under
inversions.

Additional conserved quantities can be generated by discrete
symmetries. For example, theories invariant under spatial inversion
$x^{\prime i} = -x^i$ ($i=1,2,3$) conserve parity. It is far less
straightforward, however, to determine directly the consequences of
invariance of a general action (\ref{srmatteraction}) under the
(conformal) inversion (\ref{eq:inversion}). Nonetheless, one may gain 
some insight by considering the large-parameter limit of the
finite SCT (\ref{eq:finitesct}), which is given by \cite{Ho11}
\be 
x^{\prime\mu} =
\frac{c^\mu}{c^2}+\frac{1}{c^2}{I^\mu}_\nu(\hat{c})\frac{x^\nu}{x^2} +
O\left(\frac{1}{c^3}\right).
\ee
Thus, a large-$c$ SCT consists of the composition of an inversion
(\ref{eq:inversion}), a reflection in the hyperplane perpendicular to
$\hat{c}$, a scale transformation by $1/c^2$ and a translation by
$c^\mu/c^2$. An action that is invariant under translations,
scale transformations and SCTs, as we considered above, must therefore
also be
invariant under the combination of just an inversion and a reflection
in the hyperplane perpendicular to $\hat{c}$. Hence, if the action is
invariant under an inversion alone, it must also be invariant under
reflection in an arbitrary hyperplane, which provides a covariant
generalisation of the (one-dimensional) parity and time-reversal
transformations \cite{Miller68}.

\section{Previous approaches to conformal gauging}
\label{sec:lci}

As noted in the Introduction, there are strong theoretical reasons to
consider gauging the conformal group with a view to constructing
theories of the gravitational interaction that are invariant under
local conformal transformations. Previous approaches have focussed on
infinitesimal transformations and hence considered
gauging only the elements of the conformal group that are connected to
the identity (namely translations, Lorentz rotations, dilations and
SCTs). This is achieved, in principle, by allowing the 15 parameters
of the group $a$, $\omega$, $\rho$ and $c$ to become
independent arbitrary functions of position.

\subsection{Gauging a spacetime symmetry group}

Some early approaches \cite{Utiyama56,Sciama64} to constructing a gauge
theory of a spacetime symmetry group $\mathcal{G}$ (in our case the
conformal group) encountered complications arising from attempting to
draw too close an analogy with the gauge theories of internal
symmetries \cite{Ivanenko83,Lord86a}. In modern terminology, the
corresponding gauge fields (or Yang--Mills potentials) were introduced
as the components of a connection on a principal fiber bundle with
spacetime as base space and $\mathcal{G}$ as fiber. If the whole of a
spacetime group $\mathcal{G}$ is gauged in the Yang--Mills sense,
however, the gauged `internal translations' prevent the identification
of the translational gauge fields with a vierbein in the geometric
interpretation of the gauge theory.

It was the gauging of the Poincar\'e group by Kibble \cite{Kibble61}
that first revealed how to achieve a meaningful gauging of groups that
act on the points of spacetime as well as on the components of
physical fields. The essence of Kibble's approach was to note that
when the parameters of the Poincar\'e group become
independent arbitrary functions of position, this leads to a complete
decoupling of the translational parts from the rest of the group, and
the former are then interpreted as arising from a general coordinate
transformation (GCT; or spacetime diffeomorphisms, if interpreted
actively). Thus the action of the gauged Poincar\'e group was
considered as a GCT $x^\mu \to x^{\prime\mu}$, together with the local
action of its Lorentz subgroup $\mathcal{H}$ on the orthonormal tetrad
basis vectors $\hat{\vect{e}}_a(x)$ that define local Lorentz
reference frames, where we adopt the common convention that Latin
indices (from the start of the alphabet) refer to anholonomic local
tetrad frames, while Greek indices refer to holonomic coordinate
frames.  This approach to gauging can be straightforwardly extended to more
general spacetime symmetry groups \cite{Harnad76,Lord86a,Blagojevic02,eWGTpaper}.

The physical model envisaged in Kibble's approach is an underlying
Minkowski spacetime in which a set of matter fields $\vpsi_i$ is
distributed continuously. The field dynamics are described by a matter
action $S_{\rm M} = \int L_{\rm M}(\vpsi_i,\partial_\mu \vpsi_i)\,d^4x$
that is invariant under the global action of $\mathcal{G}$. One then
gauges the group $\mathcal{G}$ by demanding that the matter action be
invariant with respect to (infinitesimal, passively interpreted) GCT
and the local action of the subgroup $\mathcal{H}$, obtained by
setting the translation parameters of $\mathcal{G}$ to zero (which
leaves the origin $x^\mu=0$ invariant), and allowing the remaining
group parameters to become independent arbitrary functions of
position. One is thus led to the introduction of new field variables,
which are interpreted as gravitational gauge fields. These are used to
assemble a covariant derivative ${\cal D}_a\vpsi$ that transforms in
the same way under the action of the gauged group $\mathcal{G}$ as
$\partial_\mu\vpsi$ does under the global action of $\mathcal{G}$.
The matter action in the presence of gravity is then typically
obtained by the minimal coupling procedure of replacing partial
derivatives in the special-relativistic matter Lagrangian by covariant
ones, to obtain $S_{\rm M} = \int h^{-1} L_{\rm M}(\vpsi_i,{\cal
  D}_a\vpsi_i)\,d^4x$ , where the factor containing $h \equiv
\mbox{det}({h_a}^\mu)$ (here ${h_a}^\mu$ is the translational gauge
field) is required to make the integrand a scalar density rather than
a scalar.

In addition to the matter action, the total action must also contain
terms describing the dynamics of the free gravitational gauge fields.
Following the normal procedure used in gauging internal symmetries,
Kibble first constructed covariant field-strength tensors for the
gauge fields by commuting covariant derivatives, i.e. by considering
$[{\cal D}_a,{\cal D}_b]\vpsi$. The free gravitational action then
takes the form $S_{\rm G} = \int h^{-1} L_{\rm G}\,d^4x$, where
$L_{\rm G}$ is some Lagrangian that depends on the field strengths and
is such that $S_{\rm G}$ is invariant under the action of the gauged
group $\mathcal{G}$. The total action is taken as the sum of the
matter and gravitational actions, and variation of the total action
with respect to the gauge fields leads to coupled gravitational field
equations.

Following Kibble's work, several other approaches to gauging a
spacetime symmetry group have been proposed, in which, for example,
the transformations are interpreted actively, or one considers finite
rather than infinitesimal transformations
 \citep{Hehl76,Wiesendanger96,Mukunda89,Lasenby98}, but in 
terms of the final locally valid field equations that these
formulations reach, given an initial total action, they are equivalent to
Kibble's original method.

Finally, it is worth noting that Kibble's gauge approach to
gravitation is most naturally interpreted as a field theory in
Minkowski spacetime \cite{Wiesendanger96,Lasenby98}, in the same way as
the gauge field theories describing the other fundamental
interactions, and this is the viewpoint that we shall adopt in this
paper. It is more common, however, to reinterpret the mathematical
structure of such gauge theories geometrically, where in particular
the translational gauge field ${h_a}^\mu$ is considered as the
components of a vierbein system in a more general spacetime
 \cite{Hehl76}. These issues are discussed in more detail elsewhere
 \cite{Blagojevic02,eWGTpaper}.

More recent approaches to gauging a spacetime symmetry group adopt
the geometric interpretation wholeheartedly, and are usually
expressed in the language of fiber bundles. In this view, it is clear
from the discussion above that only the subgroup $\mathcal{H}$ should
act on the fibers, not the whole of $\mathcal{G}$ (i.e.\ no `internal
translation'). The simplest and most natural translation of the scheme
into fiber bundle language consists of expressing the gauge theory of
a spacetime symmetry group $\mathcal{G}$ in terms of the group
manifold $\mathcal{G}$; specifically, in terms of the principal fiber
bundle $\mathcal{G}(\mathcal{G}/\mathcal{H},\mathcal{H})$, where the
coset space $\mathcal{G}/\mathcal{H}$ is
spacetime \cite{MacDowell77,Lord86a}.

Indeed, this viewpoint is embodied in the so-called quotient manifold
method \cite{Neeman78a,Neeman78b}, which may be considered as an
inversion of Kibble's approach, and is usually expressed in the
language of differential forms as follows. Consider some Lie group
$\mathcal{G}$ possessing a Lie subgroup $\mathcal{H}$. The
Maurer--Cartan structure equations for $\mathcal{G}$ read
\be
\mathbf{d}\omega^A-{f_{BC}}^A\omega^B\wedge\omega^C = 0,
\ee
where ${f_{BC}}^A$ are the structure constants of the algebra of the
group $\mathcal{G}$. These equations constitute an integrability
condition that give a 1-form $\omega^A$ on $\mathcal{G}$ that carries
the basic infinitesimal information about the group's structure. One
may thus define the exponential map of the corresponding Lie algebra
and hence a local group action. One then takes the quotient
$\mathcal{G}/\mathcal{H}$, which is necessarily a manifold
$\mathcal{M}$ (usually interpreted as spacetime), and the 1-forms
provide its connection. The result is a principal fiber bundle with
local $\mathcal{H}$ symmetry and base manifold $\mathcal{M}$.  This
structure is then modified by generalizing the manifold, and by
changing the connection. Changing the manifold has no effect on the
local structure, but changing the connection modifies the
Maurer--Cartan equations (to yield the Cartan equations), resulting in
curvature 2-forms
\be \mathcal{R}^A =
\mathbf{d}\omega^A-{f_{BC}}^A\omega^B\wedge\omega^C.
\ee
Two restrictions are placed on these curvatures \cite{Wheeler14}.
First, the curvatures must characterize the manifold only, which
requires them to be `horizontal', i.e. bilinear in the connections.
Second, one requires integrability of the Cartan equations, which
leads to the Bianchi identities satisfied by the curvatures.

Thus, once one has made the choice of $\mathcal{G}$ and $\mathcal{H}$,
this approach determines: the physical arena $\mathcal{M}$, the local
symmetry group $\mathcal{H}$, the relevant field-strength tensors
$\mathcal{R}^A$, and any structures inherited from
$\mathcal{G}$ \cite{Wheeler13}. While other structures may be imposed,
such as additional (compensator) scalar fields, it is usual to
consider only those arising directly from properties of the gauge
group.  To complete a gravity theory, one finally constructs a locally
$\mathcal{H}$-invariant action from the available tensors
$\mathcal{R}^A$, together with the invariant metric $\eta_{ab}$ and
Levi--Civita tensor $\epsilon_{abcd}$, and any desired matter fields.
The remaining, broken, group transformations of $\mathcal{G}$ are
replaced by diffeomorphisms on $\mathcal{M}$. Each gauge field in the
connection form $\omega^A$ is then varied independently to find the
field equations. A key advantage of the approach is that it keeps the
curvatures and action expressed in terms of the gauge fields, making
the variation straightforward.

The quotient manifold method, although very powerful, may appear
somewhat `rarified' to most physicists. Fortunately, Kibble's original
approach may be used to arrive at precisely the same gauge theories as
those obtained using the quotient manifold method, although this is
rarely demonstrated in the literature. We will therefore primarily
adopt Kibble's approach below, since it is more familiar to
physicists. We will also maintain the more unorthodox viewpoint,
albeit hinted at in Kibble's original paper, of considering the gauge
fields as fields in Minkowski spacetime, without attaching any
geometric interpretation to them. Consequently, we will adopt a global
Cartesian inertial coordinate system $x^\mu$ in our Minkowski
spacetime, which greatly simplifies calculations, but more general
coordinate systems may be straightforwardly accommodated, if required
\cite{eWGTpaper}.

\subsection{Auxiliary conformal gauging}
\label{sec:auxiliary}

As mentioned in the Introduction, the standard approach to gauging the
conformal group leads to ACGT. This is usually performed in the
literature using the quotient manifold method. Moreover, the handful of
accounts that instead make use of Kibble's approach are somewhat terse
and lacking in detail on the transformation properties of the gauge
fields, as well as the structure of the resulting gauge field
strengths and the action. In order to facilitate a detailed comparison
of ACGT with our discussion of eWGT in Section~\ref{sec:ewgt} (and
also with established accounts of PGT and WGT in the literature), we
therefore describe here in a transparent manner how Kibble's approach
is applied, and indeed adapted, to arrive at ACGT.



In this standard approach to gauging the conformal group
$\mathcal{G}=C(1,3)$ (or, more precisely, the elements of it that are
connected to the identity), the subgroup $\mathcal{H}$ is that
obtained by setting the translation parameters $a^\mu=0$, which
leaves the origin $x^\mu=0$ invariant (i.e. the little group), and has
the 11 generators $\Sigma_{ab}$, $\Delta$ and $\kappa_a$ that
satisfy the commutation relations (\ref{eq:hgenerators}). 

Following Lord \cite{Lord85,Lord86a,Lord86b}, under the simultaneous
action of an infinitesimal GCT $x^\mu \to x^\mu +\xi^\mu(x)$ and the
infinitesimal local action of $\mathcal{H}$, by analogy with
(\ref{eq:gctformvar2}), the form variation of a (primary) field is
given by
\be \delta_0\vpsi(x) = -\xi^\mu(x)\partial_\mu\vpsi(x) + \varepsilon(x)\vpsi(x), 
\label{eq:d0psi}
\ee
where
$\varepsilon(x)\equiv\tfrac{1}{2}\varpi^{ab}(x)\Sigma_{ab}
+\varrho(x)\Delta+ c^a(x) \kappa_a$ is an element of the
localised (little) subgroup $\mathcal{H}$, and
$\varpi^{ab}(x)$, $\varrho(x)$, $c^a(x)$ and
$\xi^\mu(x)$ are now independent arbitrary functions of position.
Consequently, the transformation law of the derivative
$\partial_\mu\vpsi(x)$ is no longer given by (\ref{eq:cgtformvard}).

The construction of a covariant derivative that transforms like
(\ref{eq:cgtformvard}) under the gauged conformal group is typically
achieved in two steps. First, one defines the `$\mathcal{H}$-covariant'
derivative
\be
\bar{D}_\mu\vpsi(x) \equiv [\partial_\mu + \bar{\Gamma}_\mu(x)]\vpsi(x),
\label{eq:Dbarmudef}
\ee
where $\bar{\Gamma}_\mu(x)$ is a linear combination of the generators
of $\mathcal{H}$ that depends on the gauge fields corresponding to
local Lorentz rotations, local dilations, and local special conformal
transformations, respectively (the bars appearing in these definitions
are to distinguish the corresponding quantities from others to be
defined later). In the second step, we define a
`generalised $\mathcal{H}$-covariant' derivative, linearly related to
$\bar{D}_\mu\vpsi$, by
\be
\bar{\cal D}_a\vpsi(x) \equiv  {h_a}^\mu(x)\bar{D}_\mu\vpsi(x),
\label{weylgencovdef}
\ee
where we have introduced the translational gauge field ${h_a}^\mu(x)$.
It is assumed that ${h_a}^\mu(x)$ has an inverse, usually denoted by
${b^a}_\mu(x)$, such that ${h_a}^\mu {b^a}_\nu = \delta_\nu^\mu$ and
${h_a}^\mu {b^c}_\mu = \delta_a^c$ (where, for brevity, we henceforth
typically drop the explicit $x$-dependence).

Under the action of an infinitesimal GCT $x^{\prime\mu} =
x^\mu +\xi^\mu$ and the infinitesimal local action of $\mathcal{H}$,
with the associated field transformation law (\ref{eq:d0psi}), we
require $\bar{\cal D}_a\vpsi$ to have an analogous transformation law
to (\ref{eq:cgtformvard}), namely
\bea 
\delta_0(\bar{\cal D}_a\vpsi) & = & -\xi^\mu\partial_\mu\bar{\cal
  D}_a\vpsi + \varepsilon\,\bar{\cal
  D}_a\vpsi
-({\varpi^b}_{a}+\varrho\,\delta^b_a)\bar{\cal
  D}_b\vpsi \nonumber\\
&& \hspace{2.5cm} +2(c^b\Sigma_{ba}-c_a\Delta)\vpsi,
\label{eq:cgtformvard2}
\eea
but where $\varpi^{ab}$, $\varrho$, $c^a$ and $\xi^\mu$ are
independent arbitrary functions of position. This requirement leads 
uniquely to the transformation laws
\begin{subequations}
\label{eq:hgammatrans}
\bea
\delta_0{h_a}^\mu & = & -\xi^\nu\partial_\nu {h_a}^\mu + {h_a}^\nu\partial_\nu\xi^\mu
 - ({\varpi^b}_{a}
+\varrho\,\delta^b_a){h_b}^\mu,\phantom{AAA} \label{eq:acgthtrans}\\
\delta_0\bar{\Gamma}_\mu & = & 
-\xi^\nu\partial_\nu\bar{\Gamma}_\mu - \bar{\Gamma}_\nu\partial_\mu\xi^\nu
 -\partial_\mu\varepsilon 
-[\bar{\Gamma}_\mu,\varepsilon] \nonumber \\
&& \hspace{2.7cm} +2{b^a}_\mu(c^b\Sigma_{ba}-c_a\Delta).
\label{eq:gammatrans}
\eea
\end{subequations}
One sees that ${h_a}^\mu$ transforms as a GCT vector, a local Lorentz
4-vector, and has Weyl weight $w=-1$ under local dilations (its
inverse ${b^a}_\mu$ transforms in an analogous way, but has $w=1$).
Similarly, the quantity $\bar{\Gamma}_\mu$ transforms as a covariant
GCT vector, but the last term on the RHS of (\ref{eq:gammatrans})
shows that $\bar{\Gamma}_\mu$ is {\it not} the connection for the
gauge group $\mathcal{H}$. Indeed, it was already apparent from the
corresponding final term in (\ref{eq:cgtformvard2}) that $\bar{\cal
  D}_a\vpsi$ is not an $\mathcal{H}$-covariant derivative in the usual
sense; its transformation law is linear but inhomogeneous. As
mentioned earlier, this behaviour originates in the final term of the
transformation law (\ref{eq:cgtformvard}) for $\partial_\mu\vpsi$
under the action of a global conformal transformation, and can be
traced to the fact that translations do not form an {\it invariant}
subgroup of the conformal group (whereas they do for the Weyl group,
obtained by setting $c^\mu=0$). Nonetheless, since $\bar{\cal
  D}_a\vpsi$ was constructed to have an analogous transformation law
to that of $\partial_\mu\vpsi$ in (\ref{eq:cgtformvard}), one can
still construct a matter action that is fully invariant under the
gauged conformal group from one that is invariant under global
conformal transformations by employing the usual minimal coupling
procedure of replacing partial derivatives by covariant ones to obtain
$S_{\rm M} = \int h^{-1} L_{\rm M}(\vpsi_i,\bar{\cal
  D}_a\vpsi_i)\,d^4x$. As discussed in Section~\ref{sec:globalcft},
the set of fields $\vpsi_i$ may include a scalar compensator field
(denoted also by $\phi$).

It is usual to assume the linear combination $\bar{\Gamma}_\mu$ of the
generators of $\mathcal{H}$ to have the form
\be
\bar{\Gamma}_\mu \equiv {\textstyle\frac{1}{2}}{A^{ab}}_\mu
\Sigma_{ab} + B_\mu\Delta + {f^a}_\mu\kappa_a,
\label{eq:gammabardef}
\ee
where ${A^{ab}}_\mu(x)$, $B_\mu(x)$ and ${f^a}_\mu(x)$ are the gauge
fields corresponding to local Lorentz rotations, local dilations, and
local special conformal transformations, respectively. It is worth
pointing out that this assumed form for $\bar{\Gamma}_\mu$ constitutes a
{\it choice} of how to include the gauge fields, and leads directly
to their required transformation laws
\begin{widetext}
\begin{subequations}
\label{eq:abftrans}
\bea
\delta_0 {A^{ab}}_\mu & = &-\xi^\nu\partial_\nu{A^{ab}}_\mu
-{A^{ab}}_\nu \partial_\mu\xi^\nu
-2{{\varpi^{[a}}_{c}A^{b]c}}_\mu 
-\partial_\mu\varpi^{ab}
- 4 {b^{[a}}_\mu c^{b]}, \\
\delta_0 B_\mu & = &  -\xi^\nu\partial_\nu B_\mu -B_\nu\partial_\mu\xi^\nu
-\partial_\mu\varrho -2{b^a}_\mu c_a, \\
\delta_0 {f^a}_\mu & = & -\xi^\nu\partial_\nu {f^a}_\mu -{f^a}_\nu\partial_\mu\xi^\nu
 - ({\varpi_b}^{a}
+\varrho\,\delta^a_b){f^b}_\mu - (\partial_\mu c^a - B_\mu c^a
-{A^{ab}}_\mu c_b), 
\eea
\end{subequations}
\end{widetext}
which are obtained by substituting (\ref{eq:gammabardef}) into 
(\ref{eq:gammatrans}) and equating coefficients of 
$\Sigma_{ab}$, $\Delta$ and $\kappa_a$, respectively.

We note that if one sets $c^a=0$, the transformation laws of
${h_a}^\mu$, ${A^{ab}}_\mu$ and $B_\mu$ in (\ref{eq:hgammatrans}) and
(\ref{eq:abftrans}) are precisely the infinitesimal versions
(i.e.\ first order in the group parameters) of those obtained for
these gauge fields under finite transformations in Weyl gauge theory
(WGT) \cite{eWGTpaper}. In the case $c^a=0$, one also sees that
${f^a}_\mu$ has similar transformation properties to ${h_a}^\mu$ in
(\ref{eq:acgthtrans}), since it transforms as a (covariant) GCT
vector, a local Lorentz 4-vector, and has Weyl weight $w=-1$ under
local dilations. In the general case $c^a \neq 0$, however, the
transformation laws of all the gauge fields ${A^{ab}}_\mu$, $B_\mu$
and ${f^a}_\mu$ involve terms containing $c^a$. Indeed, these terms
lead to the inclusion of the (inverse) translational gauge field
${b^a}_\mu$ in the transformation laws of ${A^{ab}}_\mu$ and $B_\mu$,
and also cause the transformation law of ${f^a}_\mu$ to depend on
${A^{ab}}_\mu$ and $B_\mu$. Hence, the action of (local) SCTs leads to
considerable differences between the gauge theory of the conformal
group and those of its Weyl or Poincar\'e subgroups.

The total action must also contain terms describing the dynamics of
the free gravitational gauge fields. These terms are constructed from
the gauge field strengths, which are usually defined in terms of
the commutator of covariant derivatives. Considering first the
$\mathcal{H}$-covariant derivative, one finds
\be
[\bar{D}_\mu,\bar{D}_\nu]\vpsi =
(\tfrac{1}{2}{R^{ab}}_{\mu\nu}\Sigma_{ab} 
+ H_{\mu\nu}\Delta
+ {S^{\ast a}}_{\mu\nu}\kappa_a)\vpsi,
\label{dstarmucomm}
\ee
where we have defined the `$\mathcal{H}$-rotational',
`$\mathcal{H}$-dilational' and the `$\mathcal{H}$-special conformal'
field-strength tensors, respectively (the reason for notating the last
of these with an asterisk will become clear shortly). In terms of the
gauge fields ${A^{ab}}_\mu$, $B_\mu$ and ${f^a}_\mu$, the field
strengths have the forms
\begin{subequations}
\label{eq:sfsdef}
\bea
{R^{ab}}_{\mu\nu} \! & \equiv & 2(\partial_{[\mu} {A^{ab}}_{\nu]} +
\eta_{cd}{A^{ac}}_{[\mu}{A^{db}}_{\nu]}),
\\
H_{\mu\nu} \!& \equiv & 2\partial_{[\mu} B_{\nu]},
\\
{S^{\ast a}}_{\mu\nu}\! & \equiv & 
2(\partial_{[\mu}{f^a}_{\nu]} \!+\!
{A^a}_{c[\mu}{f^c}_{\nu]} \! -\! B_{[\mu}{f^a}_{\nu]})
\!= \! 2D^\ast_{[\mu}{f^a}_{\nu]}.\phantom{AA...}
\eea
\end{subequations}
For the sake of brevity, in the final expression we have introduced the
derivative operator $D^\ast_\mu \equiv \partial_\mu +
{\textstyle\frac{1}{2}}{A^{ab}}_\mu \Sigma_{ab} + wB_\mu$, familiar
from WGT \cite{Blagojevic02,eWGTpaper}, where $w$ is the Weyl weight of
the field on which it acts.  All three field strengths in
(\ref{eq:sfsdef}) transform covariantly under GCT and local Lorentz
rotations in accordance with their respective index structures, and
also under local dilations with the Weyl weights
$w({R^{ab}}_{\mu\nu})=0$, $w(H_{\mu\nu})=0$
and $w({S^{\ast a}}_{\mu\nu})=-1$, respectively, but {\it
  none} of them transforms covariantly under local SCTs, as we discuss
further below.

Before doing so, however, we next consider the commutator of two
`generalised $\mathcal{H}$-covariant' derivatives.  Since $\bar{\cal
  D}_a\vpsi ={h_a}^\mu \bar{D}_\mu\vpsi$, this commutator differs from
(\ref{dstarmucomm}) by an additional term containing the derivatives
of ${h_a}^\mu$, and reads
\be
[\bar{\cal D}_c,\bar{\cal D}_d]\vpsi =(
\tfrac{1}{2}{{\cal R}^{ab}}_{cd}\Sigma_{ab}
+ {\cal H}_{cd}\Delta
+{{\cal S}^{\ast a}}_{cd}\kappa_a
- {{\cal T}^{\ast a}}_{cd}\bar{\cal D}_a)\vpsi,
\label{weylfsdefs}
\ee
where ${{\cal R}^{ab}}_{cd} \equiv
{h_c}^{\mu}{h_d}^{\nu}{R^{ab}}_{\mu\nu}$, ${\cal
  H}_{cd} = {h_c}^\mu {h_d}^\nu H_{\mu\nu}$ and
${{\cal S}^{\ast a}}_{cd} \equiv {h_c}^{\mu}{h_d}^{\nu}
{S^{\ast a}}_{\mu\nu}$, and the `$\mathcal{H}$-translational'
field strength of the gauge field ${h_a}^\mu$ is given by
\be
{{\cal T}^{\ast a}}_{cd} \equiv {h_c}^{\mu}{h_d}^{\nu}
  {T^{\ast a}}_{\mu\nu} \equiv
      2{h_c}^{\mu}{h_d}^{\nu}D^\ast_{[\mu} {b^a}_{\nu]},
\label{eq:cstardef}
\ee
which clearly has the same form as 
${{\cal S}^{\ast a}}_{cd}$, 
but with ${f^a}_\mu$ replaced by ${b^a}_\mu$.  It is
worth noting that ${{\cal R}^{ab}}_{cd}$, ${\cal H}_{cd}$
and ${{\cal T}^{\ast a}}_{cd}$ have the same functional
forms of the gauge fields as the rotational, dilational and
translational gauge field strengths, respectively, in
WGT \cite{eWGTpaper}.

It is straightforward to show that ${{\cal R}^{ab}}_{cd}$,
${\cal H}_{cd}$, ${{\cal S}^{\ast a}}_{cd}$ and ${{\cal
  T}^{\ast a}}_{cd}$ are GCT scalars and transform
covariantly under local Lorentz rotations and under local dilations,
with weights $w({{\cal R}^{ab}}_{cd})=w({\cal H}_{cd})=-2$,
$w({{\cal S}^{\ast a}}_{cd})=-3$ and $w({{\cal T}^{\ast
  a}}_{cd})=-1$, respectively. As one might expect,
however, the transformation laws under local SCTs are more
complicated, and are given by
\begin{subequations}
\label{eq:fsbartrans}
\bea
\delta_0{{\cal R}^{ab}}_{cd} & = & 4c^{[a}{{\cal
    T}^{\ast b]}}_{cd} + 8\delta^{[a}_{[c} {\cal
      D}^\ast_{d]}c^{b]} 
\\
\delta_0{\cal H}_{cd} & = & -2c_a {{\cal
    T}^{\ast a}}_{cd} + 4{\delta^a}_{[c}{\cal
    D}^\ast_{d]}c_a 
\\
\delta_0{{\cal S}^{\ast a}}_{cd} & = & -2c_b({{\cal
  R}^{ab}}_{cd}+8\delta^{[a|}_{[c}{h_{d]}}^\mu
  {f^{|b]}}_\mu) \nonumber\\
&& \hspace{1.85cm} +2c^a({\cal H}_{cd} + 4 f_{[c|\mu}{h_{|d]}}^\mu),\phantom{A}
\\
\delta_0{{\cal T}^{\ast a}}_{cd} & = & 0,
\eea
\end{subequations}
where the action of $\mathcal{D}^\ast_a$ assumes that $w(c^a) =
-1$. Thus, there is a `mixing' of the transformation laws of the field
strengths, which arises from mixing of the transformation laws of the
gauge fields themselves, as described above. Moreover, one sees that
the transformation laws (\ref{eq:fsbartrans}) also depend on the gauge
fields directly, rather than just through the field strengths. Indeed,
it is only the $\mathcal{H}$-translational field strength ${{\cal
  T}^{\ast a}}_{cd}$ that transforms covariantly
(indeed, invariantly) under local SCTs.

The transformation laws (\ref{eq:fsbartrans}) mean that the only
combination of terms containing field strengths that may be included
in the total Lagrangian to obtain an action that is invariant under
local conformal transformations is $\phi^2(\beta_1 {\cal
  T^\ast}_{abc}{\cal T}^{\ast abc} + \beta_2 {\cal T^\ast}_{abc}{\cal
  T}^{\ast bac} + \beta_3 {\cal T^\ast}_a{\cal T}^{\ast a})$,
where the $\beta_i$ are dimensionless parameters and $\phi$ is some
(compensator) scalar field with Weyl weight $w(\phi)=-1$.  This
behaviour differs markedly from that encountered in WGT or PGT, in
which all the fields strengths transform covariantly under all the
localised transformations. It is therefore necessary to adapt Kibble's
approach slightly \cite{Lord85}, as we now outline, to reduce the complications
arising from the gauging of SCTs.

The above complications arise, in part, from the fact that the
quantity $\bar{\Gamma}_\mu$ is not the connection for the gauge group
$\mathcal{H}$, as is apparent from its transformation law
(\ref{eq:gammatrans}).  The nature of $\bar{\Gamma}_\mu$ can be
better understood by considering a purely internal $SO(2,4)$ symmetry,
to which the conformal group $C(1,3)$ is isomorphic, with generators
$\pi_a$, $\Sigma_{ab}$, $\Delta$ and $\kappa_a$ satisfying a set of
commutation rules analogous to (\ref{eq:cgcomms}), where $\pi_a$ is
the translational generator. One then defines the quantity
\be
\widetilde{\Gamma}_\mu \equiv {b^a}_\mu\pi_a + \bar{\Gamma}_\mu,
\ee
which transforms under the simultaneous local action of $\mathcal{H}$
and a GCT as
\be
\delta_0\widetilde{\Gamma}_\mu  =  
-\xi^\nu\partial_\nu\widetilde{\Gamma}_\mu
- \widetilde{\Gamma}_\nu\partial_\mu\xi^\nu
 -\partial_\mu\varepsilon 
- [\widetilde{\Gamma}_\mu,\varepsilon],
\ee
and is hence a connection for the group $SO(2,4)$. Moreover, the
transformation laws for the two parts of $\widetilde{\Gamma}_\mu$
correspond precisely those found in (\ref{eq:hgammatrans}) (where we
recall that $b^a_\mu$ is the inverse of ${h_a}^\mu$). Thus, ${b^a}_\mu$
and $\bar{\Gamma}_\mu$ together constitute a connection for the group
$SO(2,4)$.

If one then defines the new `$\mathcal{G}$-covariant' derivative
operator $\widetilde{D}_\mu \equiv \partial_\mu +
\widetilde{\Gamma}_\mu$, the resulting commutator reads
\be
[\widetilde{D}_\mu,\widetilde{D}_\nu]\vpsi =
(\tfrac{1}{2}
\widetilde{R}^{\,ab}_{\phantom{ab}\mu\nu}
\Sigma_{ab} 
+ \widetilde{H}_{\mu\nu}\Delta
+ {S^{\ast a}}_{\mu\nu}\kappa_a+{T^{\ast a}}_{\mu\nu}\pi_a)\vpsi,
\label{dfullmucomm}
\ee
where the new rotational and dilational `$\mathcal{G}$-covariant'
field strengths are given in terms of those defined in
(\ref{eq:sfsdef}) as
\begin{subequations}
\label{eq:gcovfs}
\bea
\widetilde{R}^{\,ab}_{\phantom{ab}\mu\nu} & = & R^{ab}_{\phantom{ab}\mu\nu} 
+ 8{b^{[a}}_{[\mu}{f^{b]}}_{\nu]}, \\
\widetilde{H}_{\mu\nu} & = & H_{\mu\nu} + 4{b^{a}}_{[\mu|}f_{a|\nu]}.
\eea
\end{subequations}
One may define the corresponding GCT scalar field strengths
$\widetilde{\cal R}^{ab}_{\phantom{ab}cd}$ and $\widetilde{\cal
  H}_{cd}$, in an analogous manner to that used above. The resulting
set of field strengths again transform covariantly under local Lorentz
rotations and local dilations (with the same weights as given
previously), but now transform under local SCTs as
\begin{subequations}
\label{eq:fsnobartrans}
\bea
\delta_0\widetilde{\cal R}^{ab}_{\phantom{ab}cd} & = & 4c^{[a}
{\cal T}^{\ast b]}_{\phantom{\ast b]}cd} 
\\
\delta_0\widetilde{\cal H}_{cd} & = & -2c_a 
{\cal T}^{\ast a}_{\phantom{\ast a,}cd} 
\\
\delta_0{\cal S}^{\ast a}_{\phantom{\ast a,}cd} & = & -2c_b\widetilde{\cal
  R}^{ab}_{\phantom{ab}cd}+2c^a\widetilde{\cal H}_{cd},
\\
\delta_0{\cal T}^{\ast a}_{\phantom{\ast a,}cd} & = & 0.
\eea
\end{subequations}
Once again, there is `mixing' between these transformation laws,
although now they depend only on the field strengths (and on the
parameters of the local SCT, as expected).

\subsubsection{Auxiliary conformal gauge theory with non-zero torsion}

Given the transformation laws (\ref{eq:fsnobartrans}), the most
general, parity-even free-gravitational action (containing no
compensator scalar fields) that is invariant under local conformal
transformations is {\it uniquely} determined (up to an overall
multiple), and given by \cite{Wheeler14}
\be
S_{\rm G} = \alpha \int h^{-1} (\widetilde{\cal
  R}_{abcd}\widetilde{\cal R}^{bacd} 
+ 4{\cal
  T}^\ast_{abc}{\cal S}^{\ast abc}
+2\widetilde{\cal H}_{ab}\widetilde{\cal H}^{ab})\,d^4x. 
\label{eq:ACGTaction}
\ee
The phenomenology of the resulting gravity theory remains to be fully
explored, but one can show \cite{Wheeler91} that the gauge field
${f^a}_\mu$ corresponding to local SCTs acts as an {\it auxiliary}
field (hence the name for this approach), since its field equation may
be used to eliminate it from the action (\ref{eq:ACGTaction}).  Hence
it appears that the symmetry reduces back to the local Weyl group. In
principle, the total Lagrangian could again also include the
combination $\phi^2(\beta_1 {\cal T^\ast}_{abc}{\cal T}^{\ast abc} +
\beta_2 {\cal T^\ast}_{abc}{\cal T}^{\ast bac} + \beta_3 {\cal
  T^\ast}_a{\cal T}^{\ast a})$, where the $\beta_i$ are dimensionless
parameters and $\phi$ is some (compensator) scalar field with Weyl
weight $w(\phi)=-1$, together possibly with an additional kinetic term
for $\phi$, but these additional terms appear not to have been
considered previously.

\subsubsection{Auxiliary conformal gauge theory with vanishing torsion}

One sees from (\ref{eq:fsnobartrans}) that, since ${\cal
  T}^{\ast a}_{\phantom{\ast a,}cd}$ transforms covariantly under the
full gauged conformal group, one can consistently set it to zero, if
desired. In this case, $\widetilde{\cal R}^{ab}_{\phantom{ab}cd}$ and
$\widetilde{\cal H}_{cd}$ then also become fully covariant, and so the
number of terms that may be included in a total action that remains
invariant under local conformal transformations is considerably
increased.  Moreover, the condition ${\cal T}^{\ast a}_{\phantom{\ast
    a,}cd}=0$ can be used to eliminate the rotational gauge field
${A^{ab}}_\mu$ by writing it in terms of the translational and
dilational gauge fields ${h_a}^\mu$ and $B_\mu$.  This torsionless
special case of auxiliary conformal gauge theory has been studied more
extensively \cite{Crispim77,Kaku77}. The only admissible Lagrangian
term that is linear in the gauge field strengths is
$\phi^2\widetilde{\cal R}$, but it may be shown that variation of the
resulting action with respect to the gauge field ${f^a}_\mu$ leads to
inconsistencies \cite{Wheeler13}. Attention has therefore focussed on
Lagrangians consisting of (arbitrary) linear combinations of terms
quadratic in the field strengths $\widetilde{\cal
  R}^{ab}_{\phantom{ab}cd}$ and $\widetilde{\cal H}_{cd}$ and their
contractions (and without compensator scalar fields).  It may be
shown, however, that in every such case, the gauge field ${f^a}_\mu$
may again be eliminated from the original action by its own field
equation \cite{Kaku77,Kaku78,Crispim78,Wheeler91}, such that the
resulting action depends only on ${h_a}^\mu$ and $B_\mu$. Indeed,
every such action is found to be equivalent to \cite{Wheeler91}
\be
S_{\rm G} = \int h^{-1} (\alpha\czero{{\cal C}}_{abcd}\czero{{\cal C}}^{abcd} + \beta
{\cal H}_{ab}{\cal H}^{ab})\,d^4x,
\ee
where $\czero{{\cal C}}_{abcd}$ is the conformal tensor, defined by
\be
\czero{{\cal C}}_{abcd} = \czero{{\cal R}}_{abcd} -
\eta_{c[a}\czero{{\cal R}}_{b]d} + \eta_{d[a}\czero{{\cal R}}_{b]c} +
\tfrac{1}{3} \eta_{c[a}\eta_{b]d}\czero{{\cal R}},
\ee
in which $\czero{{\cal R}}^{ab}_{\phantom{ab}cd} = 2{h_c}^\mu
{h_d}^\nu(\partial_{[\mu} {\zero{A}^{ab}}_{\nu]} +
{\zero{A}^a}_{c[\mu}{\zero{A}^{cb}}_{\nu]})$ is the gauge theory
equivalent of the Riemann tensor, obeying all the usual symmetries and
identities, and the quantities
${\zero{A}^{ab}}_{\mu}$ are the Ricci rotation coefficients
\be 
\zero{A}_{ab\mu} = {h_a}^\nu\partial_{[\mu|} b_{b|\nu]}
-{h_b}^\nu\partial_{[\mu|} b_{a|\nu]} -{b^c}_\mu {h_{[a}}^\lambda
  {h_{b]}}^\nu \partial_\lambda b_{c\nu},
\ee
which depend entirely on the translational gauge field $h_a^\mu$ (and
its inverse) \cite{eWGTpaper}. Thus, with the elimination of
${f^a}_\mu$, the symmetry appears once again to have reduced back to
the local Weyl group.

\bigskip
We conclude this section by noting that the auxiliary conformal gauge
theories that we have constructed using (a slight generalisation
of) Kibble's approach \cite{Lord85} are identical to those obtained using
the quotient manifold method of gauging, in which the Lie group
$\mathcal{G}$ is the conformal group (more precisely, the elements of
it that are connected to the identity) with the 15 generators
$\{P_a,M_{ab},D,K_a\}$ given in (\ref{eq:gcggenerators}),
$\mathcal{H}$ is the inhomogeneous Weyl group with 11 generators
$\{M_{ab},D,K_a\}$, and the quotient $\mathcal{G}/\mathcal{H}$ is thus
a homogeneous four-dimensional manifold $\mathcal{M}$ (interpreted as
spacetime). In particular, this approach leads to field-strength
tensors that agree precisely with those used to construct the action
(\ref{eq:ACGTaction}) \cite{Kaku77,Wheeler91}.

\subsection{Ungauging the conformal group}
\label{sec:ungaugingcft}

In our discussion in Section~\ref{sec:ewgt} below of the relative
merits of eWGT and WGT as putative gauge theories of the conformal
group, one of our key considerations will be the `ungauged' limit of
each theory. We therefore give an account here of the process
typically used for determining this limit and propose some
modifications of it, before applying our adapted version to ACGT as an
exemplar.

According to Lord \cite{Lord86b}, in order to justify that the local
action of the (little) subgroup $\mathcal{H}$, together with general
diffeomorphisms (or GCT) on $\mathcal{M}$, does indeed constitute a
true gauge theory of a spacetime group $\mathcal{G}$, one must show
that the limiting case of `ungauged' transformations does in fact
correspond to the correct global action of $\mathcal{G}$ on
$\mathcal{M}$ and on fields in $\mathcal{M}$. Lord 
demonstrates that this holds for the auxiliary conformal gauge theory
(ACGT) described above, and also for analogous gauge theories based on
the de Sitter group and (by Wigner--Inonii contraction of the de
Sitter case) the Poincar\'e group; the `ungauged' limit of Poincar\'e
gauge theory is also considered by Hehl \cite{Hehl78}.

The `ungauged' limit corresponds to vanishing gauge field strengths.
For ACGT, one thus requires $\widetilde{\cal
  R}^{ab}_{\phantom{ab}cd}$, $\widetilde{\cal H}_{cd}$, ${\cal
  S}^{\ast a}_{\phantom{\ast a,}cd}$ and ${\cal T}^{\ast
  a}_{\phantom{\ast a,}cd}$ to vanish. In this limit, the coordinate
system and ${\cal H}$-gauge can be chosen such that
\be
{h_a}^\mu(x) = \delta_a^\mu,\quad {A^{ab}}_\mu(x)=0, \quad B_\mu(x)=0, 
\quad {f^a}_\mu(x)=0.
\label{eq:nogauge}
\ee
In this reference system, the first condition means that the
distinction between Latin and Greek indices is
lost. It is important, however, to retain this distinction (and that
between calligraphic and non-calligraphic quantities) when considering
behaviour under any subsequent GCT and ${\cal H}$-gauge
transformation.

From the transformation laws (\ref{eq:hgammatrans}) and
(\ref{eq:abftrans}) of the gauge fields, Lord notes simply that in
order for any subsequent GCT and ${\cal H}$-gauge transformation to
{\it preserve} the relations (\ref{eq:nogauge}), one requires
\begin{subequations}
\label{eq:recovergct}
\bea
\partial_\beta\xi^\alpha & = & {\varpi^\alpha}_{\beta}
+ \varrho\delta^\alpha_\beta, \\
\partial_\mu\varpi^{\alpha\beta} & = &
4c^{[\alpha}\delta^{\beta]}_\mu, \\
\partial_\mu\varrho & = & -2c_\mu, \\
\partial_\mu c^\alpha & = & 0.
\eea
\end{subequations}
Successive integration of these equations (from the last to the first)
is straightforward and yields an expression for $\xi^\alpha(x)$ of
the form (\ref{eq:gct}) for an infinitesimal global conformal
transformation. Similarly, the transformation law (\ref{eq:d0psi})
reduces to that given in (\ref{eq:gctformvar2}) for the action of an
infinitesimal global conformal transformation on a (primary) physical
field. Hence, Lord concludes that ACGT has the correct `ungauged'
limit.

It is not clear from Lord's discussion, however, {\it why} the
`ungauged' limit should be derived by requiring that the relations
(\ref{eq:nogauge}) be preserved. Although preserving these
relations ensures that the covariant derivative remains equal to the
simple partial derivative, it is certainly unnecessary for the field
strength tensors to remain zero, to which the `ungauged' limit
corresponds. Indeed, since these tensors are GCT scalars and transform
covariantly under local Lorentz rotations and local dilations, and
according to (\ref{eq:fsnobartrans}) under local SCTs, they will
remain zero under {\it any} subsequent GCT and ${\cal H}$-gauge
transformation, which in general will {\it not} preserve the relations
(\ref{eq:nogauge}).

Moreover, as we now show, the final three relations in
(\ref{eq:nogauge}) are superfluous for identifying global conformal
transformations as the `ungauged' limit of ACGT.  Requiring only that
the {\it first} relation in (\ref{eq:nogauge}) be preserved (which
ensures the equivalence of Latin and Greek indices before and after
the transformation) leads immediately to the first equation in
(\ref{eq:recovergct}), which is simply a consequence of demanding that
$\delta_0{h_a}^\mu = 0$.  It is straightforward to show, however, that
the first relation in (\ref{eq:recovergct}) is both a necessary and
sufficient condition for $\xi^\alpha(x)$ to satisfy the four-dimensional
conformal Killing equation in Minkowski spacetime given in
(\ref{killingeq}), from which it follows that the most general
solution for $\xi^\alpha(x)$ has the form (\ref{eq:gct}) of an
infinitesimal global conformal transformation. The remaining three
conditions in (\ref{eq:recovergct}) then follow automatically, which
in turn means that the final three relations in (\ref{eq:nogauge}) are
also preserved.  Thus, for ACGT, the requirement that these further
relations be preserved is superfluous, and the correct `ungauged'
limit can be identified by requiring only that the first relation in
(\ref{eq:nogauge}) is preserved.

It is clear, however, that imposing this reduced requirement on other
gauge theories will not in general isolate the correct `ungauged'
limit. Consider WGT, for example, for which ${\cal G}$ is the
inhomogeneous Weyl group and ${\cal H}$ is the homogeneous Weyl
group. The structure of WGT is easily obtained from that of ACGT by
setting $c^a\equiv 0$ and ${f^a}_\mu \equiv 0$ throughout
Section~\ref{sec:auxiliary}. The transformation law
(\ref{eq:acgthtrans}) for the translational gauge field therefore
holds unchanged in WGT. Thus, requiring only the first relation in
(\ref{eq:nogauge}) to be preserved and following the same argument as
above leads to the erroneous conclusion that the `ungauged' limit of
WGT also corresponds to global conformal transformations, rather than
global Weyl transformations, for which $c^\alpha=0$. This further
condition can be obtained only by requiring the relations
${A^{ab}}_\mu=0$ and $B_\mu=0$ in (\ref{eq:nogauge}) are also
preserved (recall that ${f^a}_\mu \equiv 0$ in WGT). That this does
indeed lead to the correct `ungauged' limit of global Weyl
transformations can be seen immediately from the transformation laws
for ${A^{ab}}_\mu$ and $B_\mu$ in WGT, which are given by
(\ref{eq:abftrans}) with $c^a=0$.

Given the lack of a clear rationale for identifying the `ungauged'
limit of a gravitational gauge theory by imposing Lord's condition
that (the appropriate subset of) the relations (\ref{eq:nogauge})
should be preserved, it is of interest to investigate an alternative
prescription.  This may be motivated most naturally by considering
more closely the identification of the `ungauged' limit with the
requirement that field-strength tensors should vanish.

To this end, let us consider the ACGT covariant derivative of some
matter field $\vpsi(x)$, which from (\ref{eq:Dbarmudef}),
(\ref{weylgencovdef}) and (\ref{eq:gammabardef}) is given by
\be
\bar{\cal D}_c\vpsi = {h_c}^\mu(\partial_\mu + 
{\textstyle\frac{1}{2}}{A^{ab}}_\mu
\Sigma_{ab} + B_\mu\Delta + {f^a}_\mu\kappa_a)\vpsi.
\ee
It is clear that the dynamics of the matter field will be sensitive to
the translational gauge field ${h_c}^\mu$, irrespective of the nature
of $\vpsi$. This is not the case, however, for the other gauge fields
${A^{ab}}_\mu$, $B_\mu$ and ${f^a}_\mu$. Depending on the nature of
$\vpsi$, the dynamics of the matter field may be insensitive to one or
more of these gauge fields. To establish the `ungauged' limit, one
should therefore consider the `subsidiary' field-strength tensors
 obtained from $\widetilde{\cal R}^{ab}_{\phantom{ab}cd}$,
$\widetilde{\cal H}_{cd}$, ${\cal S}^{\ast a}_{\phantom{\ast a,}cd}$
and ${\cal T}^{\ast a}_{\phantom{\ast a,}cd}$ by including only those
terms that depend on ${A^{ab}}_\mu$, $B_\mu$ or ${f^a}_\mu$,
respectively (or, equivalently, by setting the other gauge fields in
each case to be identically zero).

Thus, starting from the relations (\ref{eq:nogauge}), one should
demand that under subsequent GCT and ${\cal H}$-gauge transformations
that preserve the first relation (such that $\delta_0{h_a}^\mu = 0$),
{\it all} `subsidiary' field-strength tensors remain zero. Such tensors are
still covariant under GCTs, but (typically) not so under general
${\cal H}$-gauge transformations, and hence will not automatically
remain zero, even if one starts from the set of relations
(\ref{eq:nogauge}). Thus, demanding that they do so can impose {\it further}
constraints on the allowed nature of the subsequent GCT and ${\cal
  H}$-gauge transformations, {\it beyond} the requirement imposed by
$\delta_0{h_a}^\mu = 0$ that the most general solution for
$\xi^\alpha(x)$ is the infinitesimal global conformal transformation
(\ref{eq:gct}).

A straightforward way of imposing this requirement is to demand that,
under subsequent GCT and ${\cal H}$-gauge transformations satisfying
$\delta_0{h_a}^\mu = 0$, the change in each `full' field strength
$\widetilde{\cal R}^{ab}_{\phantom{ab}cd}$, $\widetilde{\cal H}_{cd}$,
${\cal S}^{\ast a}_{\phantom{\ast a,}cd}$ and ${\cal T}^{\ast
  a}_{\phantom{\ast a,}cd}$ arising from the change in each gauge
field ${A^{ab}}_\mu$, $B_\mu$ or ${f^a}_\mu$ should vanish {\it
  separately}. It is clear that the condition $\delta_0{h_a}^\mu = 0$
guarantees that the variation in the field-strength tensors vanishes
if the variation in their non-calligraphic counterparts does.  The
transformation laws of the latter (assuming $\delta_0{h_a}^\mu = 0$)
are given in terms of the transformations of the other gauge fields by
\begin{subequations}
\label{eq:acgtfstrans}
\bea 
\delta_0\widetilde{R}^{ab}_{\phantom{\alpha\beta}\mu\nu} & = &
2\partial_{[\mu}\delta_0{A^{ab}}_{\nu]} + 8
\delta^{[a}_{[\mu}\delta_0{f^{b]}}_{\nu]},\\ 
\delta_0\widetilde{H}_{\mu\nu}
& = & 2\partial_{[\mu}\delta_0 B_{\nu]} + 4
\eta_{a[\mu}\delta_0{f^a}_{\nu]},\\ 
\delta_0 S^{\ast
  a}_{\phantom{\ast\alpha}\mu\nu} & = &
2\partial_{[\mu}\delta_0{f^{a}}_{\nu]},\\ 
\delta_0T^{\ast
  a}_{\phantom{\ast\alpha}\mu\nu} & = &
2\delta_0{A^a}_{b[\mu}\delta^b_{\nu]} +
2\delta_0B_{[\mu}\delta^a_{\nu]}.  
\eea
\end{subequations}
We thus require that each term on the RHS of these equations should
vanish separately.  From the transformation laws (\ref{eq:abftrans})
of these gauge fields, this requirement is satisfied only if
$\delta_0{A^{ab}}_{\mu}$, $\delta_0B_{\mu}$ and
$\delta_0{f^{a}}_{\nu}$ each vanish. This is, however, equivalent
merely to the final three relations in (\ref{eq:nogauge}) being
preserved, which follows automatically from our initial requirement
that the first relation in (\ref{eq:nogauge}) is preserved. Thus, no
further conditions apply and one correctly deduces that the most
general solution for $\xi^\alpha(x)$ has the form (\ref{eq:gct}) of an
infinitesimal global conformal transformation.

Let us now repeat the above process for WGT. In this case, the field
strengths are ${\cal R}^{ab}_{\phantom{ab}cd}$, ${\cal H}_{cd}$ and
${\cal T}^{\ast a}_{\phantom{\ast a,}cd}$, where the first two may be
obtained from their counterparts in ACGT by setting ${f^a}_\mu\equiv
0$, as is clear from (\ref{eq:gcovfs}). Thus, if one again starts from
the conditions (\ref{eq:nogauge}) (again recalling that
${f^a}_\mu\equiv 0$ in WGT) and demands only that the first condition
is preserved under subsequent GCT and ${\cal H}$-gauge
transformations, the transformation laws of the WGT field strengths
are given in terms of the transformations of the WGT gauge fields
${A^{ab}}_\mu$ and $B_\mu$ by the corresponding expressions in
(\ref{eq:acgtfstrans}) with $\delta_0{f^a}_\mu \equiv 0$.  From the
transformation laws of the WGT gauge fields ${A^{ab}}_\mu$ and
$B_\mu$, which may be obtained from (\ref{eq:abftrans}) by setting
$c^a\equiv 0$, one may show that our additional requirement is
satisfied only if $\delta_0{A^{ab}}_{\mu}$ and $\delta_0B_{\mu}$ each
vanish. These further conditions correspond to the second and third
relations in (\ref{eq:nogauge}) being preserved, which in turn
requires $c^a=0$. Thus, one correctly deduces that the most general
solution for $\xi^\alpha(x)$ has the form of a global Weyl
transformation.

Finally, it is a simple matter to verify that an analogous procedure
applied to PGT leads to the correct identification of the
corresponding `ungauged' limit as global Poincar\'e transformations.
Indeed, for PGT, WGT and ACGT, this procedure is found to be
equivalent to requiring that (the appropriate subset of) the relations
(\ref{eq:nogauge}) are preserved, as Lord originally suggested.  As we
will see in Section~\ref{sec:ungaugingewgt}, however, these two
approaches are not always equivalent.

\subsection{Biconformal gauging}
 
Before moving on to discuss our new approach for gauging the conformal
group in Section~\ref{sec:newapproach}, we conclude this section with
a brief discussion of an existing alternative scheme, known as
biconformal gauging, which leads to a very different conformal gauge
theory to ACGT, with several interesting properties. 

As discussed in Section~\ref{sec:auxiliary}, in the standard approach
to gauging the conformal group, one may eliminate the gauge field
${f^a}_\mu$ corresponding to local SCTs, which implies that the
symmetry has reduced back to the local Weyl group. As pointed out by
Wheeler \cite{Wheeler91}, however, this reduction is rather curious,
since in addition to the elimination of ${f^a}_\mu$, the symmetry
between the generators $P_a$ and $K_a$ in the Lie algebra of the
conformal group has also been lost. Indeed, the elimination of
${f^a}_\mu$ occurs because one chooses to identify the translational
gauge field ${h_a}^\mu$ with the vierbein (at least in the geometrical
interpretation), as is done in PGT. In gauging the conformal group,
however, one can make the alternative choice of identifying the SCT
gauge field ${f^a}_\mu$ with the vierbein, in which case the
translational gauge field ${h_a}^\mu$ may be eliminated instead. Thus,
in the case of the conformal group, there is an additional symmetry
between the two generators $P_a$ and $K_a$, which is broken by one's
(arbitrary) choice for identifying the vierbein.

An alternative approach to gauging the conformal group, which
preserves the symmetry between $P_a$ and $K_a$ by construction, is the so-called
biconformal gauging \cite{Ivanov82a,Ivanov82b}. Expressed in terms of
the quotient manifold method, in biconformal gauging the Lie group
$\mathcal{G}$ is again the conformal group (only the elements
connected to the identity) with the 15 generators
$\{P_a,M_{ab},D,K_a\}$, but the Lie subgroup $\mathcal{H}$ is now the
homogeneous Weyl group with the seven generators $\{M_{ab},D\}$. The
resulting quotient $\mathcal{G}/\mathcal{H}$ is thus an eight-dimensional
manifold, called biconformal space, which has a number of very
interesting properties \cite{Wheeler98,Wehner99,Hazboun12,Wheeler19},
as we now briefly describe.

Biconformal space is spanned by the basis 1-forms $\vect{h}_a$ and
$\vect{f}^a$, derived from the translations and special conformal
transformations, respectively. If the dilational
curvature vanishes, then the non-degenerate 2-form $\vect{h}_a \wedge
\vect{f}^a$ is also closed (i.e. an exact differential), and hence
symplectic. Moreover, the Killing metric is non-degenerate when
restricted to the base manifold, so that the group structure
determines a metric, rather than imposing one by hand.  If the basis
1-forms $\vect{h}_a$ and $\vect{f}^a$ are separately in involution and
orthogonal, then biconformal space can be considered as a form of
relativistic phase space, consisting of separate configuration and
momentum metric submanifolds. Moreover, the signatures of these
submanifolds are severely limited and, in particular, the notion of
time emerges naturally, since the configuration space must be
Lorentzian and is therefore interpreted as spacetime. It is usually
argued that the full biconformal space should be interpreted as
representing `the world', since both classical and quantum mechanics
take their most elegant forms in phase space and, moreover, a phase
space is required to formulate the uncertainty principle.

To define a dynamical theory, one writes down an action on the full
biconformal space.  Its symplectic structure means that the volume
element is dimensionless, and so an action linear in the curvatures
can be conformally invariant, without introducing any
additional (compensating) fields. If one assumes the
torsion to vanish and the momentum subspace is flat, the
theory reduces to general relativity on the spacetime tangent bundle.

It is clear that biconformal gauge theory (BCGT) has some very
interesting properties and is worthy of continued investigation, but
we will not pursue it further here.

\section{New approach to conformal gauging}
\label{sec:newapproach}

In an earlier paper \cite{eWGTpaper}, we introduced a novel alternative
to standard Weyl gauge theory, in which we proposed an `extended' form
for the transformation law of the rotational gauge field under finite local
dilations, given by
\be
{A'^{ab}}_\mu = {A^{ab}}_\mu + \theta ({b^a}_\mu{\cal Q}^b-{b^b}_\mu{\cal Q}^a),
\label{eweylatrans}
\medskip
\ee
where ${\cal Q}_a \equiv {h_a}^\mu Q_\mu$, $Q_\mu \equiv \partial_\mu
\varrho$, and $\theta$ is an arbitrary parameter that can take any
value.  We noted further there that this extended transformation law
implements Weyl scaling in a novel way that may be related to gauging
of the full conformal group. 
We now discuss this issue in more detail.

The proposal (\ref{eweylatrans}) was motivated by the observation that
the WGT (and PGT) matter actions for the massless Dirac field and the
electromagnetic field are invariant under local dilations even if one
assumes this `extended' transformation law for the rotational gauge
field, which includes its `normal' transformation law in WGT as the
special case $\theta=0$. Moreover, under a global scale
transformation, the two transformation laws clearly coincide.

A complementary motivation for introducing the extended transformation
law (\ref{eweylatrans}) is that under local dilations it places the
transformation properties of the PGT rotational gauge field strength
(or `curvature') ${{\cal R}^{ab}}_{cd}$ and translational gauge field
strength (or `torsion') ${{\cal T}^a}_{bc}$ on a more equal footing
with one another than is the case under the standard WGT
transformation law with $\theta=0$.  Indeed, assuming the extended
transformation law, they transform under
local dilations as
\begin{widetext}
\begin{subequations}
\label{eqn:rttransgeneral}
\begin{eqnarray}
{{\cal R}^{\prime ab}}_{cd} & = & e^{-2\varrho}\{{{\cal R}^{ab}}_{cd} +
2\theta\delta^{[a}_d({\cal D}_c-\theta{\cal Q}_c){\cal Q}^{b]} -
2\theta\delta^{[a}_c({\cal D}_d-\theta{\cal Q}_d){\cal Q}^{b]} -
2\theta{\cal Q}^{[a}{{\cal T}^{b]}}_{cd} -
2\theta^2\delta^{[a}_c\delta^{b]}_d {\cal Q}^e {\cal Q}_e \},
\label{rspecialt} \\
{{\cal T}^{\prime a}}_{bc} & = & e^{-\varrho}\{{{\cal T}^a}_{bc}+2(1-\theta){\cal
  Q}_{[b}\delta_{c]}^a\}.\label{eqn:tortransgeneral}
\end{eqnarray}
\end{subequations}
\end{widetext}
Thus, for general values of $\theta$, neither ${{\cal R}^{ab}}_{cd}$
nor ${{\cal T}^{a}}_{bc}$ transforms covariantly. For
$\theta=0$, however, one recovers the `normal' transformation law
for the $A$-field, such that ${{\cal R}^{ab}}_{cd}$ transforms covariantly 
under local dilations, but ${{\cal T}^a}_{bc}$ transforms
inhomogeneously. By contrast, for $\theta=1$ one obtains 
%
a covariant transformation law for ${{\cal T}^a}_{bc}$,
but an inhomogeneous one for ${{\cal R}^{ab}}_{cd}$. The extended
$A$-field transformation law (\ref{eweylatrans}) accommodates
these extreme cases in a balanced manner.

\medskip
We therefore developed our so-called `extended' Weyl gauge theory
(eWGT), which is based on the construction of a new form of covariant
derivative ${\cal D}^\dagger_a\vpsi$ that transforms in the same way
under local Weyl transformations as $\partial_\mu\vpsi$ does under the
global Weyl transformations, but where the rotational gauge field
introduced is assumed to transform under local dilations as
(\ref{eweylatrans}). The resulting theory has a number of interesting
features, which we summarise briefly below.


\subsection{Extended Weyl gauge theory (eWGT)}
\label{sec:ewgt}

In eWGT, the spacetime group ${\cal G}$ under consideration is the
inhomogeneous Weyl group and its subgroup ${\cal H}$ is the
homogeneous Weyl group, as in standard WGT. It is also assumed that
any physical (matter) fields $\vpsi(x)$ belong to an irreducible
representation of the Lorentz group (as is usually the case in
physical theories). Consequently, as discussed in
Sections~\ref{sec:infgct} and \ref{sec:finitegct}, the generator of
dilations takes the simple form $\Delta=wI$, where $w$ is the Weyl
weight of $\vpsi$.  Adopting Kibble's general methodology, the gauged
action of ${\cal G}$ on such a field is considered as a GCT $x^\mu \to
x^{\prime\mu}$, together with the local action of ${\cal H}$, such
that (in finite form) one obtains a local version of (\ref{eq:gwt}), namely
\be
\vpsi'(x') = e^{w\varrho(x)}\matri{S}(\omega(x))\vpsi(x).
\label{eq:finitelwt}
\ee

\subsubsection{Covariant derivative}

Following the usual approach, the construction of the covariant
derivative in eWGT is achieved in two steps. First, one defines the
`$\mathcal{H}$-covariant' derivative
\be
D^\dagger_\mu\vpsi(x) \equiv [\partial_\mu + \Gamma^\dagger_\mu(x)]\vpsi(x),
\label{eq:Ddaggermudef}
\ee
where $\Gamma^\dagger_\mu(x)$ is a linear combination of the
generators of $\mathcal{H}$ that depends on the gauge fields. Second,
one constructs the `generalised $\mathcal{H}$-covariant' derivative,
linearly related to $D^\dagger_\mu\vpsi$ by
\be
{\cal D}^\dagger_a\vpsi(x) \equiv  {h_a}^\mu(x)D^\dagger_\mu\vpsi(x),
\label{ewgtgencovdef}
\ee
where ${h_a}^\mu(x)$ is the translational gauge field, again assumed
to have the inverse ${b^a}_\mu(x)$.

In eWGT, however, one does not adopt the standard approach of
introducing each gauge field in $\Gamma^\dagger_\mu(x)$ as the linear
coefficient of the corresponding generator, such as in
(\ref{eq:gammabardef}), since this would lead directly to the standard
WGT transformation laws for the rotational and dilational gauge fields
(given in infinitesimal form by the first two relations in
(\ref{eq:abftrans}) with $c^a\equiv 0$). Rather, in order to
accommodate our proposed extended transformation law
(\ref{eweylatrans}) under local dilations, one is led to introduce the
`rotational' gauge field ${A^{ab}}_\mu(x)$ and the `dilational' gauge
field\footnote{We denote the `dilational' gauge field here by $B_\mu$
  and also use a different sign convention in order to harmonise our
  notation with that of WGT. To recover the notation of our original
  paper \cite{eWGTpaper}, one should make the replacements $B_\mu \to
  -V_\mu$, $H_{\mu\nu} \to -H_{\mu\nu}$ and $\zeta^\mu \to
  -\zeta^\mu$, and similarly for their counterparts carrying only Latin
  indices and/or daggers.} $B_\mu(x)$ 
in a very different way, so
that
\be
\Gamma^\dagger_\mu = {A^{\dagger ab}}_\mu\Sigma_{ab} +(B_\mu-\tfrac{1}{3}T_\mu)\Delta,
\label{eq:gammadaggerdef}
\ee
in which $T_\mu = {b^a}_\mu {\cal T}_a$, where ${\cal T}_a \equiv
{{\cal T}^b}_{ab}$ is the trace of the PGT torsion, and we have introduced
the modified $A$-field
\begin{equation}
{A^{\dagger ab}}_\mu  \equiv  {A^{ab}}_\mu + ({b^a}_\mu{\cal B}^b
- {b^b}_\mu{\cal B}^a),
\label{adaggerdef}
\end{equation}
where ${\cal B}_a = {h_a}^\mu B_\mu$. It is worth noting that
${A^{\dagger ab}}_\mu$ is not considered to be a fundamental field,
but merely a shorthand for the above combination of the gauge fields
${h_a}^\mu$ (or its inverse), ${A^{ab}}_\mu$ and $B_\mu$. Similarly,
$T_\mu$ is merely a shorthand for the corresponding function of the
gauge fields ${h_a}^\mu$ (or its inverse) and ${A^{ab}}_\mu$.  

It is straightforward to show that, if $\vpsi$ has Weyl weight $w$,
then (\ref{ewgtgencovdef}) does indeed transform covariantly 
with Weyl weight $w-1$, as required, under the gauged (finite) action
of ${\cal G}$, such that
\begin{equation}
{\cal D}^{\dagger\prime}_a\vpsi'(x') =
e^{(w-1)\varrho(x)}{\Lambda_a}^b(x) 
\matri{S}(\omega(x)) {\cal D}^\dagger_b\vpsi(x),
\label{weylgencovtrans}
\end{equation}
if the gauge fields transform according to\footnote{The derivative (\ref{ewgtgencovdef}) does in
  fact transform covariantly as in (\ref{weylgencovtrans}) under the
  much {\it wider} class of gauge field transformations in which $\theta
  Q_\mu(x)$ is replaced in (\ref{eq:ewgtatrans}--\ref{eqn:vtransform})
  by an {\it arbitrary} vector field $Y_\mu(x)$. Indeed, if one also
  makes this replacement in (\ref{eweylatrans}), one finds that the
  WGT (and PGT) matter actions for the massless Dirac field and the
  electromagnetic field are still invariant under local dilations,
  although the discussion regarding the transformation properties of
  ${{\cal R}^{ab}}_{cd}$ and ${{\cal T}^a}_{bc}$ following
  (\ref{eqn:rttransgeneral}) requires appropriate
  modification. The covariance of ${\cal
    D}^\dagger_a\vpsi$ under this wider class of transformations
  allows one to identify a further gauge symmetry of eWGT, namely
  under the simultaneous transformations ${A^{ab}}_\mu \to
  {A^{ab}}_\mu + {b^a}_\mu{\cal Y}^b - {b^b}_\mu{\cal Y}^b$ and $B_\mu
  \to B_\mu - Y_\mu$, where ${\cal Y}_a = {h_a}^\mu Y_\mu$ and $Y_\mu$
  is an arbitrary vector field. Under this symmetry, both ${A^{\dagger
      ab}}_\mu$ and $B_\mu - \tfrac{1}{3}T_\mu$ remain unchanged and
  thus ${\cal D}^\dagger_a\vpsi$ is invariant, as too are the eWGT
  field strengths and action. One may make use of this symmetry of
  eWGT to self-consistently choose a gauge in which either $B_\mu$ or
  $T_\mu$ is set to zero, which can considerably simplify subsequent
  calculations.}
\begin{widetext}
\begin{subequations}
\label{eq:ewgtgftrans}
\bea
{h'_a}^{\mu}(x^\prime) & = &
{X^\mu}_\nu\,e^{-\varrho(x)}{\Lambda_a}^b(x){h_b}^\nu(x), \label{eq:ewgthtrans}\\
{{A}^{\prime ab}}_\mu(x^\prime)  &=& {X_\mu}^\nu
[{\Lambda^a}_c(x){\Lambda^b}_d(x) {A^{cd}}_\nu(x) + {\Lambda^{[a}}_c(x)
\partial_\nu\Lambda^{b]c}(x) + 
2\theta {b^{[a}}_\nu(x){\cal Q}^{b]}(x)],\label{eq:ewgtatrans}\\
B^\prime_\mu(x^\prime) & = & {X_\mu}^\nu[B_\nu(x) -\theta Q_\nu(x)],
\label{eqn:vtransform}
\eea
\end{subequations}
\end{widetext}
where ${X^\mu}_\nu \equiv \partial{x'}^\mu/\partial{x}^\nu$ are the
elements of the GCT transformation matrix and ${X_\mu}^\nu \equiv
\partial{x^\nu}/\partial{x'}^\mu$ are the elements of its inverse.
Hence, we have achieved our goal of accommodating the $A$-field
transformation (\ref{eweylatrans}) under local dilations, while
recovering the full transformation law in WGT for the special case
$\theta=0$.  By contrast, the transformation law for $B_\mu$ reduces
to that in WGT for the special case $\theta=1$. Unlike the transformation laws for ${A^{ab}}_\mu$ and
$B_\mu$, the covariant derivative (\ref{ewgtgencovdef}) does not
explicitly contain the parameter $\theta$. Consequently, it does {\it
  not} reduce to the standard WGT covariant derivative
$D_\mu^\ast\vpsi$ in either special case $\theta=0$ or $\theta=1$,
while retaining the transformation law (\ref{weylgencovtrans}) for
{\it any} value of $\theta$.

It is clear that the structure of the `connection' $\Gamma^\dagger$ in
(\ref{eq:gammadaggerdef}) is very different to that normal adopted,
such as (\ref{eq:gammabardef}) in ACGT. In particular, whereas each
generator in (\ref{eq:gammabardef}) is multiplied purely by the
corresponding gauge field, each generator in (\ref{eq:gammadaggerdef})
is multiplied by a (non-linear) function of all the gauge fields,
including the translational gauge field ${h_a}^\mu$ (or its inverse),
which is completely absent from (\ref{eq:gammabardef}). This results
in eWGT having a fundamentally different structure to standard gauge
theories. In particular, the eWGT field strengths depend very
differently on the gauge fields from their counterparts in other gauge
theories, as we now describe.

\subsubsection{Field strengths}

The eWGT gauge field strengths are defined in the usual way in terms
of the commutator of the covariant derivatives.  Considering first the
eWGT ${\cal H}$-covariant derivative, one finds that
\begin{equation}
[D^\dagger_\mu,D^\dagger_\nu]\vpsi = (\tfrac{1}{2}{R^{\dagger
ab}}_{\mu\nu}\Sigma_{ab} + H^\dagger_{\mu\nu}\Delta)\vpsi,
\label{rdaggerdefine}
\end{equation}
which is of an analogous form to the corresponding result in WGT
(which may be obtained from (\ref{dstarmucomm}) by setting 
${f^a}_\mu=0$ and hence ${S^{\ast a}}_{\mu\nu}=0$), but
the eWGT field strengths have very different dependencies on the gauge
fields. In particular, one finds
\bea
{R^{\dagger ab}}_{\mu\nu}  & \equiv & 2(\partial_{[\mu} {A^{\dagger ab}}_{\nu]} +
\eta_{cd}{A^{\dagger ac}}_{[\mu}{A^{\dagger db}}_{\nu]}),\nonumber \\
H^\dagger_{\mu\nu}
& \equiv & 2\partial_{[\mu} (B_{\nu]}-\tfrac{1}{3}T_{\nu]}),
\eea
both of which transform covariantly under GCT and local Lorentz
rotations in accordance with their respective index structures, and
are invariant under local dilations.

Considering next the commutator of two `generalised ${\cal H}$-covariant'
derivatives, one finds
\begin{equation}
[{\cal D}^\dagger_c,{\cal D}^\dagger_d]\vpsi = (\tfrac{1}{2}{{\cal R}^{\dagger
ab}}_{cd}\Sigma_{ab} + {\cal H}^\dagger_{cd}\Delta
- {{\cal T}^{\dagger a}}_{cd}{\cal D}^\dagger_a) \vpsi,
\label{ewgtdacomm}
\end{equation}
where ${{\cal R}^{\dagger ab}}_{cd}
= {h_c}^\mu {h_d}^\nu {R^{\dagger ab}}_{\mu\nu}$ and
${\cal H}^\dagger_{cd} = {h_c}^\mu {h_d}^\nu H^\dagger_{\mu\nu}$, and
the translational field strength is given by
\begin{equation}
{{\cal T}^{\dagger a}}_{bc} \equiv {h_b}^\mu {h_c}^\nu
{T^{\dagger a}}_{\mu\nu} \equiv 2{h_b}^\mu {h_c}^\nu
D^\dagger_{[\mu} {b^a}_{\nu]}.
\label{ewgttorsiondef}
\end{equation}
We note that 
${{\cal R}^{\dagger ab}}_{cd}$ and ${{\cal T}^{\dagger a}}_{bc}$
are given in terms of their counterparts ${{\cal R}^{ab}}_{cd}$
and ${{\cal T}^{a}}_{bc}$ in PGT by
\bea
{{\cal R}^{\dagger ab}}_{cd} & \!\!=\! & \!{{\cal R}^{ab}}_{cd}
\!+\!4\delta^{[b}_{[c}(\!{\cal D}_{d]} \!-\! {\cal B}_{d]}\!){\cal B}^{a]}
\!-\!2 {\cal B}^2 \delta^{[a}_c \delta^{b]}_d
\!-\!2 {\cal B}^{[a}{{\cal T}^{b]}}_{cd},\nonumber \\
{{\cal T}^{\dagger a}}_{bc}  & = &  {{\cal T}^a}_{bc}
+\tfrac{2}{3}\delta^a_{[b} {\cal T}_{c]},
\eea
where ${\cal B}^2 \equiv {\cal B}^a {\cal B}_a$ and for brevity we
have introduced the derivative operator ${\cal D}_a \equiv {h_a}^\mu
D_\mu \equiv {h_a}^\mu(\partial_\mu +
\tfrac{1}{2}{A^{ab}}_\mu\Sigma_{ab})$ familiar from
PGT.  It is particularly important to note that the
trace of the eWGT torsion vanishes identically, namely ${\cal
  T}^\dagger_b \equiv {{\cal T}^{\dagger a}}_{ba} = 0$, so that
${{\cal T}^{\dagger a}}_{bc}$ is completely trace free (contraction on
any pair of indices yields zero). ${{\cal
    R}^{\dagger ab}}_{cd}$, ${\cal H}^\dagger_{cd}$ and ${{\cal
    T}^{\dagger a}}_{cd}$ are GCT scalars and transform covariantly
under local Lorentz transformations and under local dilations with
weights $w({{\cal R}^{\dagger ab}}_{cd})=w({\cal H}^\dagger_{cd})=-2$
and $w({{\cal T}^{\dagger a}}_{cd})=-1$ respectively.

\subsubsection{Action}

As in other gravitational gauge theories, the total action in eWGT
consists typically of kinetic terms for any matter field(s) $\vpsi$,
terms describing the coupling of the matter field(s) to the
gravitational gauge fields (and possibly to each other), and (kinetic)
terms describing the dynamics of the free gravitational gauge
fields. 

Since ${\cal D}^\dagger_a\vpsi$ is constructed to have an
analogous transformation law under extended local Weyl transformations
to that of $\partial_\mu\vpsi$ under global Weyl transformations, one
may immediately construct a matter action that is fully invariant
under the extended gauged Weyl group from one that is invariant under
global Weyl transformations by employing the usual minimal coupling
procedure of replacing partial derivatives by covariant ones to
obtain\footnote{It should be noted, however, that this minimal
  coupling procedure can (as in other gravitational gauge
  theories \cite{Lasenby98}) result in an action $S_{\rm M}$ that no
  longer satisfies other (required) invariance properties of the
  original one; indeed, this occurs for the Faraday action of the
  electromagnetic field, for which the minimal coupling procedure
  destroys electromagnetic gauge invariance. In such cases, the gauged
  action must be modified to restore the original (required)
  invariances, so that $S_{\rm M}$ does not have the form given above.}
\be
S_{\rm M} =
\int h^{-1} L_{\rm M}(\vpsi_i,{\cal D}^\dagger_a\vpsi_i)\,d^4x.
\ee
As mentioned previously, the set of fields $\vpsi_i$ may already
include a scalar compensator field (denoted also by $\phi$) with Weyl
weight $w=-1$, for example in a Yukawa coupling term of the form
$\mu\phi\bar{\psi}\psi$ with a massless Dirac field $\psi$ (since this
allows for the Dirac field to acquire a mass dynamically upon adopting
the Einstein gauge $\phi=\phi_0$) \cite{Dirac73,eWGTpaper}, together perhaps with
kinetic and quartic potential terms for $\phi$ of the form $\nu{\cal
  D}^\dagger_a\phi\,{\cal D}^{\dagger a}\phi -\lambda\phi^4$ (where
$\mu$, $\nu$ and $\lambda$ are dimensionless parameters).

The terms in the total action that describe the dynamics of the free
gravitational gauge fields are constructed from the gauge field
strengths. In contrast to ACGT, the eWGT field strengths all transform
covariantly under the full group of localised transformations, and so
may be used straightforwardly to construct the free-gravitational
action. The requirement of local scale invariance requires the
free-gravitational Lagrangian $L_{\rm G}$ to be a relative scalar with
Weyl weight $w(L_{\rm G})=-4$, which may therefore contain an
arbitrary linear combination $L_{{\cal R}^{\dagger 2}}$ of the six
distinct terms quadratic in ${\cal R}^\dagger_{abcd}$ and its
contractions, and a term $L_{{\cal H}^{\dagger 2}} \propto {\cal
  H}^\dagger_{ab}{\cal H}^{\dagger ab}$. In principle, one could also
include quartic terms in ${\cal T}^\dagger_{abc}$ (which has no
non-trivial contractions, unlike its counterparts in PGT, WGT and
ACGT), or cross terms such as ${\cal R}^{\dagger [ab]}{\cal
  H}^{\dagger}_{ab}$, but these are not usually considered. Thus, one
typically has
\be
S_{\rm G} = \int h^{-1}(L_{{\cal R}^{\dagger 2}} + L_{{\cal H}^{\dagger 2}})\,d^4x,
\ee
where any parameters in the action are dimensionless.

In particular, $L_{\rm G}$ cannot contain the linear Einstein--Hilbert
analogue term $L_{{\cal R}^\dagger} \equiv -\tfrac{1}{2}a {\cal
  R}^\dagger$ (where ${\cal R}^\dagger \equiv {{\cal R}^{\dagger
    ab}}_{ab}$ and the factor of $-1/2$ is conventional) or $L_{{\cal
    T}^{\dagger 2}} \equiv \beta_1{\cal T}^\dagger_{abc}{\cal
  T}^{\dagger abc} + \beta_2 {\cal T}^\dagger_{abc}{\cal T}^{\dagger
  bac}$. Nonetheless, such terms can be included in the total
Lagrangian if they are multiplied by a compensator scalar field term
$\phi^2$ \cite{Dirac73,Padmanabhan85}. Such combinations are
therefore usually considered not to belong to the {\it free}
gravitational Lagrangian and are instead added to the matter
Lagrangian $L_{\rm M}$ \cite{Mannheim06}.  Thus, the matter Lagrangian
may have an extended form, including all interactions of the matter
fields with the gravitational gauge fields, which is given by $L_{{\rm
    M}^+} \equiv L_{\rm M} + \phi^2(L_{{\cal R}^\dagger} + L_{{\cal
    T}^{\dagger 2}})$ (in which the parameters $a$, $\beta_i$ are
again dimensionless), such that the corresponding action has the
functional dependencies
\be
S_{\rm M} =
\int h^{-1} L_{{\rm M}^+}(\vpsi_i,{\cal D}^\dagger_a\vpsi_i,{\cal
  R}^{\dagger},{\cal T}^\dagger_{abc}) \,d^4x,
\label{eq:ewgtmatteraction}
\ee
where the set of fields $\vpsi_i$ includes the scalar compensator. In
any case, it is only the form of the total Lagrangian $L_{\rm T} =
L_{{\rm M}^+} + L_{\rm G}$ that is relevant for the field equations.

Finally, it is worth noting that terms containing covariant
derivatives of (contracted) field strengths, such as ${\cal
  D}^{\dagger}_a{\cal D}^{\dagger}_b{\cal R}^{\dagger ab}$ or ${\cal
  D}^{\dagger}_a{\cal D}^{\dagger a} {\cal R}^\dagger$, are of Weyl weight
$w=-4$ and so can, in principle, be included in $L_{\rm
  G}$ \cite{Wheeler91}. In eWGT, however, such terms contribute only
surface terms to the action, as a consequence of the trace ${\cal
  T}^\dagger_a$ of the eWGT torsion vanishing identically. Thus, such
terms have no effect on the resulting field equations, and so may
be omitted (at least classically); this is not true in general for other
gauges theories, such as PGT, WGT and ACGT.

\subsubsection{Field equations}

The eWGT field equations are obtained by varying the total action
$S_{\rm T}$ with respect to the gravitational gauge fields
${h_a}^\mu$, ${A^{ab}}_\mu$ and $B_\mu$, together with the matter
fields $\vpsi_i$ (which may include a scalar compensator field
$\phi$).  Defining\footnote{The definition of ${\sigma_{ab}}^\mu$ used here
  differs by a factor of 2 from that used in our original
  paper \cite{eWGTpaper}, in order to allow for a more straightforward
  comparison to be made with the canonical spin-angular-momentum
  tensor. To recover the notation of our original
  paper \cite{eWGTpaper}, one should make the replacement 
${\sigma_{ab}}^\mu \to 2{\sigma_{ab}}^\mu$, and similarly for its
  counterpart carrying only Latin indices.} ${\tau^a}_\mu\equiv
\delta{\cal L}_{\rm T}/\delta {h_a}^\mu$, ${\sigma_{ab}}^\mu\equiv
2\delta{\cal L}_{\rm T}/\delta {A^{ab}}_\mu$ and
$\zeta^\mu\equiv\delta{\cal L}_{\rm T}/\delta B_\mu$, where ${\cal
  L}_{\rm T} \equiv h^{-1}L_{\rm T}$, the set of gravitational field
equations are most naturally expressed in terms of their counterparts
carrying only Latin indices ${\tau^a}_b \equiv {\tau^a}_\mu
{h_b}^\mu$, ${\sigma_{ab}}^c \equiv {\sigma_{ab}}^\mu {b^c}_\mu$ and
$\zeta^a\equiv \zeta^\mu {b^a}_\mu$, as
\begin{subequations}
\label{eqn:eweylgenfe}
\begin{eqnarray}
{\tau^{a}}_b & = & 0, \label{eqn:eweylgenfe1}\\
{\sigma_{ab}}^c & = & 0,\label{eqn:eweylgenfe2}\\
\zeta^{a} & = & 0.\label{eqn:eweylgenfe3}
\end{eqnarray}
\end{subequations}

The quantities ${\tau^{a}}_b$, ${\sigma_{ab}}^c$ and $\zeta^{a}$ are
clearly scalars under GCT, and it is straightforward to show that each
of them also transforms covariantly under local Lorentz rotations and
local dilations, as expected, with Weyl weights $w=0$, $w=1$ and $w=1$
respectively. Moreover, with one exception, these transformation
properties also hold for the corresponding quantities obtained from
{\it any subset} of the terms in $L_{\rm T}$ that transforms
covariantly with weight $w=-4$ under local Lorentz rotations and local
dilations (for example, $L_{\rm M}$, $L_{{\rm M}^+}$  or $L_{\rm G}$ separately). The
exception relates to quantities corresponding to ${\tau^a}_b$, which
transform covariantly under local dilations only if one considers all
the terms in $L_{\rm T}$, and then only by virtue of the $A$-field
equation (\ref{eqn:eweylgenfe2}).  This unusual feature is a result of
the extended transformation law (\ref{eweylatrans}) for ${A^{ab}}_\mu$
containing one of the other gauge fields, namely ${b^a}_\mu$, and
leads one to introduce the related quantities
\be
{\tau^{\dagger a}}_{b} \equiv  {\tau^a}_{b} - {\sigma_{cb}}^a{\cal B}^c -
     {\sigma^{ca}}_{c}{\cal B}_b.
\label{eq:covemtdef}
\ee
These do transform covariantly when one considers only some subset of
the terms in $L_{\rm T}$ that themselves transform covariantly and
with weight $w=-4$ under local Lorentz rotations and local
dilations. It is therefore more convenient to replace (\ref{eqn:eweylgenfe1}) with
the alternative field equation
\be
{\tau^{\dagger a}}_b  =  0.
\label{eq:tdaggereom}
\ee
Indeed, this field equation emerges naturally if one adopts an
alternative variational principle, in which ${A^{ab}}_\mu$ is replaced
by ${A^{\dagger ab}}_\mu$ in the set of field variables; this approach
also considerably shortens the calculations involved in deriving all
the gravitational field equations.

Another unusual feature of the eWGT field equations, which also emerges
most naturally from the alternative variational principle, is that for
any total Lagrangian $L_{\rm T} $ in which the gravitational gauge
fields appear only through eWGT covariant derivatives or field
strengths (which is usually the case), one may show that
\be
\zeta^a \equiv {\sigma^{ba}}_b.
\label{eq:virialrel}
\ee
Consequently, in this generic case, the $B$-field equation
(\ref{eqn:eweylgenfe3}) is no longer independent, but merely the
relevant contraction of the $A$-field equation
(\ref{eqn:eweylgenfe2}). Moreover, the relation (\ref{eq:virialrel})
also holds for the corresponding quantities obtained from {\it any
  subset} of the terms in $L_{\rm T}$ that transforms covariantly with
weight $w=-4$ under local Lorentz rotations and local dilations.

Finally, the remaining (matter) field equations are obtained by
varying $S_{\rm T}$ with respect to the fields $\vpsi_i$ (which may
include a scalar compensator field $\phi$). In the (usual) case in
which $L_{\rm T}$ is a function of the matter fields only through $\vpsi_i$
and ${\cal D}_a^\dagger\vpsi_i$, these may be shown to have the simple
(and manifestly covariant) forms
\be
\bpd{L_{\rm T}}{\vpsi_i}-{\cal D}^\dagger_a
\left(\pd{L_{\rm T}}{({\cal D}^\dagger_a\vpsi_i)}\right)  =  0,
\ee
where $\bar{\partial}L_{\rm T}/\partial \vpsi_i \equiv [\partial L_{\rm
    T}(\vpsi_i,{\cal D}^\dagger_au)/\partial \vpsi_i]_{u=\vpsi_i}$, so that
$\vpsi_i$ and ${\cal D}^\dagger_a\vpsi_i$ are treated as independent
variables.

\subsubsection{Conservation laws}
Invariance of $S_{\rm T}$ under (infinitesimal) GCTs, local Lorentz
rotations and extended local dilations, respectively, lead to
conservation laws of the general form (\ref{eq:n2cond1}), as
discussed in Appendix~\ref{app:b2}. These can be written in the
following manifestly covariant form:
\begin{widetext}
\begin{subequations}
\label{eqn:ewgtmconsall}
\begin{eqnarray}
{\cal D}^\dagger_c(h{\tau^{\dagger c}}_d) 
+ h({\tau^{\dagger c}}_b {{\cal T}^{\dagger b}}_{cd}
-\tfrac{1}{2}{\sigma_{ab}}^c {{\cal R}^{\dagger ab}}_{cd}
- \zeta^{\dagger c} {\cal H}_{cd})
+ \frac{\delta L_{\rm T}}{\delta\vpsi_i}
{\cal D}^\dagger_d\vpsi_i & = & 0,\label{eqn:ewgtmcons1}\\
{\cal D}^\dagger_c(h{\sigma_{ab}}^c) + 2h\tau^\dagger_{[ab]}
+\frac{\delta L_{\rm T}}{\delta\vpsi_i}
\Sigma_{ab}\vpsi_i & = & 0,\\
h{\tau^{\dagger c}}_c - \frac{\delta
  L_{\rm T}}{\delta\vpsi_i} w_i\vpsi_i & = & 0,
\label{eqn:ewgtmcons4}\\
{\cal D}^\dagger_c(h\zeta^{\dagger c}) & = & 0,\label{eqn:ewgtmcons3}
\end{eqnarray}
\end{subequations}
\end{widetext}
where we have defined the quantities $\zeta^{\dagger a} \equiv
\zeta^a- {\sigma^{ba}}_b$ and ${\cal H}_{ab} = 2{h_a}^\mu
     {h_b}^\nu\partial_{[\mu} B_{\nu]}$, which are both easily
     verified to be GCT scalars and to transform covariantly under
     local Lorentz rotations and local dilations.

These conservation equations have a very different form to those in
WGT. In particular, invariance of $S_{\rm T}$ under local dilations
leads to {\it both} of the last two conservation laws.  The third
conservation law (\ref{eqn:ewgtmcons4}) is unusual in being an
algebraic condition on the trace ${\tau^{\dagger a}}_a$ in terms of
the field equations of the matter fields $\vpsi_i$. Indeed, assuming
the field equations of all non-compensator matter fields to hold, one thus
finds that the field equation for the scalar compensator field $\phi$
is no longer independent, but simply related to the trace of the
(alternative) $h$-field equation (\ref{eq:tdaggereom}).
Also, as mentioned above, for the usual forms of $S_{\rm T}$, the condition
(\ref{eq:virialrel}) holds, in which case $\zeta^{\dagger a} \equiv 0$ and so
the first conservation law (\ref{eqn:ewgtmcons1}) is simplified and the 
the final one (\ref{eqn:ewgtmcons3}) is satisfied identically. 

Finally, it is worth noting that the conservation laws
(\ref{eqn:ewgtmconsall}) also hold for corresponding quantities
obtained from {\it any subset} of terms in $L_{\rm T}$ that is
covariant under local Lorentz transformations and under local
dilations with weight $w=-4$ (e.g.\ $L_{\rm M}$, $L_{{\rm M}^+}$ or
$L_{\rm G}$ separately).



\subsection{Finite local conformal invariance}


As we noted in Sections~\ref{sec:infgct} and \ref{sec:finitegct}, for
physical fields $\vpsi_i(x)$ that belong to irreducible
representations of the Lorentz group, as is assumed in eWGT, the
action (both infinitesimal and finite) of a general element of the
conformal group that is connected to the identity corresponds to a
combination of a translation, (proper) Lorentz rotation and dilation;
in particular, a SCT corresponds  merely to a Lorentz rotation and
dilation that depend on spacetime position $x$ in a prescribed
way. Since translations, (restricted) Lorentz rotations and dilations are
already gauged in eWGT (and WGT), then so too are SCTs and hence any
element of the conformal group that is connected to the identity.

As discussed in Section~\ref{sec:finitegct}, however, the full conformal group
also includes the inversion operation (\ref{eq:inversion}), which is
finite and discrete, and hence not connected to the
identity. Moreover, it is
worth recalling that a SCT is merely the
composition of an inversion, a translation and a second inversion.  In
Section~\ref{sec:finitegct}, we demonstrated that an inversion,
together with its action on physical fields that belong to an irreducible
representation of the Lorentz group, consists of the composition of a
dilation $1/x^2$ and a reflection ${I^\mu}_\nu(\hat{x})$ in the
hyperplane perpendicular to $\hat{x}$, both of which are clearly
position dependent in a prescribed way. In particular, under an
inversion, physical fields are acted upon by ${I^\mu}_\nu(\hat{x})$
for each tensor index and by $\gamma\cdot\hat{x}$ for each 4-spinor
index.

Since dilations are already gauged in eWGT (and WGT), the only new
operation to consider is the reflection.  To our knowledge, the
gauging of reflections has not been addressed previously, but the most
natural approach is to generalise the reflection in the hyperplane
perpendicular to $\hat{x}$ at each point to a reflection in the
hyperplane perpendicular to some unit vector $\hat{n}(x)$ that can
vary arbitrarily with spacetime position $x$. As usual, this gauged
transformation should be completely decoupled from GCTs, and so we
denote the reflection matrix at each spacetime point by
${I^a}_b(\hat{n}(x))$, which operates on each Latin tensor index carried by a
field (or, equivalently, $\gamma\cdot\hat{n}(x)$ for each spinor
index).

From the discussion in Section~\ref{sec:finitegct}, however,
${I^a}_b(\hat{n}(x))$ corresponds to a finite improper Lorentz
transformation matrix at each spacetime point. Thus eWGT {\it already}
accommodates gauged reflections, without the need to introduce any
more gauge fields, provided that each occurrence of the
restricted Lorentz
transformation matrix ${\Lambda^a}_b(x)$ in the finite transformation
laws (\ref{weylgencovtrans}) and (\ref{eq:ewgtgftrans}) for the
covariant derivative and the existing gauge fields, respectively, is
extended to denote a general transformation matrix of the full Lorentz
group (which consists of proper Lorentz rotations and spacetime
reflections) and, in particular, is given by ${I^a}_b(\hat{n}(x))$
under gauged reflections. Indeed, the same holds true for WGT, for
which the finite transformation laws of the gauge fields are given by
(\ref{eq:ewgtgftrans}), with $\theta=0$ in (\ref{eq:ewgtatrans}) and
$\theta=1$ in (\ref{eqn:vtransform}).

Thus, provided all matter fields $\vpsi_i(x)$ are assumed to belong to
irreducible representations of the Lorentz group and with the above
modest extension to the transformation laws of the gauge fields, {\it
  both} WGT and eWGT accommodate {\it all} the gauged symmetries of the full
conformal group, such that actions constructed in the usual way in
each theory are invariant under (finite) local conformal
transformations. As we now demonstrate below, however, WGT {\it
  cannot} be considered as a true gauge theory of the conformal group
in the usual sense, whereas eWGT {\it can} be interpreted as such.

\subsection{Local conformal conservation laws}

As discussed in Section~\ref{sec:globalcft}, if one considers a field
theory in Minkowski spacetime that describes the dynamics of a set of
fields $\vpsi_i(x)$ that belong to irreducible representations of the
Lorentz group, then for the action (\ref{srmatteraction}) to be
invariant under global conformal transformations (that are connected
to the identity), one requires the first three conservation laws in
(\ref{eq:gctcons}) to hold `on-shell' (which together ensure
Poincar\'e and scale invariance) {\it and} the field virial
(\ref{eq:virialdef}) to vanish (which ensures the additional
invariance under SCTs), up to a total divergence. 

We now consider in more detail the forms of the conservation laws in
WGT and eWGT, both of which we have just demonstrated have actions
that are invariant under local conformal transformations. As we will
see, eWGT has very different conservation laws to WGT, which arises
primarily from the unconventional form of the eWGT covariant
derivative as compared with other gravitational gauge theories.

\subsubsection{WGT}
 
Let us first consider a matter action in WGT containing all the terms
in the total action except those that depend only on the gauge fields
and their derivatives.  This typically has the form $S_{\rm M} = \int
h^{-1} L_{{\rm M}^+}(\vpsi_i,{\cal D}^\ast_a\vpsi_i,{\cal R},{\cal
  T}^\ast_{abc})\,d^4x$, by analogy with our discussion leading to
(\ref{eq:ewgtmatteraction}) in the context of eWGT. Assuming that the
matter equations of motion are satisfied (including that of the
compensator field), such that $\delta L_{{\rm M}^+}/\delta\vpsi_i=0$, invariance
of $S_{\rm M}$ under local Weyl transformations leads to three conservation
laws of the general form (\ref{eq:n2cond1}), which can be written as
the manifestly covariant conditions \cite{eWGTpaper}
\begin{widetext}
\begin{subequations}
\label{eqn:wgtmconsall}
\bea
({\cal D}^\ast_c + {\cal T}^\ast_c)(h{\tau^c}_d)
+ h({\tau^c}_b {{\cal T}^{\ast b}}_{cd}-\tfrac{1}{2}{\sigma_{ab}}^c {{\cal R}^{ab}}_{cd}
- \zeta^c {\cal H}_{cd}) & = & 0,\label{wgtlmcons2}\\
({\cal D}^\ast_c + {\cal T}^\ast_c)(h{\sigma_{ab}}^c)
+ 2h\tau_{[ab]} & = & 0, \label{wgtlmcons1}\\
({\cal D}^\ast_c + {\cal T}^\ast_c)(h\zeta^c) -
h{\tau^c}_c & = & 0, \label{wgtlmcons3}
\eea
\end{subequations}
\end{widetext}
where ${\tau^a}_\mu\equiv \delta{\cal L}_{{\rm M}^+}/\delta {h_a}^\mu$,
${\sigma_{ab}}^\mu\equiv 2\delta{\cal L}_{{\rm M}^+}/\delta {A^{ab}}_\mu$
and $\zeta^\mu\equiv\delta{\cal L}_{{\rm M}^+}/\delta B_\mu$ (in which
${\cal L}_{{\rm M}^+} \equiv h^{-1}L_{{\rm M}^+}$), and their counterparts
carrying only Latin indices ${\tau^a}_b \equiv {\tau^a}_\mu
{h_b}^\mu$, ${\sigma_{ab}}^c \equiv {\sigma_{ab}}^\mu {b^c}_\mu$ and
$\zeta^a\equiv \zeta^\mu {b^a}_\mu$ are most naturally considered as
the (total) {\it dynamical} energy-momentum, spin-angular-momentum and
dilation current, respectively, of the matter fields. The above
conservation laws are clearly invariant under GCTs and transform
covariantly under the local action of the subgroup ${\cal H}$ of
homogeneous Weyl transformations, as expected. 

The conservation laws (\ref{eqn:wgtmconsall}) provide a natural
generalisation for localised Weyl transformations of the first three
conservation laws in (\ref{eq:gctcons}). There is not, however, any
{\it further} conservation law corresponding to the generalisation of
the condition that the field virial (\ref{eq:virialdef})
should vanish up to a total divergence, which was necessary to ensure
that the original action (\ref{srmatteraction}) be invariant under
SCTs, in addition to global Weyl transformations, and hence invariant
under global conformal transformations (connected to the
identity). The absence of such a further conservation law in WGT
demonstrates that it does not constitute a gauge theory of the
conformal group in the usual sense.

One should note, however, that the quantities in
(\ref{eqn:wgtmconsall}) are dynamical currents, whereas those in
(\ref{eq:gctcons}) are canonical. It is therefore of interest to
compare the forms of these two types of current. This comparison is
facilitated by first separating the contributions to the dynamical
matter currents resulting from each of the terms in ${\cal L}_{{\rm
  M}^+} \equiv {\cal L}_{\rm M} + \phi^2 {\cal L}_{{\cal R}} + \phi^2
{\cal L}_{{\cal T}^{\ast 2}}$, which we denote by ${\tau^a}_b =
{(\tau_{\rm M})^a}_b + {(\tau_{\cal R})^a}_b + {(\tau_{{\cal
      T}^2})^a}_b$, and similarly for ${\sigma_{ab}}^c$ and
$\zeta^a$. We then introduce the following {\it covariant canonical}
currents of the matter fields \cite{Blagojevic02}
\begin{subequations}
\label{eq:wgtcccurrents}
\bea
{\mathfrak{t}^{\ast a}}_b &\equiv& \pd{L_{\rm M}}{({\cal D}^\ast_a\vpsi_i)}{\cal
  D}^\ast_b\vpsi_i-\delta^a_b L_{\rm M},\\
{\mathfrak{s}^{\ast c}}_{ab} &\equiv& \pd{L_{\rm M}}{({\cal D}^\ast_c\vpsi_i)}
\Sigma_{ab}\vpsi_i,\\
{\mathfrak{j}^{\ast a}} &\equiv& \pd{L_{\rm M}}{({\cal D}^\ast_a\vpsi_i)}
w_i\vpsi_i,
\eea
\end{subequations}
which provide a natural generalisation of the standard canonical
currents ${t^\mu}_a$, ${s^\mu}_{\alpha\beta}$ and $j^\mu$ in
(\ref{eq:confcurrents}) and (\ref{eq:srcurrentsdef}). By considering
the form of the WGT covariant derivative ${\cal D}^\ast_a\vpsi_i$, one
may show directly that the covariant canonical currents and the 
dynamical currents derived from ${\cal L}_{\rm M}$ alone
are essentially {\it equivalent} in WGT, since $h{(\tau_{\rm M})^a}_b \equiv
{\mathfrak{t}^{\ast a}}_b$, $h{(\sigma_{\rm M})_{ab}}^c \equiv
{\mathfrak{s}^{\ast c}}_{ab}$ and $h(\zeta_{\rm M})^a \equiv \mathfrak{j}^{\ast
  a}$.  These equivalences may also be derived by demanding the
coincidence of the currents $J^\mu$ and $S^\mu$ derived from ${\cal L}_{\rm
  M}$, which are discussed in
Appendix~\ref{app:b2}.

\subsubsection{eWGT}

Let us now repeat the above analysis for a matter action of the form
$S_{\rm M} = \int h^{-1} L_{{\rm M}^+}(\vpsi_i,{\cal
  D}^\dagger_a\vpsi_i,{\cal R}^\dagger,{\cal T}^\dagger_{abc})\,d^4x$
in eWGT, as given in (\ref{eq:ewgtmatteraction}). In this case, from the
combination of (\ref{eq:virialrel}) and (\ref{eqn:ewgtmconsall})
(applied only to $S_{\rm M}$ and assuming all the matter equations of
motion to hold), one instead obtains the conditions
%
\begin{subequations}
\label{eq:ewgtcls}
\begin{eqnarray}
{\cal D}^\dagger_c(h{\tau^{\dagger c}}_d) 
+ h({\tau^{\dagger c}}_b {{\cal T}^{\dagger b}}_{cd}
-\tfrac{1}{2}{\sigma_{ab}}^c {{\cal R}^{\dagger ab}}_{cd}) & = & 0,
\\
{\cal D}^\dagger_c(h{\sigma_{ab}}^c) + 2h\tau^\dagger_{[ab]} & = & 0,
\\
h{\tau^{\dagger c}}_c & = & 0,\label{eq:ewgtcl3}
\\
h(\zeta^a-{\sigma^{ba}}_b) & = & 0, \label{eq:ewgtcl4}
\end{eqnarray}
\end{subequations}
%
where ${\tau^{\dagger a}}_b$ is defined in (\ref{eq:covemtdef}). The
conditions (\ref{eq:ewgtcls}) have a somewhat different form from
their WGT counterparts in (\ref{eqn:wgtmconsall}). In particular,
(\ref{eq:ewgtcl3}) shows that the trace of the modified dynamical
energy-momentum tensor vanishes. This is reminiscent of the vanishing
trace of the improved energy-momentum tensor (\ref{eq:improvedem}),
which encodes the invariance of theories under global scale
transformations. In (\ref{eq:ewgtcls}), however, one has not used the
Belinfante procedure to combine the translational and rotational
currents, but instead retained the distinction between them. Thus,
${\tau^{\dagger a}}_b$ remains non-symmetric, which is appropriate
when working in terms of the tetrad rather than the metric, and also
allows one straightforwardly to accommodate torsion. Most important in
eWGT, however, is the additional final condition (\ref{eq:ewgtcl4}),
which is analogous to a covariant generalisation of the condition that
the field virial (\ref{eq:virialdef}) should vanish. Thus the eWGT
conservations laws (\ref{eq:ewgtcls}) provide a natural local
generalisation of {\it all} of the usual conservation laws
(\ref{eq:gctcons}) for theories that are invariant under global
conformal transformations (and contain only fields that belong to
irreducible representations of the Lorentz group).

As was the case in our consideration of the WGT conservation laws,
however, it is also of interest to consider the relationship between
the dynamical currents in (\ref{eq:ewgtcls}) and their canonical
counterparts. This comparison is again facilitated by first separating
the contributions to the dynamical matter currents resulting from each
of the terms in ${\cal L}_{{\rm M}^+} \equiv {\cal L}_{\rm M} + \phi^2
{\cal L}_{{\cal R}^\dagger} + \phi^2 {\cal L}_{{\cal T}^{\dagger 2}}$, in a
similar manner to that used for WGT. We also define a set of eWGT
covariant canonical currents ${\mathfrak{t}^{\dagger a}}_b$,
${\mathfrak{s}^{\dagger c}}_{ab}$ and $\mathfrak{j}^{\dagger a}$ in an
analogous manner to their WGT counterparts in
(\ref{eq:wgtcccurrents}), but with each occurrence of the WGT covariant
derivative ${\cal D}^\ast_a\vpsi_i$ replaced by the eWGT covariant
derivative ${\cal D}^\dagger_a\vpsi_i$. By considering the form of the
latter, one may again directly relate the dynamical and covariant
canonical currents, but in eWGT these relationships are somewhat more
complicated than those in WGT.  In particular, one finds (after a
lengthy calculation) that
\begin{subequations}
\bea
h{(\tau_{\rm M})^{\dagger a}}_b &\equiv& {\mathfrak{t}^{\dagger a}}_b 
+ \tfrac{1}{3}({\cal D}^\dagger_b\mathfrak{j}^{\dagger a} 
- \delta^a_b{\cal D}^\dagger_c\mathfrak{j}^{\dagger c}), \\
h{(\sigma_{\rm M})_{ab}}^c &\equiv& {\mathfrak{s}^{\dagger c}}_{ab} 
+ \tfrac{1}{3}(\delta^c_a\mathfrak{j}^\dagger_b 
- \delta^c_b\mathfrak{j}^\dagger_a), \\
h(\zeta_{\rm M})^a &\equiv& \mathfrak{j}^{\dagger a} -{\mathfrak{s}^{\dagger ba}}_b.
\label{eq:canondil}
\eea
\end{subequations}
Once again, these equivalences may also be derived by demanding the
coincidence of the currents $J^\mu$ and $S^\mu$ as derived from ${\cal
  L}_{\rm M}$, as discussed in Appendix~\ref{app:b2}.  Substituting
the above expressions into (\ref{eq:ewgtcls}), for the restricted case
in which ${\cal L}_{\rm M}$ is the full matter Lagrangian density, yields
the covariant canonical conservation laws
\begin{subequations}
\label{eq:ewgtccls}
\bea
{\cal D}^\dagger_c{\mathfrak{t}^{\dagger c}}_d 
+ {\mathfrak{t}^{\dagger c}}_b {{\cal T}^{\dagger b}}_{cd}
-\tfrac{1}{2}{\mathfrak{s}^{\dagger c}}_{ab} {{\cal R}^{\dagger ab}}_{cd}
-  \mathfrak{j}^{\dagger c}{\cal H}^\dagger_{cd} & = & 0,\phantom{AA}\\
{\cal D}^\dagger_c{\mathfrak{s}^{\dagger c}}_{ab}+
2\mathfrak{t}^\dagger_{[ab]} 
& = & 0,\\
{\cal D}^\dagger_c\mathfrak{j}^{\dagger c}-{\mathfrak{t}^{\dagger
    c}}_c &=& 0,
\eea
\end{subequations}
and the final condition (\ref{eq:ewgtcl4}) is satisfied identically. 

The expressions (\ref{eq:ewgtccls}) clearly represent a natural local
generalisation of the first three conservation laws in
(\ref{eq:gctcons}). Moreover, one sees from (\ref{eq:canondil}) that,
provided $(\zeta_{\rm M})^a$ vanishes up to a total divergence, then
so too should ${\mathfrak{s}^{\dagger ba}}_b - \mathfrak{j}^{\dagger
  a}$, which provides a replacement additional condition that is a
natural generalisation of the analogous requirement on the field
virial (\ref{eq:virialdef}) for globally conformal invariant theories.
This requirement is indeed satisfied, not only by $(\zeta_{\rm M})^a$
but also by $\zeta^a$ evaluated from the extended
matter Lagrangian density ${\cal L}_{{\rm M}^+}$, which is given by
\be
h\zeta^a = h(\zeta_{\rm M})^a
-(\tfrac{1}{2}\nu + 3a) {\cal D}^{\dagger a}\phi^2,
\label{eq:ewgtvirialrel}
\ee
provided the terms in ${\cal L}_{{\rm M}^+}$ corresponding to the
non-compensator matter fields $\vpsi_i$ do not contain the dilation
gauge field $B_\mu$. This occurs naturally if the $\vpsi_i$ correspond
to the Dirac field and/or the electromagnetic
field \cite{eWGTpaper}. Thus, in terms of the covariant canonical
currents, the eWGT conservation laws once again
provide a natural local
generalisation of all of the usual conservation laws
(\ref{eq:gctcons}) for theories that are invariant under global
conformal transformations.

\subsection{Ungauging eWGT}
\label{sec:ungaugingewgt}

In Section~\ref{sec:ungaugingcft}, we considered the process of
`ungauging' ACGT, and obtained the correct limit of global conformal
transformations. We also considered `ungauging' WGT and found the
limit to correspond to global Weyl transformations, which again shows
that WGT cannot be considered as a true gauge theory of the conformal
group. In this section, we consider the `ungauged' limit of eWGT.

One begins by requiring the field-strength tensors ${{\cal R}^{\dagger
    ab}}_{cd}$, ${\cal H}^\dagger_{cd}$ and ${{\cal T}^{\dagger
    a}}_{cd}$ to vanish in the `ungauged' limit. Similarly to WGT, in
this limit, the coordinate system and ${\cal H}$-gauge can be chosen
such that
\be
{h_a}^\mu(x) = \delta_a^\mu,\quad {A^{ab}}_\mu(x)=0, \quad B_\mu(x)=0.
\label{eq:nogaugeewgt}
\ee
It is worth noting that, in this reference system, the eWGT covariant
derivative reduces to a partial derivative. Thus, the eWGT covariant
canonical currents ${\mathfrak{t}^{\dagger a}}_b$,
${\mathfrak{s}^{\dagger c}}_{ab}$ and $\mathfrak{j}^{\dagger a}$
reduce to the standard ones in (\ref{eq:confcurrents}) and
(\ref{eq:srcurrentsdef}), and the conditions (\ref{eq:ewgtccls}) and
(\ref{eq:ewgtvirialrel}) reduce, respectively, to the first three
conservation laws in (\ref{eq:gctcons}) and the vanishing of the field
virial (\ref{eq:virialdef}) up to a total divergence. Indeed, we note
further that all these reductions also occur under the less
restrictive set of conditions ${h_a}^\mu(x) = \delta_a^\mu$ and
${A^{\dagger ab}}_\mu(x)=0$, which also ensure that ${{\cal
    R}^{\dagger ab}}_{cd}$, ${\cal H}^\dagger_{cd}$ and ${{\cal
    T}^{\dagger a}}_{cd}$ vanish.

Let us now apply an analogous approach to that discussed in
Section~\ref{sec:ungaugingcft} to `ungauge' eWGT. We thus begin by
considering the eWGT covariant derivative of some matter field
$\vpsi(x)$ (here belonging to some irreducible representation of the
Lorentz group), which from (\ref{eq:Ddaggermudef}),
(\ref{ewgtgencovdef}) and (\ref{eq:gammadaggerdef}) is given by
\be
{\cal D}^\dagger_c\vpsi = {h_c}^\mu[\partial_\mu + 
{\textstyle\frac{1}{2}}{A^{\dagger ab}}_\mu
\Sigma_{ab} + (B_\mu-\tfrac{1}{3}T_\mu)\Delta]\vpsi,
\ee
where ${\cal T}_a \equiv {{\cal T}^b}_{ab}$ is the trace of the PGT
torsion and the modified $A$-field is ${A^{\dagger ab}}_\mu \equiv
{A^{ab}}_\mu + ({b^a}_\mu{\cal B}^b - {b^b}_\mu{\cal B}^a)$, as
defined in (\ref{adaggerdef}). As previously, the dynamics of the
matter field will be sensitive to the translational gauge field
${h_c}^\mu$, irrespective of the nature of $\vpsi$. In eWGT, however,
depending on the nature of $\vpsi$, the dynamics of the matter field
may be insensitive to one or both of the {\it combinations}
${A^{\dagger ab}}_\mu$ and $B_\mu-\tfrac{1}{3}T_\mu$ of the gauge
fields.  Thus, following the reasoning presented in
Section~\ref{sec:ungaugingcft}, to establish the `ungauged' limit one
should consider the `subsidiary' field-strength tensors obtained from ${{\cal R}^{\dagger ab}}_{cd}$, ${\cal
  H}^\dagger_{cd}$ and ${{\cal T}^{\dagger a}}_{cd}$ by including only
those terms that depend on the combinations ${A^{\dagger ab}}_\mu$ or
$B_\mu-\tfrac{1}{3}T_\mu$ of the gauge fields, respectively.

\subsubsection{Infinitesimal transformations}

We first consider infinitesimal transformations, as we did in
Section~\ref{sec:ungaugingcft}. Thus, starting from the relations
(\ref{eq:nogaugeewgt}), one should demand that under subsequent GCT
and ${\cal H}$-gauge transformations that preserve the first relation
(such that $\delta_0{h_a}^\mu = 0$ and so ensuring the equivalence of
Latin and Greek indices before and after the transformation), all
`subsidiary' field-strength tensors remain zero.  The infinitesimal
form of the eWGT transformation law for ${h_a}^\mu$ is easily obtained
from its finite form in (\ref{eq:ewgtgftrans}), and is identical to
that given in (\ref{eq:hgammatrans}) for ACGT (and WGT). Thus,
following the argument given in Section~\ref{sec:ungaugingcft}, the
most general solution for $\xi^\alpha(x)$ has the form (\ref{eq:gct})
of an infinitesimal global conformal transformation. One now has to
check, however, if any further conditions apply to this solution by
our requirement on the behaviour of the `subsidiary' field-strength
tensors.


By analogy with the discussion in Section~\ref{sec:ungaugingcft}, a
straightforward way of imposing this requirement is to demand that,
under subsequent GCT and ${\cal H}$-gauge transformations satisfying
$\delta_0{h_a}^\mu = 0$, the change in each `full' field strength
${{\cal R}^{\dagger ab}}_{cd}$, ${\cal H}^\dagger_{cd}$ and ${{\cal
    T}^{\dagger a}}_{cd}$ arising from the change in either combination
of gauge fields ${A^{\dagger ab}}_\mu$ or $B_\mu-\tfrac{1}{3}T_\mu$
should vanish {\it separately}. Unlike the cases considered in
Section~\ref{sec:ungaugingcft}, however, the quantities ${A^{\dagger
    ab}}_\mu$ and $B_\mu-\tfrac{1}{3}T_\mu$ are not
independent. Indeed, starting from (\ref{adaggerdef}) and assuming
$\delta_0{h_a}^\mu = 0$, it is easily shown that
$\delta_0(B_\mu-\tfrac{1}{3}T_\mu)=\tfrac{1}{3}\delta_0{A^{\dagger
    a}}_{\mu a}$. Thus, one need consider only the changes in 
${{\cal R}^{\dagger ab}}_{cd}$, ${\cal H}^\dagger_{cd}$ and ${{\cal
    T}^{\dagger a}}_{cd}$ arising from the change in the combination
${A^{\dagger ab}}_\mu$ of the gauge fields.


As previously, the condition $\delta_0{h_a}^\mu = 0$ guarantees that
the variation in the field-strength tensors vanishes if the variation
in their non-calligraphic counterparts does so.  The transformation
laws of the latter (assuming $\delta_0{h_a}^\mu = 0$) are given very
simply in terms of the transformations of the quantities ${A^{\dagger
    ab}}_\mu$ by
\begin{subequations}
\label{eq:ewgtfstrans}
\bea 
\delta_0 R^{\dagger ab}_{\phantom{\alpha\beta}\mu\nu} & = &
2\partial_{[\mu}\delta_0{A^{\dagger ab}}_{\nu]},\label{eq:ewgtrtrans}\\ 
\delta_0H^\dagger_{\mu\nu} & = & 
\tfrac{2}{3}\partial_{[\mu}\delta_0{A^{\dagger b}}_{\nu]b},\label{eq:ewgthfstrans}\\ 
\delta_0T^{\dagger  a}_{\phantom{\ast\alpha}\mu\nu} 
& = & 
2\delta_0{A^{\dagger a}}_{b[\mu}\delta^b_{\nu]}
-\tfrac{2}{3}\delta^a_{[\mu}\delta_0{A^{\dagger b}}_{\nu]b}.
\label{eq:ewgtttrans}
\eea
\end{subequations}
Since these expressions depend solely on $\delta_0{A^{\dagger
    ab}}_\mu$, our procedure is equivalent to demanding only that the
variation in each field-strength tensor vanishes, but this is
satisfied by construction. Alternatively, one may show this directly
by making use of the transformation law $\delta_0 {A^{\dagger ab}}_\mu =
-2\delta^{[a}_\mu\partial^{b]}\varrho$, which may be derived by taking
the infinitesimal limits of (\ref{eq:ewgtgftrans}) and assuming
$\delta_0{h_a}^\mu = 0$. Substituting this form for $\delta_0
{A^{\dagger ab}}_\mu$ into (\ref{eq:ewgtfstrans}) one finds that
(\ref{eq:ewgthfstrans}) and (\ref{eq:ewgtttrans}) vanish identically,
and (\ref{eq:ewgtrtrans}) vanishes by virtue of the condition
(\ref{eq:gctinterp2}).  Hence no further conditions apply to this
solution, which is sufficient to show that the `ungauged' limit of eWGT
corresponds to global conformal transformations; this differs markedly
from WGT, for which we showed in Section~\ref{sec:ungaugingcft} that
the `ungauged' limit corresponds to global Weyl transformations.

It is worth noting that under the global conformal transformation
(\ref{eq:gct}), the second and third conditions in
(\ref{eq:nogaugeewgt}) are {\it not} preserved. Indeed, one finds
\be
\delta_0{A^{ab}}_\mu = 4(1-\theta)\delta_\mu^{[a}c^{b]},\qquad 
\delta_0B_\mu=2\theta c_\mu,
\label{eq:ewgtugab} 
\ee
which in turn lead to $\delta_0{A^{\dagger ab}}_\mu =
4\delta_\mu^{[a}c^{b]}$.  Thus, as anticipated in
Section~\ref{sec:ungaugingcft}, although applying our `ungauging'
approach to WGT is equivalent to requiring that all the conditions in
(\ref{eq:nogaugeewgt}) are preserved, this equivalence does not hold
when it is applied to eWGT. Indeed, as is clear from
(\ref{eq:ewgtugab}), the latter requirement in eWGT would lead to the
condition $c_\mu=0$, which corresponds to a global Weyl
transformation.

\subsubsection{Finite transformations}

We may extend our discussion to finite transformations, which also
include inversions. Starting again from the relations
(\ref{eq:nogaugeewgt}), one should demand that under subsequent finite
GCT and ${\cal H}$-gauge transformations that preserve the first
relation (such that ${h'_a}^\mu(x') = \delta_a^\mu$),
all `subsidiary' field-strength tensors remain zero.

It is straightforward to show that ${h'_a}^\mu(x') = \delta_a^\mu$ is
a necessary and sufficient condition for the coordinate transformation
matrix to satisfy (\ref{eq:fgctmat}); this in turn satisfies
(\ref{eq:ctdef}), from which it follows that the most general form for
the transformation is a finite global conformal transformation
satisfying the conditions (\ref{eq:fgctrel}), as described in
Section~\ref{sec:finitegct}. It therefore remains to check if any
further conditions apply arising from our requirement on the behaviour
of the `subsidiary' field-strength tensors.

By analogy with the infinitesimal case, we demand that, under
subsequent GCT and ${\cal H}$-gauge transformations satisfying
${h'_a}^\mu(x') = \delta_a^\mu$, the change in each `full' field
strength ${{\cal R}^{\dagger ab}}_{cd}$, ${\cal H}^\dagger_{cd}$ and
${{\cal T}^{\dagger a}}_{cd}$ arising from the change in either
combination of gauge fields ${A^{\dagger ab}}_\mu$ or
$B_\mu-\tfrac{1}{3}T_\mu$ should vanish {\it separately}. Starting
from (\ref{adaggerdef}) and assuming ${h'_a}^\mu(x') = \delta_a^\mu$,
one may show that
$B'_\mu(x')-\tfrac{1}{3}T'_\mu(x')=\tfrac{1}{3}{A^{\dagger\prime
    a}}_{\mu a}(x')$, and so again one need only consider changes in
${{\cal R}^{\dagger ab}}_{cd}$, ${\cal H}^\dagger_{cd}$ and ${{\cal
    T}^{\dagger a}}_{cd}$ arising from the change in the combination
${A^{\dagger ab}}_\mu$ of the gauge fields.

The condition ${h'_a}^\mu(x') = \delta_a^\mu$ guarantees that the
transformed field-strength tensors vanish if their
non-calligraphic counterparts do so.  The transformation laws of the
latter (assuming ${h'_a}^\mu(x') = \delta_a^\mu$) are given in
terms of ${A^{\dagger\prime ab}}_\mu(x')$ by
\begin{subequations}
\label{eq:ewgtfstransf}
\bea 
R^{\dagger\prime ab}_{\phantom{\alpha\beta}\mu\nu} & = &
2\partial'_{[\mu}{A^{\dagger\prime ab}}_{\nu]} 
+2{A^{\dagger\prime a}}_{e[\mu} {A^{\dagger\prime
      eb}}_{\nu]},\label{eq:ewgtrtransf}\\ 
H^{\dagger\prime}_{\mu\nu} & = & 
\tfrac{2}{3}\partial'_{[\mu}{A^{\dagger\prime b}}_{\nu]b},\label{eq:ewgthfstransf}\\ 
T^{\dagger\prime  a}_{\phantom{\ast\alpha}\mu\nu} 
& = & 
2{A^{\dagger\prime a}}_{b[\mu}\delta^b_{\nu]}
-\tfrac{2}{3}\delta^a_{[\mu}{A^{\dagger\prime b}}_{\nu]b}.
\label{eq:ewgtttransf}
\eea
\end{subequations}
As in the infinitesimal case, since these expressions depend solely on
${A^{\dagger\prime ab}}_\mu$, our procedure is equivalent to demanding
only that each transformed field-strength tensor vanishes, but this is
again satisfied by construction.  Alternatively, one may show this
directly by making use of the transformation law ${A^{\dagger\prime
    ab}}_\mu = -2\delta^{[a}_\mu\partial^{b]}\varrho$, which may be
derived from (\ref{eq:ewgtgftrans}) with the assumption
${h'_a}^\mu(x') = \delta_a^\mu$. Substituting this form for
${A^{\dagger\prime ab}}_\mu$ into (\ref{eq:ewgtfstransf}) one finds
that (\ref{eq:ewgthfstransf}) and (\ref{eq:ewgtttransf}) vanish
identically, and (\ref{eq:ewgtrtransf}) vanishes by virtue of the
condition (\ref{eq:fgctrel2}). Hence no further conditions apply to
the solution, so that the `ungauged' limit of eWGT corresponds to
finite global conformal transformations, including inversions.

\section{Conclusions}
\label{sec:conc}

We have reconsidered the process of gauging of the conformal group and
the resulting construction of gravitational gauge theories that are
invariant under local conformal transformations. The standard approach
leads to auxiliary conformal gauge theories (ACGT), so called because
they suffer from the problem that the gauge field corresponding to
special conformal transformations can be eliminated from the theory
using its own equation of motion, so that the symmetry appears to
reduce back to the local Weyl group. Such theoretical difficulties
with AGCT have led to the development of an alternative biconformal
gauging and the construction of its associated biconformal gauge field
theories (BCGT). Although these theories possess some very interesting and
promising properties, their physical interpretation is complicated by
their requirement of an eight-dimensional base manifold. Thus, the
role played by local conformal invariance in gravitational gauge
theories remains uncertain.

We have therefore revisited the recently proposed extended Weyl gauge
theory (eWGT), which was previously noted to implement Weyl scaling in
a novel way that may be related to gauging of the full conformal
group. We demonstrated this relationship here by first showing that,
provided any physical matter fields belong to an irreducible
representation of the Lorentz group, eWGT is indeed invariant under
the full set of (finite) local conformal transformations, including
inversions. This property is, however, also shared by standard WGT, as
might be expected from the theoretical shortcomings of
ACGT. Nonetheless, we also showed that eWGT has two further properties
not shared by WGT. First, the conservation laws of eWGT provide a
natural local generalisation of those satisfied by field theories with
global conformal invariance, in particular that the field virial
should vanish; this is the key criterion for an action to be invariant
under SCTs, in addition to the remainder of the global conformal group
(connected to the identity). Second, we showed that the `ungauged' limit
of eWGT corresponds to global conformal transformations, rather than
global Weyl transformations.  These findings suggest that eWGT can be
regarded as a valid alternative gauge theory of the conformal group,
despite not having been derived by direct consideration of the
localisation of its group parameters.  Therefore, eWGT might be
considered as a `concealed' conformal gauge theory (CCGT).

\begin{acknowledgments}
The authors thank Will Barker for useful comments. 
\end{acknowledgments}
\vspace*{-0.2cm}
\appendix

\bigskip
\section{Direct derivation of finite global conformal transformations}
\label{app:a}

For a finite coordinate transformation $x^{\prime\mu} = f^\mu(x)$ in
$n$-dimensional Minkowski spacetime to satisfy the defining condition
(\ref{eq:ctdef}) to be conformal, one immediately requires
\be
(\partial_\alpha f_\gamma)(\partial_\beta f^\gamma) 
= \tfrac{1}{n}(\partial_\nu f_\mu)(\partial^\nu f^\mu)\eta_{\alpha\beta},
\label{eq:A1}
\ee
where $(\partial_\nu f_\mu)(\partial^\nu f^\mu) = n\Omega^2$ and 
$\partial_\nu f^\mu = \partial x^{\prime\mu}/\partial x^\nu =
{X^\mu}_\nu$ is the coordinate transformation matrix. Equation
(\ref{eq:A1}) is the finite version of the conformal Killing
equation (\ref{killingeq}), to which it reduces in the infinitesimal
limit $x^{\prime\mu} \approx x^\mu + \xi^\mu(x)$.

Acting on (\ref{eq:A1}) with $\partial_\lambda$, cyclically permuting
the indices $\lambda$, $\alpha$ and $\beta$ to obtain two further
equivalent equations and subtracting the first equation from the sum
of the other two, one obtains
\be
2(\partial_\alpha\partial_\beta f^\gamma)(\partial_\lambda f_\gamma) = 
\tfrac{1}{n}(\eta_{\lambda\alpha}\partial_\beta + \eta_{\beta\lambda}\partial_\alpha
-\eta_{\alpha\beta}\partial_\lambda)\Omega^2,
\label{eq:A2}
\ee
of which we will make use shortly.

Another useful equation may be obtained by first acting on
(\ref{eq:A1}) with $\partial^\beta$, then acting on the resulting
equation with $\partial_\beta$, symmetrising on $\alpha$ and $\beta$
and finally using (\ref{eq:A1}) again. This yields
\bea
[\eta_{\alpha\beta}\square^2 +
  (n-2)\partial_\alpha\partial_\beta]\Omega^2
&+&2(\partial_\alpha\partial_\beta f^\gamma)\square^2f_\gamma\phantom{AAA(\partial_\alpha\partial_\lambda f^\gamma)}\nonumber\\
&-&2(\partial_\alpha\partial_\lambda f^\gamma)
(\partial_\beta\partial^\lambda f_\gamma)=0,
\label{eq:A3}
\eea
from which one may derive two further useful equations. 
First,
contracting (\ref{eq:A3}) with $\eta^{\alpha\beta}$, one obtains
\be
(n-1)\square^2\Omega^2 + (\square^2 f^\gamma)(\square^2 f_\gamma)
-(\partial_\tau\partial_\lambda f^\gamma)
(\partial^\tau\partial^\lambda f_\gamma)=0.
\label{eq:A4}
\ee
Then, multiplying (\ref{eq:A4}) by $\eta^{\alpha\beta}$ and
subtracting the result from $(n-1)$ times (\ref{eq:A3}) gives
\begin{widetext}
\be
(n-1)(n-2)\partial_\alpha\partial_\beta\Omega^2 
+ 2(n-1)[(\partial_\alpha\partial_\beta f^\gamma)\square^2f_\gamma
-(\partial_\alpha\partial_\lambda f^\gamma)
(\partial_\beta\partial^\lambda f_\gamma)]
-\eta_{\alpha\beta}[(\square^2 f^\gamma)(\square^2 f_\gamma)
-(\partial_\tau\partial_\lambda f^\gamma)
(\partial^\tau\partial^\lambda f_\gamma)]=0.
\label{eq:A5}
\ee
\end{widetext}

As discussed in Section~\ref{sec:finitegct}, one may write the
transformation matrix of a smooth conformal transformation in the form
(\ref{eq:fgctmat}), such that
\be
\partial_\nu f^\mu = \Omega(x){\Lambda^\mu}_\nu(x),
\label{eq:A6}
\ee
where ${\Lambda^\mu}_\nu(x)$ is, in general, a position-dependent
Lorentz rotation matrix (either proper or improper). First,
substituting (\ref{eq:A6}) into (\ref{eq:A2}), one finds that
$\Omega(x)$ and ${\Lambda^\mu}_\nu(x)$ must satisfy the relation
\be
\partial_\mu{\Lambda^\gamma}_\beta = ({\Lambda^\gamma}_\mu\partial_\beta
- \eta_{\mu\beta}{\Lambda^\gamma}_\lambda\partial^\lambda)\ln\Omega,
\label{eq:A7}
\ee
from which one may straightforwardly obtain the result
(\ref{eq:fgctrel1}), namely
\be
{\Lambda_\gamma}^\alpha\partial_\mu\Lambda^{\gamma\beta} - 
2\delta^{[\alpha}_\mu\partial^{\beta]}\ln\Omega =  0.
\label{eq:A8}
\ee
Then, substituting (\ref{eq:A6}) into (\ref{eq:A4}) and (\ref{eq:A5}),
respectively, and using the result (\ref{eq:A7}), one finds
\begin{widetext}
\bea (n-1)[2\Omega\square^2\Omega +
  (n-4)(\partial_\gamma\Omega)(\partial^\gamma\Omega)] &=& 0, \label{eq:A9}\\
(n-1)(n-2)[2\Omega\partial_\alpha\partial_\beta\Omega
+\eta_{\alpha\beta}(\partial_\gamma\Omega)(\partial^\gamma\Omega)
-4(\partial_\alpha\Omega)(\partial_\beta\Omega)]&=&0,\label{eq:A10}
\eea
\end{widetext}
where the second result matches (\ref{eq:fgctrel2}) for $n \ge 3$.

In order to solve (\ref{eq:A8}--\ref{eq:A10}) for $\Omega$ and
${\Lambda^\mu}_\nu$, it is in fact more convenient work in terms of
the reciprocal dilation $\sigma \equiv 1/\Omega$, for which
(\ref{eq:A8}--\ref{eq:A10}) become
\bea
{\Lambda_\gamma}^\alpha\partial_\mu\Lambda^{\gamma\beta} +
2\delta^{[\alpha}_\mu\partial^{\beta]}\ln\sigma &=&  0,
\label{eq:A8sigma} \\
(n-1)[n(\partial_\gamma\sigma)(\partial^\gamma\sigma)-2\sigma\square^2\sigma]&=&0,
\label{eq:sigmarel1}\\
(n-1)(n-2)[\eta_{\alpha\beta}(\partial_\gamma\sigma)(\partial^\gamma\sigma)-
2\sigma\partial_\alpha\partial_\beta\sigma]&=&0.\phantom{AAA}\label{eq:sigmarel2}
\eea

Assuming $n \ge 3$, (\ref{eq:sigmarel2}) immediately implies that
$\partial_\alpha\partial_\beta\sigma=0$ for $\alpha\ne\beta$. One thus
requires $\sigma(x) = q_\mu(x^\mu)$, i.e. the sum of $n$ functions,
each of which is a function only of the corresponding coordinate (and
possibly a constant). Substituting this form back into
(\ref{eq:sigmarel2}) with $\alpha=\beta$ and adopting the signature
$\eta_{\alpha\beta}=\mbox{diag}(1,-1,\ldots,-1)$, one finds that
\be q_0^{\prime\prime} = -q_i^{\prime\prime} =
\tfrac{1}{2}\sigma(\partial_\gamma\ln\sigma)
(\partial^\gamma\ln\sigma),
\label{eq:qdprimerel}
\ee
where primes denote differentiation with respect to the function
argument and the index $i$ runs from $1$ to $n-1$. Thus, up to a sign,
each $q_\mu^{\prime\prime}$ must be equal to the same
constant. Consequently, $\sigma(x)$ must have the form
\be
\sigma(x) = a + 2c_\mu x^\mu + bx^2,
\ee
where $a$, $b$ and $c_\mu$ are constants. Finally, substituting this
form back into (\ref{eq:qdprimerel}) or (\ref{eq:sigmarel1}), yields
the condition that $ab=c^2$.

Let us first assume that the vector $c_\mu$ is non-null. In this case,
there are three non-trivial possibilities: 
\begin{enumerate}[label=(\roman*)]
\item $a\ne 0$ and
$b=0=c_\mu$, so that $\sigma = a$; 
\item $b\ne 0$ and $a=0=c_\mu$, so
that $\sigma=bx^2$; 
\item $a$, $b$ and at least one component of $c_\mu$ are non-zero, so
  that $\sigma=a(1+2\bar{c}_\mu x^\mu + \bar{c}^2x^2)$, where
  $\bar{c}_\mu\equiv c_\mu/a$.
\end{enumerate}
Turning then to the case where the vector $c_\mu$ is non-zero but
null, one requires at least one of $a$ and $b$ to be zero. Hence, 
there are three further possibilities:
\begin{enumerate}[label=(\roman*),start=4]
\item $a\ne 0$ and $b=0$, so that $\sigma = a + 2c_\mu x^\mu$; 
\item $a=0$ and $b\neq 0$, so that $\sigma = 2c_\mu x^\mu + bx^2$;
\item $a=0=b$, so that $\sigma= 2c_\mu x^\mu$.
\end{enumerate}

For each of the above possible forms for the reciprocal scale factor
$\sigma(x)$, one may now use (\ref{eq:A8sigma}) to determine the form of
the corresponding (proper or improper) Lorentz transformation
${\Lambda^\mu}_\nu(x)$, and hence the full transformation matrix
${X^\mu}_\nu(x) = [1/\sigma(x)]{\Lambda^\mu}_\nu(x)$ in each case. Before
proceeding, however, it should be noted that (\ref{eq:A8sigma})
is insufficient to determine ${\Lambda^\mu}_\nu(x)$ fully, since
if ${\Lambda^\mu}_\nu(x)$
satisfies (\ref{eq:A8sigma}), then so too does 
${\Lambda^\mu}_\lambda(x){\Lambda^\lambda}_\nu$, where ${\Lambda^\lambda}_\nu$
may be any position-independent Lorentz transformation matrix. With
this caveat in mind, we now consider each of the possible forms for
$\sigma(x)$ listed above.
\begin{enumerate}[label=(\roman*)]
\item For $\sigma=a$, one requires ${\Lambda^\mu}_\nu =
  \mbox{constant}$. Thus ${X^\mu}_\nu$ corresponds to a combination of
  a position-independent scaling $1/a$, (proper or improper) Lorentz
  transformation and translation (namely a global Weyl
  transformation).
\item For $\sigma=bx^2$, (\ref{eq:A8sigma}) is solved by
  ${\Lambda^\mu}_\nu(x) = {I^\mu}_\nu(\hat{x})$, which corresponds to
  a reflection in the hyperplane perpendicular to the unit vector
  $\hat{x}$. Thus ${X^\mu}_\nu$ corresponds an inversion followed by a
  position-independent scaling $1/b$.
\item For $\sigma=a(1+2\bar{c}_\mu x^\mu + \bar{c}^2x^2)$,
  (\ref{eq:A8sigma}) is solved by ${\Lambda^\mu}_\nu(x) =
  {I^\mu}_\lambda(\hat{x}'){I^\lambda}_\nu(\hat{x})$, where
  $x^{\prime\mu}$ is given by (\ref{eq:finitesct}) with $c_\mu$
  replaced by $\bar{c}_\mu$. Thus ${X^\mu}_\nu$ corresponds to a SCT,
  in which the intermediate translation is through the vector
  $\bar{c}_\mu$, followed by a position-independent scaling $1/a$ (see
  Section~\ref{sec:finitegct}).
\item For $\sigma = a + 2c_\mu x^\mu = a(1+2\bar{c}_\mu x^\mu)$,
  (\ref{eq:A8sigma}) is solved by ${\Lambda^\mu}_\nu(x) =
  {I^\mu}_\lambda(\hat{x}'){I^\lambda}_\nu(\hat{x})$, where
  $x^{\prime\mu}$ is given by (\ref{eq:finitesct}) with $c_\mu$
  replaced by $\bar{c}_\mu$, and for which $\bar{c}^2=0$.  Thus
  ${X^\mu}_\nu$ corresponds to a SCT, in which the intermediate
  translation is through the null vector $\bar{c}_\mu$, followed by a
  position-independent scaling $1/a$.
\item For $\sigma = 2c_\mu x^\mu + bx^2 = b(2\tilde{c}_\mu x^\mu +
  x^2)$, where $\tilde{c}_\mu \equiv c_\mu/b$, (\ref{eq:A8sigma}) is
  solved by ${\Lambda^\mu}_\nu(x) = {I^\mu}_\nu[\hat{n}(x)]$, where
  the unit vector $\hat{n}(x)$ has components
\be
\hat{n}^\mu(x) = 
\frac{x^\mu + \tilde{c}^\mu}{\sqrt{x^2+2\tilde{c}\cdot x}}.
\ee
It is straightforward to show that the resulting ${X^\mu}_\nu$
corresponds to a translation through $\tilde{c}^\mu$, followed by an
inversion, followed by a position-independent scaling $1/b$.
\item For $\sigma= 2c_\mu x^\mu$, (\ref{eq:A8sigma}) is solved by
  ${\Lambda^\mu}_\nu = {I^\mu}_\nu(\hat{c})$. In a similar way to
  case (v), it is straightforward to show that the resulting
  ${X^\mu}_\nu$ corresponds to a translation through $c^\mu$ in the
  limit $c^\mu \to \infty$, followed by an inversion.

\end{enumerate}

In deriving the above solutions, we have made use of the following
results. First, if ${\Lambda^\mu}_\nu(x) = {I^\mu}_\nu[\hat{n}(x)]$,
which corresponds to a reflection in the hyperplane perpendicular to a
position-dependent unit vector $\hat{n}(x)$, the first term in
(\ref{eq:A8sigma}) may be written as
\be
{\Lambda_\gamma}^\alpha\partial_\mu\Lambda^{\gamma\beta} = 4 \hat{n}^{[\alpha}\partial_\mu\hat{n}^{\beta]}.
\ee
By then considering the identity
$\hat{n}^{[\gamma}\hat{n}^{\alpha}\partial_\mu\hat{n}^{\beta]} = 0$,
one quickly finds that (\ref{eq:A8sigma}) implies that $\hat{n}_\mu
\propto \partial_\mu\sigma$. Second, if ${\Lambda^\mu}_\nu(x) =
{I^\mu}_\lambda[\hat{n}(x)]{I^\lambda}_\nu[\hat{m}(x)]$, which corresponds to a
reflection in the hyperplane perpendicular to a position-dependent
unit vector $\hat{m}(x)$, followed by a reflection in the hyperplane
perpendicular to $\hat{n}(x)$ (which together constitute a local
rotation in the hyperplane defined by $\hat{m}(x)$ and $\hat{n}(x)$,
through twice the angle between them), then 
the first term in (\ref{eq:A8sigma}) may be written as
\bea
{\Lambda_\gamma}^\alpha\partial_\mu\Lambda^{\gamma\beta} 
= 4(\hat{m}^{[\alpha}\partial_\mu\hat{m}^{\beta]} &+& 
\hat{n}^{[\alpha}\partial_\mu\hat{n}^{\beta]}
+ 2\hat{m}^{[\alpha}\hat{n}^{\beta]}\,
\hat{m}_\gamma\partial_\mu\hat{n}^\gamma \nonumber \\
&-&2\hat{m}\cdot\hat{n}\,\hat{m}^{[\alpha}\partial_\mu\hat{n}^{\beta]}).
\eea

As expected, cases (i)-(vi) contain only the four distinct finite elements
of the conformal group, namely position-independent translations,
rotations and scalings, together with inversions.


\section{Global and local symmetries in field theory}
\label{app:b}

Consider a Minkowski spacetime ${\cal M}$, labelled using Cartesian
inertial coordinates $x^\mu$, in which the dynamics of some set of fields
$\chi(x) = \{\chi_i(x)\}$ ($i=1,2,\ldots$) is described by the action
\begin{equation}
S = \int {\cal L}(\chi,\partial_\mu\chi)\,d^4x.
\label{eq:genmatteraction}
\end{equation}
It should be understood here that the index $i$ merely labels
different matter fields, rather than denoting the tensor or spinor
components of individual fields (which are suppressed throughout). It
is worth noting that, in general, the fields $\chi_i(x)$ may include
matter fields $\vpsi_i(x)$ and gauge fields $g_i(x)$.

Invariance of the action (\ref{eq:genmatteraction}) under an
infinitesimal coordinate transformation $x^{\prime\mu} = x^\mu +
\xi^\mu(x)$ and form variations $\delta_0\chi_i(x)$ in the fields
(where the latter do not necessarily result solely from the coordinate
transformation), implies that, up to a total divergence of any
quantity that vanishes on the boundary of the integration region, one
has
\be
\delta_0 {\cal L} + \partial_\mu(\xi^\mu {\cal L}) = 0,
\label{eq:genactioninv}
\ee
where the form variation of the Lagrangian is given by
\be
\delta_0 {\cal L} = \pd{{\cal L}}{\chi_i}\delta_0\chi_i + 
\pd{{\cal L}}{(\partial_\mu\chi_i)}\delta_0(\partial_\mu\chi_i),
\label{eq:lagformvar}
\ee
and, according to the usual summation convention, there is an
implied sum on the index $i$.

The invariance condition (\ref{eq:genactioninv}) can alternatively be
rewritten as
\be
\frac{\delta {\cal L}}{\delta\chi_i}\delta_0\chi_i + \partial_\mu J^\mu
= 0,
\label{eq:genactioninv2}
\ee
where $\delta {\cal L}/\delta\chi_i$ denotes the standard variational
derivative
and the Noether current $J^\mu$ is given by
\be
J^\mu = \momc{\mu}{i}\delta_0\chi_i + \xi^\mu
{\cal L}.
\label{eq:noethercurrent}
\ee
If the field equations $\delta L/\delta\chi_i=0$ are satisfied, then
(\ref{eq:genactioninv2}) reduces to the (on-shell) conservation law
$\partial_\mu J^\mu=0$, which is the content of Noether's first
theorem and applies both to global and local symmetries.

\subsection{Global symmetries}
\label{app:b1}

Let us first consider an action invariant under a global symmetry. In
the context of constructing gauge theories, it is usual first to
consider an action of the form
\begin{equation}
S = \int L(\vpsi,\partial_\mu\vpsi)\,d^4x,
\label{eq:globalmatteraction}
\end{equation}
where the Lagrangian density ${\cal L}$ and the Lagrangian $L$
coincide and depend only on a set of matter fields $\vpsi(x) =
\{\vpsi_i(x)\}$ and their first derivatives. Moreover, we will
consider only the case for which the action of the global symmetry on
the coordinates and fields can be realised linearly.

In this case, the coordinate transformation
and the resulting form variations of the fields that leave the action
invariant can be written as, respectively,
\be
\xi^\mu(x) = \lambda^j \xi_j^\mu(x),\qquad
\delta_0\vpsi_i(x) = \lambda^j G_j\vpsi_i(x)
\ee
where $\lambda^j$ are a set of constant parameters, $\xi_j^\mu(x)$ are
given functions and $G_j$ are the generators of the global
symmetry corresponding to the representation to which $\vpsi_i$
belongs. Note that, for each value of $j$, the parameter $\lambda^j$
typically represents a set of infinitesimal constants carrying one or
more coordinate indices; for example, if one considers global
conformal invariance, then
$\{\lambda^1,\lambda^2,\lambda^3,\lambda^4\} =
\{a^\alpha,\omega^{\alpha\beta},\rho,c^\alpha\}$.

The Noether current (\ref{eq:noethercurrent}) then takes the form
\be
J^\mu = \lambda^j\left(\mom{\mu}{i}G_j\vpsi_i + \xi_j^\mu L\right).
\label{eq:noetherjdef}
\ee
Since the parameters $\lambda^j$ are constants, the (on-shell)
conservation law $\partial_\mu J^\mu = 0$ hence leads to a separate
condition for each value of $j$, given by
\be
\partial_\mu\left(\mom{\mu}{i}G_j\vpsi_i + \xi_j^\mu L\right) = 0,
\label{eq:globalconslaws}
\ee 
which again hold up to a total divergence of any quantity that
vanishes on the boundary of the integration region of the action
(\ref{eq:globalmatteraction}).

\subsection{Local symmetries}
\label{app:b2}

We now consider an action of the form (\ref{eq:genmatteraction})
that is invariant under a local symmetry. In particular, we focus on the
(usual) case in which the form variations of the fields can be written
as
\be 
\delta_0\chi_i = \lambda^j f_{ij}(\chi,\partial\chi) +
(\partial_\mu\lambda^j) f_{ij}^\mu(\chi,\partial\chi),
\label{eq:localformvar}
\ee
where now $\lambda^j=\lambda^j(x)$ are a set of independent arbitrary
functions of position, and $f_{ij}(\chi,\partial\chi)$ and
$f_{ij}^\mu(\chi,\partial\chi)$ are two sets of given functions that,
in general, may depend on all the fields and their first derivatives.
The general form (\ref{eq:localformvar}) typically applies only when
$\chi_i = g_i$ is a gauge field, whereas if $\chi_i = \vpsi_i$ is a
matter field, then $f_{ij}(\chi,\partial\chi) = G_j\vpsi_i$ and
$f_{ij}^\mu(\chi,\partial\chi) = 0$, as for a global symmetry. By
analogy with our discussion above, for each value of $j$, the function
$\lambda^j(x)$ typically represents a set of infinitesimal functions
carrying one or more coordinate or local Lorentz frame indices; for
example, if one considers local conformal invariance, then
$\{\lambda^1(x),\lambda^2(x),\lambda^3(x),\lambda^4(x)\} =
\{a^\alpha(x),\omega^{ab}(x),\rho(x),c^a(x)\}$, where $a^\alpha(x)$ is
interpreted as an infinitesimal general coordinate transformation
(GCT) and is usually denoted instead by $\xi^\alpha(x)$.

Using the expression (\ref{eq:localformvar}), and after performing an
integration by parts, the corresponding variation of the action
(\ref{eq:genmatteraction}) is given by (suppressing functional
dependencies for brevity)
\be
\delta S = \int \lambda^j \left[f_{ij}\frac{\delta{\cal
      L}}{\delta\chi_i}-\partial_\mu\left(f_{ij}^\mu\frac{\delta{\cal
      L}}{\delta\chi_i}\right)\right] + \partial_\mu (J^\mu - S^\mu)\,d^4x,
\ee
where we define the new current $S^\mu \equiv -\lambda^j
f_{ij}^\mu \delta{\cal L}/\delta\chi_i$. Since the $\lambda^j$ are
arbitrary functions, for the action to be invariant one requires
the separate conditions
\bea
f_{ij}\frac{\delta{\cal
      L}}{\delta\chi_i}-\partial_\mu\left(f_{ij}^\mu\frac{\delta{\cal
      L}}{\delta\chi_i}\right) &=& 0,\label{eq:n2cond1}\\
\partial_\mu(J^\mu -S^\mu) & = & 0,\label{eq:n2cond2}
\eea
where the former hold for each value of $j$ separately and the latter
holds up to a total divergence of a quantity than vanishes on the
boundary of the integration region. 

The first set of conditions (\ref{eq:n2cond1}) are usually interpreted
as conservation laws, which are covariant under the local symmetry,
although not manifestly so in the form given above. The condition
(\ref{eq:n2cond2}) implies that $J^\mu = S^\mu + \partial_\nu
Q^{\nu\mu}$, where $Q^{\nu\mu} = -Q^{\mu\nu}$, so the two currents
coincide up to a total divergence. By contrast with the case of a
global symmetry, if the field equations $\delta{\cal
  L}/\delta\chi_i = 0$ are satisfied, then the conservation
laws (\ref{eq:n2cond1}) hold identically and $S^\mu$
vanishes.  Thus, the conditions (\ref{eq:n2cond1}--\ref{eq:n2cond2})
effectively contain no information on-shell, which is essentially the
content of Noether's second theorem \cite{Avery16}.

Nonetheless, the on-shell condition that {\it all} the field equations
$\delta{\cal L}/\delta\chi_i = 0$ are satisfied can only be imposed if
${\cal L}$ is the {\it total} Lagrangian density, and not if ${\cal
  L}$ corresponds only to some {\it subset} thereof (albeit one for
which the corresponding action should still be invariant under the
local symmetry).  In particular, suppose one is considering a field
theory for which the total Lagrangian density ${\cal L}_{\rm T} =
{\cal L}_{\rm M} + {\cal L}_{\rm G}$, where ${\cal L}_{\rm G}$
contains every term that depends {\it only} on the gauge fields $g_i$
and/or their derivatives, and ${\cal L}_{\rm M}$ contains all the
remaining terms. Thus, if ${\cal L} = {\cal L}_{\rm M}$, then only the
matter field equations $\delta{\cal L}/\delta\vpsi_i = 0$ can be
imposed, whereas if ${\cal L} = {\cal L}_{\rm G}$ none of the
field equations can be imposed.  In either
case, the surviving terms in (\ref{eq:n2cond1}--\ref{eq:n2cond2}) do
contain information \cite{Forger04}.




\begin{thebibliography}{99}

\bibitem{Crispim77}
J.~Crispim-Romao, A.~Ferber, P.G.O.~Freund, Nucl.~Phys.~B {\bf 126}, 429 (1977)

\bibitem{Crispim78}
J.~Crispim-Romao, Nuc.~Phys.~B {\bf 145}, 535 (1978)

\bibitem{Kaku77}
M.~Kaku, P.K.~Townsend, P.~van Nieuwenhuizen, Phys.~Lett.~B {\bf 69}, 304 (1977)

\bibitem{Kaku78}
M.~Kaku, P.K.~Townsend, P.~van Nieuwenhuizen, Phys.~Rev.~D {\bf 17},
3179 (1978)

\bibitem{Lord85}
E.A.~Lord, P.~Goswami, Pramana J.~Phys. {\bf 25}, 635 (1985)

\bibitem{Wheeler91} 
J.T.~Wheeler, Phys.~Rev.~D {\bf 44}, 1769 (1991)


\bibitem{Ivanov82a}
E.A.~Ivanov, J.~Niederle, Phys.~Rev.~D {\bf 25}, 976 (1982)


\bibitem{Ivanov82b}
E.A.~Ivanov, J.~Niederle, Phys.~Rev.~D {\bf 25}, 988 (1982)

\bibitem{Wheeler98}
J.T.~Wheeler, J.~Math.~Phys. {\bf 39}, 299 (1998)

\bibitem{Wehner99}
A.~Wehner, J.T.~Wheeler, Nuc.~Phys.~B {\bf 557}, 380 (1999)

\bibitem{Hazboun12}
J.S.~Hazboun, J.T.~Wheeler, J.~Phys.~Conf.~Ser. {\bf 360}, 012013 (2012)

\bibitem{Wheeler13}
J.T.~Wheeler, J.~Phys.~Conf.~Ser. {\bf 462}, 012059 (2013)

\bibitem{Wheeler14}
J.T.~Wheeler, Phys.~Rev.~D {\bf 90}, 025027 (2014)

\bibitem{Wheeler19}
J.T.~Wheeler, Nucl.~Phys.~B {\bf 943}, 114624 (2019)

\bibitem{Cunningham10}
E.~Cunningham, Proc.~London Math.~Soc. {\bf 8}, 77 (1910)

\bibitem{Bateman10a}
H.~Bateman, Proc.~London Math.~Soc. {\bf 8}, 223 (1910)

\bibitem{Bateman10b}
H.~Bateman, Proc.~London Math.~Soc. {\bf 8}, 469 (1910)

\bibitem{Schouten36}
J.A.~Schouten, J.~Haantjes, Kon.~Ned.~Akad.~Wet.~Proc. {\bf 39}, 1059 (1936)

\bibitem{eWGTpaper}
A.N.~Lasenby, M.P.~Hobson, J.~Math.~Phys. {\bf 57}, 092505 (2016)

\bibitem{Hehl78}
F.W.~Hehl, in {\it Proceedings of the 6th course of the International
  School of Cosmology and Gravitation}, edited by P.G.~Bergmann and 
V.~de Sabbata (Plenum, New York, 1978)

\bibitem{Lord86b}
E.A.~Lord, P.~Goswami, J.~Math.~Phys. {\bf 27}, 3051 (1986)


\bibitem{Dubrovin79}
B.~Dubrovin, S.~Novikov, A.~Fomenko, \textit{Sovremennaya Geometriya}
(Nauka, Moscow, 1979)


\bibitem{deWitt03}
B.S.~deWitt, \textit{The Global Approach to Quantum Field Theory},
(Clarendon Press, Oxford, 2003)

\bibitem{Dirac73}
P.A.M.~Dirac, Proc.~R.~Soc.~A {\bf 333}, 403 (1973)

\bibitem{Ho11}
S.-H.~Ho, R.~Jackiw, S.-Y. Pi, J.~Phys.~A: Math.~Theor. {\bf 44}, 225401 (2011)

\bibitem{Coleman71}
S.~Coleman, R.~Jackiw, Ann.~Phys. {\bf 67}, 552 (1971)

\bibitem{Blagojevic02}
M.~Blagojevic, \textit{Gravitation and Gauge Symmetries},
(IOP Publishing, Bristol, 2002)

\bibitem{Belinfante40}
F.J.~Belinfante, Physica {\bf 7}, 449 (1940)

\bibitem{Callan70}
C.~Callan, S.~Coleman, R.~Jackiw, Ann.~Phys. {\bf 59}, 42 (1970)

\bibitem{Miller68}
J.T.~Miller, G.N.~Fleming, Phys.~Rev. {\bf 174}, 1625 (1968)

\bibitem{Utiyama56}
R.~Utiyama, Phys.~Rev. {\bf 101}, 1597 (1956)

\bibitem{Sciama64}
D.~Sciama, Rev.~Mod.~Phys. {\bf 36}, 463 (1964)

\bibitem{Ivanenko83}
D.~Ivanenko, G. Sardanashvily, Phys.~Rep. {\bf 94}, 1 (1983)

\bibitem{Lord86a}
E.A.~Lord, P.~Goswami, J.~Math.~Phys. {\bf 27}, 2415 (1986)

\bibitem{Kibble61}
T.W.B.~Kibble, J.~Math.~Phys. {\bf 2}, 212 (1961)

\bibitem{Harnad76}
J.P.~Harnad, R.B.~Pettitt, J.~Math.~Phys. {\bf 17}, 1827 (1976)

\bibitem{Hehl76}
F.W.~Hehl, P.~von der Heyde, G.D.~Kerlick and J.M.~Nester,
Rev.~Mod.~Phys. {\bf 48}, 393 (1976)

\bibitem{Wiesendanger96}
C.~Wiesendanger, Class.~Quant.~Grav. {\bf 13}, 681 (1996)

\bibitem{Mukunda89}
N.~Mukunda, in {\it Gravitation, Gauge Theories and the Early
Universe}, edited by B.R.~Iyer, N.~Mukunda and C.V.~Vishveshwara
(Kluwer, Dordrecht, 1989), p.~467

\bibitem{Lasenby98}
A.N.~Lasenby., C.~Doran and S.F.~Gull,
Phil.~Trans.~R.~Soc.\ Lond.~A {\bf 356}, 487 (1998)

\bibitem{MacDowell77}
S.~MacDowell, R.~Mansouri, Phys.~Rev.~Lett. {\bf 38}, 739 (1977)

\bibitem{Neeman78a} 
Y.~Ne’eman, T.~Regge, La Rivista del Nuovo Cimento (1978-1999) 
{\bf 1(5)}, 1 (1978)

\bibitem{Neeman78b} 
Y.~Ne’eman, T.~Regge, Phys.~Lett.~B {\bf 74}, 54 (1978)


\bibitem{Padmanabhan85}
T.~Padmanabhan, Class.~Quant.~Grav. {\bf 2}, L105 (1985)


\bibitem{Mannheim06}
P.D.~Mannheim, Prog.~Part.~ Nucl.~Phys. {\bf 56}, 340 (2006)


\bibitem{Avery16}
S.G.~Avery, B.U.W.~Schwab, JHEP, {\bf 02}, 031 (2016)


\bibitem{Forger04}
M.~Forger, H.~Romer,  Annals~Phys.~{\bf 309}, 306 (2004)

\end{thebibliography}
\end{document}